# Designing for Cooperative Grain Boundary Segregation in Multicomponent Alloys

Malik Wagih[1,2], Yannick Naunheim[1]†, Tianjiao Lei[1]†, and Christopher A. Schuh[1,3]*

[1] Department of Materials Science and Engineering, Massachusetts Institute of Technology, 77 Massachusetts Avenue, Cambridge, MA 02139, USA

[2] Materials Science Division, Lawrence Livermore National Laboratory, 7000 East Avenue, Livermore, CA 94550, USA.

[3] Department of Materials Science and Engineering, Northwestern University, 2145 Sheridan Road, Evanston, IL 60208, USA

* Corresponding author. Email address: schuh@northwestern.edu
† These authors contributed equally to this work.

## Abstract

Tailoring the nanoscale distribution of chemical species at grain boundaries is a powerful method to dramatically influence the properties of polycrystalline materials. However, classical approaches to the problem have tacitly assumed that only competition is possible between solute species. In this paper, we show that solute elements can cooperate in the way they segregate to grain boundaries: in properly targeted alloys, the different chemical species cooperate to each fill complementary grain boundary sites disfavored by the other. By developing a theoretical "spectral" approach to this problem based on quantum-accurate grain boundary site distributions, we show how grain boundaries can be cooperatively alloyed, whether by depletion or enrichment. We provide machine-learned co-segregation information for over 700 ternary aluminum-based alloys, and experimentally validate the concept in one ternary alloy where co-segregation is not expected by prior models, but is expected based on the cooperative model.

Most of the inorganic materials we use daily are polycrystals, to which alloying elements, or dopants, or more generally "solutes", are routinely added to optimize their properties. The boundaries between crystals in such materials, i.e., grain boundaries, are atomically disordered by nature and thus provide a variety of local atomic environments (*1*). Many of these grain boundary sites can be more favorable for the solute atoms to occupy than the ordered bulk lattice (*2*), and as a result, the solute concentration at grain boundaries can be locally amplified by orders of magnitude over the bulk (*3, 4*). Thus a very small quantity of alloying element can have a dramatically inflated influence. This phenomenon, termed grain boundary segregation, often dominates the structural and functional properties of inorganic materials, including metals (*5*), ceramics (*6, 7*), semiconductors (*8*) and superconductors (*9*).

Over more than 70 years of research (*10–13*), the collective body of knowledge on grain boundary segregation is largely confined to binary metal alloys, in which a single solute species segregates to the boundaries. Although there are many practical cases where solutes segregate in multicomponent systems (*14, 15*), the level of theoretical understanding of that more complex problem is lagging—even though most practical alloys are rarely binary. This limitation primarily stems from the very complex array of local atomic environments that exist at the grain boundaries and their interactions with solute atoms, and the fact that different solute species can interact differently with these grain boundary sites. This complex spectrum of sites and interactions is an unexplored opportunity for future alloy design and optimization. Here, we theoretically develop and experimentally validate a new mode of "cooperative" co-segregation that is only possible to leverage with a full spectral view of segregation at each site (*16–18*).

The classical approach to segregation—in the vein of McLean-style isotherms (*10, 12, 19*)—treats the complex grain boundary network as a single entity, and as a result, reduces all grain boundary sites to a single site-type. This simplification ignores the fact that grain boundaries have a wide variety of local atomic environments that interact with solute atoms with a wide range of energetics that can range from attractive to repulsive; even for a single solute species, the grain boundaries have a spectrum of segregation energies (*2, 16, 17, 20–24*). Leaving this spectrum out of the classical treatment removes an important physical factor for multi-solute systems: *whether different solute species compete for the same grain boundary sites or target different ones.*

In fact, by virtue of its single site-type limitation, the classical approach comprises a physically incorrect tacit assumption: it presumes that different solute species compete for exactly the same grain boundary sites. Co-segregation, if it happens, can only be explained in terms of favorable solute-solute interactions that lead the two species to co-exist at the boundaries (*12, 25–27*). We elaborate on this point more in Fig. 1, where we show the predicted segregation behavior for the simplest hypothetical case of a multi-solute system: a ternary A(B, C) alloy with no solute-solute interactions. We solve for the equilibrium state at 5 K by performing Monte Carlo simulations (which allow us to capture both entropic and enthalpic effects) for a hypothetical polycrystal, in which the fraction of grain boundary sites, $f^{gb}$, is equal to the total solute content of either solute species (i.e., the system has enough grain boundary sites to fully accommodate either solute species when present in the system).

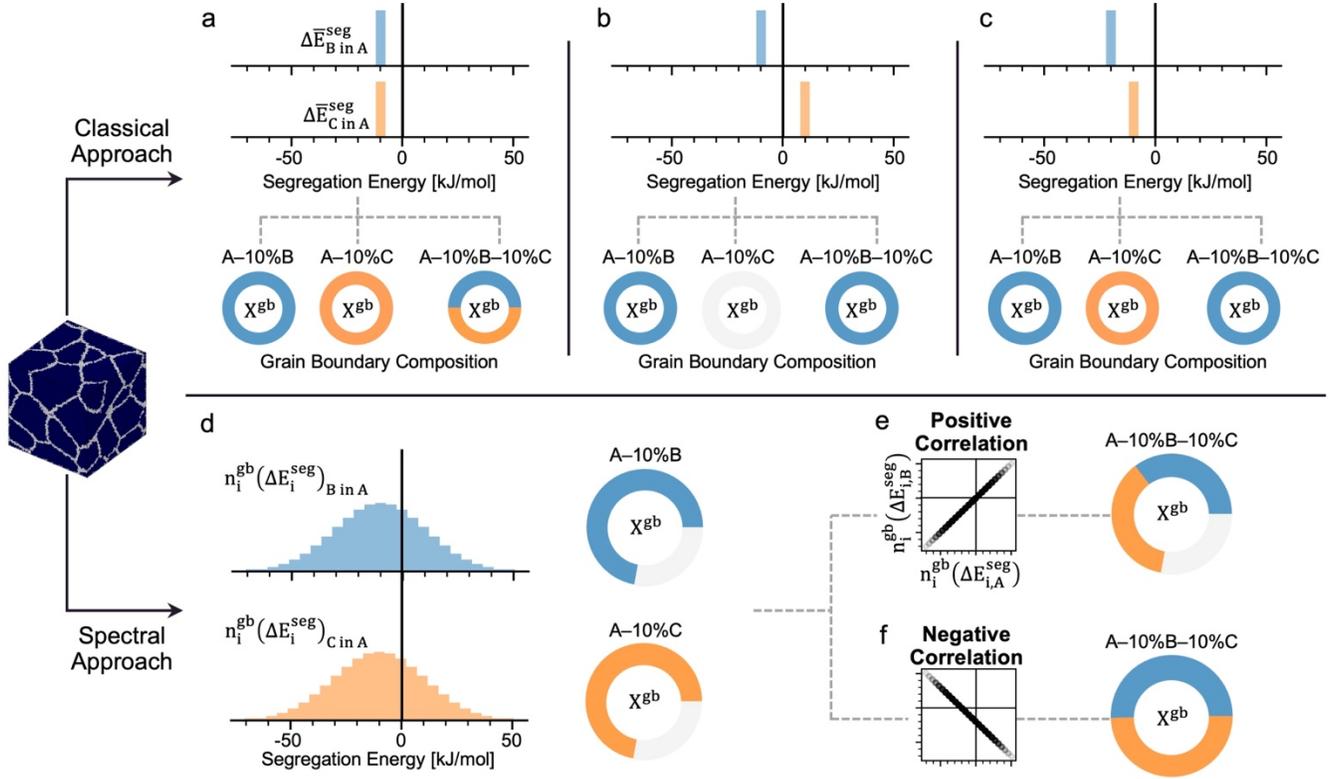

Fig. 1: Comparison between the classical and spectral approach to segregation in multi-solute alloys. For a hypothetical polycrystal with $f^{gb}$=10%, we solve for the equilibrium segregation state for binary and ternary combinations at 5K using Monte Carlo simulations for, first, the classical approach assuming (a) elements B, and C have the same segregation strength in A with $\Delta\overline{E}_{B\,in\,A}^{seg} = \Delta\overline{E}_{C\,in\,A}^{seg} = -10$ kJ/mol (b) B being a segregant with $\Delta\overline{E}_{B\,in\,A}^{seg} = -10$ kJ/mol, and C being an anti-segregant with $\Delta\overline{E}_{C\,in\,A}^{seg} = 10$ kJ/mol, (c) B being a stronger segregant than C, with $\Delta\overline{E}_{B\,in\,A}^{seg} = -20$ kJ/mol, and $\Delta\overline{E}_{C\,in\,A}^{seg} = -10$ kJ/mol; and, second, the spectral approach assuming (d) C and B having the same spectrum of segregation energies in A that is normal with a mean of –10 kJ/mol (the same as the classical average in panel (a)), and a standard deviation of 20 kJ/mol. In (e) and (f), we solve for the cases where B and C have a positive and negative site-wise correlation at the grain boundaries, respectively. $X^{gb}$ is the solute concentration at the grain boundaries.

In the classical approach, the energetic drive for a solute species to segregate at the grain boundaries is quantified by a single segregation energy, $\Delta\overline{E}^{seg}$. In Fig. 1, we show three different cases for the classical approach: first, in Fig. 1 (a), both B and C segregate in A, and have the same segregation strength $\Delta\overline{E}_{B\,in\,A}^{seg} = \Delta\overline{E}_{C\,in\,A}^{seg} = -10$ kJ/mol; second, in Fig. 1 (b), B segregates with $\Delta\overline{E}_{B\,in\,A}^{seg} = -10$ kJ/mol, but C is a mild anti-segregant with $\Delta\overline{E}_{C\,in\,A}^{seg} = 10$ kJ/mol; and, third, in Fig. 1 (c), both B and C segregate, but B has a stronger segregation drive, $\Delta\overline{E}_{B\,in\,A}^{seg} = -20$ kJ/mol, versus $\Delta\overline{E}_{C\,in\,A}^{seg} = -10$ kJ/mol. In the first case, B and C have the potential to saturate the whole grain boundary network, as seen in their binaries, A–10%B and A–10%C. Thus, when combined in ternary, A–10%B–10%C, they compete for the same grain boundary sites, and since their segregation strength is similar, they end up with equal compositions in the boundaries. In the second case, since C is a weak anti-segregant, none of it ends up at the boundaries either in its binary or ternary with B. In the third case, since B is a slightly stronger segregant than C, it dominates C completely in the ternary and pushes it out of the boundaries—a different outcome from the first case where they shared the boundary sites. The limitation of the classical single site-type representation is clear here: co-segregation, if possible, leads to direct competition for the same sites (*12*, *25*, *26*), with grain boundary composition determined by the relative magnitudes of $\Delta\overline{E}^{seg}$ for the different solute species.

The spectral approach, on the other hand, removes this implicit assumption of competition and allows for new, more physically relevant modes of co-segregation. In Fig. 1(d), we consider a hypothetical grain boundary network that has a normal distribution of segregation energies with a standard deviation of 20 kJ/mol around the same average as the classical case in Fig. 1(a) at –10 kJ/mol. Though the distribution has the same average value as the classical one, spectrality leads to a completely different system; due to the spread in energies, we now see it is possible to have a sizable fraction of sites that are enthalpically inaccessible to segregation, i.e., $\Delta E_i^{seg} > 0$ kJ/mol (where the subscript i refers to site-type i). In this case, the fraction of inaccessible sites is ~28%. If we assume that solutes B and C have the same exact distribution of segregation energies in A, $n_i(\Delta E_i^{seg})_{B\ in\ A} = n_i(\Delta E_i^{seg})_{C\ in\ A}$, we see in Fig. 1(d) that in their binaries with A, both solutes are unable to access those unfavorable grain boundary sites. However, when combined into a ternary, there is a multitude of outcomes depending on the site-wise correlation between the two solutes.

To illustrate, we consider two scenarios: First, solutes B and C have a one-to-one positive site-wise correlation, meaning that the most attractive grain boundary sites for solute B (i.e., sites that fall on the left-hand tail of the segregation spectrum in Fig. 1(d)) are also the sites with the lowest segregation energies for solute C. In this case, since both B and C have the same segregation energetics, they compete for every single accessible site in the ternary, resulting in the sharing of accessible grain boundary sites as shown in Fig. 1(e). This result is similar to the classical case of Fig. 1(a), but this time as a result of a competition between multiple site types, and with the difference that 28% of grain boundary sites remain inaccessible to segregation.

The more interesting case to which spectral segregation opens the door is depicted schematically in the second scenario, Fig. 1(f). Here we assume B and C to have a *negative* site-wise correlation; they both have the same spectrum of energetic opportunities to segregate, but they favor different sites. In other words, the grain boundary sites with the lowest segregation energies for solute B (i.e., left-hand tail of B's spectrum) are those with the highest segregation energies for solute C (i.e., right-hand tail of C's spectrum). This case portends a new mode of "cooperative segregation" in which the species largely do not need to compete for any sites and yet can both segregate to their own preferred sites. The idea of targeting complementary pairs of alloying elements is in line with many prior works in alloy design in which, for example, a 'large' and a 'small' alloying element are added to conceptually target atomic sites of different volume or strain, as applied to fault planes (*28*), grain boundaries (*29–31*) and amorphous structures (*32*, *33*). The present approach takes this concept to a quantitative extreme, and the proposed mode of cooperative segregation is impossible within the classical treatment of segregation. The spectral view of the grain boundary network explicitly enables this approach, and furthermore suggests that there is a large opportunity for materials design at the level of site-wise correlations, which have never been cataloged before. We quantitatively pursue those correlations in what follows.

Prior work has developed extensive catalogs of segregation enthalpy spectra for binary alloys ([20, 21]). We propose here that using the same basic computational workflow from those studies, and by retaining information about individual sites for multiple solutes, the site-wise correlations that dictate correlation or competition between solutes can be fully charted for many alloys (see Methods). Using the site-wise correlation between the segregation energies of two binaries, such as A(B) and A(C), we are able to reveal their mode of co-segregation in their multi-solute combination behaves in the dilute limit, such as their ternary combination A(B, C). To quantify this correlation, we use the Pearson correlation coefficient as a simple screening metric that ranges from –1 to +1. A coefficient value of +1 indicates site competition, and a coefficient value of –1 indicates anti-competition (i.e., cooperation); the two hypothetical cases highlighted in Fig. 1 (e) and (f) represent the two extreme ends of +1 to –1, respectively. We expect real alloys to fall between the two extremes.

In Fig. 2, we show the main components of our framework to compute this coefficient and screen for the co-segregation modes in ternary alloys of a base metal. For panels (a) to (c), we refer the reader to Ref. ([21]) for a detailed description of the learning framework, developed recently by Wagih and Schuh ([21]), to compute the segregation spectra for dilute binary alloys with quantum accuracy; we briefly reiterate its main features here. For a thermally annealed 20x20x20 $nm^3$ aluminum polycrystal ([34–39]), we obtain the atomic "fingerprints" of its grain boundary network using the smooth overlap of atomic positions algorithm ([40, 41]), which encodes the local atomic environment of every grain boundary site into a feature vector. Through applying dimensionality reduction ([42, 43]) and clustering techniques ([44, 45]) on this feature matrix, we are then able to identify optimal grain boundary sites that can serve as training data points for the learning algorithm. These are the sites for which the segregation energy of a particular solute species needs to be calculated. We then use a hybrid quantum mechanics/molecular mechanics (QM/MM) algorithm ([46, 47]) to compute ([48, 49]), at the level of accuracy of density functional theory ([50–52]), the segregation energies for those sites. A fitted regression model is then used to predict the full spectrum of segregation energies for the solute species at all grain boundary sites. Using steps (a) to (b) in Fig. 2, we generated binary segregation data for dozens of elements in aluminum as illustrated in Fig. 2(c), which are presented in a prior publication ([21]). For the present work, this same procedure is employed to develop full cross-correlation data for each possible pair of elements added in ternary combinations to Al, giving a total of 780 ternary aluminum-based alloys.

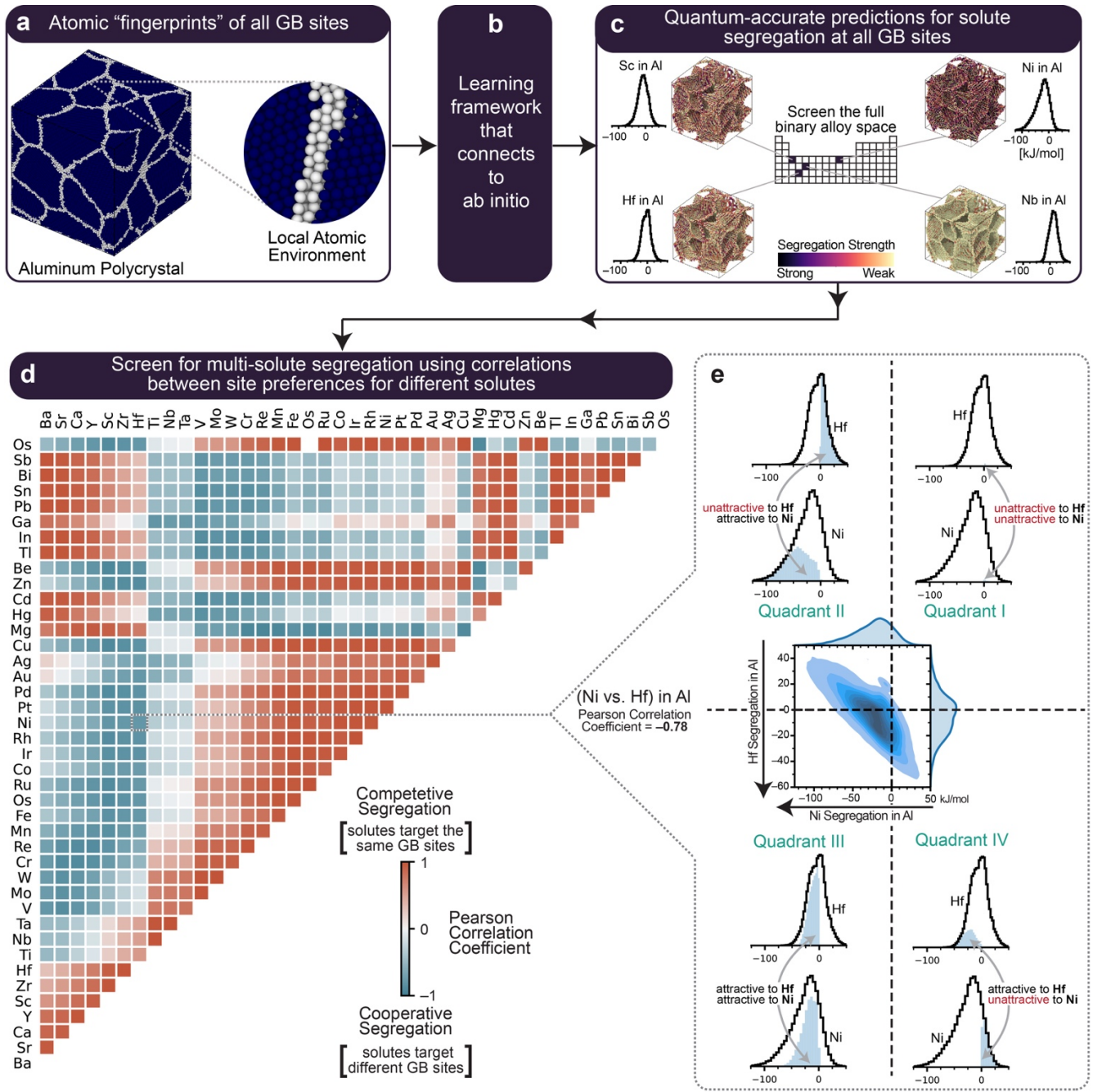

Fig. 2: A framework to map higher-order correlations for multi-solute segregation in alloys. Using aluminum as a model base metal, for (a) a thermally annealed polycrystal, we feed all the local atomic environments of its grain boundary sites into (b) the learning framework detailed in Ref. (21), which is able to learn and predict the full spectrum of segregation energies for a dilute binary solute species in the base metal with quantum accuracy. Using the binary segregation information, we are able to map the ternary correlation for their site-wise preference for segregation as quantified in (d) using the Pearson correlation coefficient. In (e) we zoom in on Al(Ni, Hf) and show the details of its site-wise correlations, which can be divided into four quadrants that detail the segregation tendencies for each element, and whether they compete or cooperate for segregation. The detailed correlation panel in (e) is plotted for all 780 ternary combinations shown in panel (d) in Data S1.

For every alloy we have computed, Fig. 2(d) provides a single relevant datapoint, i.e., the Pearson correlation coefficient for ternary segregation. Each data point in Fig. 2(d) subsumes a wealth of spectral information into a simple screening metric, as illustrated in Fig. 2(e), where we expand out details for the ternary system Al(Ni, Hf). Fig. 2(e) can be divided into four quadrants: I, both repulsive, III both attractive, II and IV, attractive to one species and repulsive to the other. The full spectrum information overlaid on these quadrants idealizes what a physically-realistic strongly cooperative segregating alloy looks like; the competitive quadrants I and especially III represent 44.8% of sites, while the cooperative quadrants II and IV are 55.2% of sites. If only Ni were present, just 84.4% of grain boundary sites would accept segregant, and if only Hf were present, 59.7%. Together, in a cooperative mode, the site occupancy can be increased to 99.6% (and the fraction of deeply favorable sites with energies less than –25 kJ/mol from 44% to 55%), with all the attendant energy reduction that implies. Since such cooperative segregation specifically offsets all the most energetic grain boundary sites, it is a pathway to far greater thermal stability of the boundaries. Thus, a multi-solute system with a negative correlation, as illustrated in Fig. 2(e), enables us to design for (i) higher magnitudes of solute segregation in a base metal, and (ii) segregated polycrystals that are more resistant to desegregation at elevated temperatures. We plot the detailed correlation figure for all 780 aluminum-based ternary alloys in Data S1.

The full value of this cooperative segregation scheme can be revealed by comparison to a hypothetical control case, in which the same alloy system is considered but the site-wise correlation is unknown. In the Al(Ni, Hf) system considered in detail in Fig. 2(e), based on their binary spectra alone, Ni appears to be a much stronger segregator than Hf. Ni's spectrum of segregation energies spans between approximately –150 and 50 kJ/mol, with only ~16% of grain boundary sites being enthalpically inaccessible to it, whereas the spectrum for Hf spans between approximately –50 and 50 kJ/mol, with 40% of sites being inaccessible to Hf. If there were no correlation amongst the segregation energies at the individual sites, the classical approach in the segregation literature would be to naively assume site competition at the grain boundaries (as in Fig. 1(a-d)). If this were the case, then only Ni should segregate in the Al(Ni, Hf) system. (It is important to note here that, according to the classical Miedema-based ($53$–$55$) values ($56$) for $\Delta\overline{E}^{seg}$, Ni and Hf are not expected to segregate at all at the grain boundaries in Al, with $\Delta\overline{E}^{seg}_{Ni \, in \, Al} \approx 16$ kJ/mol, and $\Delta\overline{E}^{seg}_{Hf \, in \, Al} \approx 45$ kJ/mol, respectively—i.e., fundamentally opposite predictions to our computed spectra in Fig. 2(e). Hf, to the best of our knowledge, has not been previously experimentally examined for segregation in Al).

We have carefully quantified these correlation effects by performing Monte Carlo simulations to predict the extent of segregation at equilibrium. The details of these computations, as well as more commentary on the results, are presented in Supplementary Text S1, but Fig. 3 summarizes the main point: the two simulation sets (classical assumption of competition vs. true spectral approach) produce completely opposite results. Whereas the classical approach sees Ni "win" the competition at all sites and completely suppress Hf segregation, the true spectral model permits their cooperation and a substantial enrichment of Hf is expected. This large distinction remains even if attractive Ni-Hf interactions are considered, as discussed in detail in Supplementary Text S2; in the classic model such attractive interactions represent the only physical mechanism to explain co-segregation, and their effect in this system is ~5 times smaller than the cooperative effect of the segregation spectra. (This disparity in segregation predictions is expected to be even stronger for systems with repulsive solute-solute interactions as shown in Supplementary Text S2).

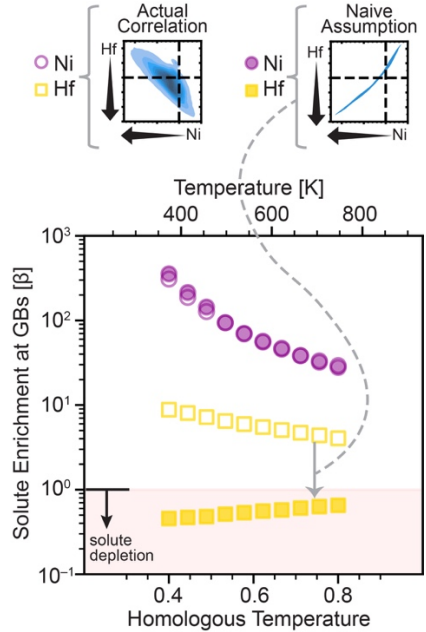

Fig. 3: The importance of site-wise correlations for predicting co-segregation behavior. For Al(Ni, Hf), we show the results of Monte Carlo simulations for two sets of simulations: (i) using the true site-wise correlation at the grain boundaries between Ni and Hf obtained for the polycrystal in Fig. 2(a), and (ii) naively assuming a positive site-wise correlation and assigning Ni and Hf segregation energies in order of increasing magnitude. For both sets of simulations, we fix the total concentration of solute atoms to be equal to the volume fraction of grain boundary sites, and use a Ni-to-Hf ratio of 1:1 in the system. We define enrichment as $\beta = X^{gb}/X^{bulk}$, where $X^{gb}$, $X^{bulk}$ are the solute concentrations at the grain boundaries, and bulk (intra-grain) lattice, respectively.

Thus, the classical competitive model and the present correlated-spectral model give diametrically opposing predictions for Hf segregation in Al(Ni, Hf). This alloy thus provides an ideal experimental test case for our proposed mechanism of "cooperative" co-segregation and the importance of site-wise correlations. The key differentiating question is: *Do Ni and Hf co-segregate in Al?* To answer this question, we fabricated a dilute Al–3 atomic % Ni–1 atomic % Hf nanocrystalline alloy by mechanically milling high purity elemental powders in an Ar atmosphere, followed by thermal annealing at 300 °C for 1 hour under a reducing atmosphere of Ar–4%H, which equilibrates the system to a grain size of less than 50 nm, allowing us to characterize multiple grain boundaries in a small volume. The combination of high homologous temperature (~60% of the solidus) and short diffusion distances to the grain boundaries in the nanocrystalline structure achieves three things. First, it relaxes the grain boundaries and their local atomic environments from the initial milled structure. Second, it enables equilibrium secondary phases, if any, to precipitate; in this system, Al₃Hf and Al₃Ni are equilibrium intermetallic phases(*57, 58*), and we do observe Ni- and Hf- rich precipitates in the bulk lattice. Third, it enables solute atoms to equilibrate across the structure and occupy their preferred sites at the grain boundaries.

In Fig. 4, we show micrographs and chemical maps for two representative grain boundaries characterized using scanning transmission electron microscopy (STEM) and energy-dispersive x-ray spectroscopy (EDS); three additional boundaries are shown in Fig. S2. Every one of these five boundaries show clear co-segregation of Hf and Ni. A closer examination of grain boundary regions using EDS line scans reveals a rich variety of segregation behaviors that range from Ni enrichment with Hf depletion, to Hf enrichment with Ni depletion, to Ni and Hf co-enrichment, as shown in Fig. 4 (c), (d), and (f), respectively.

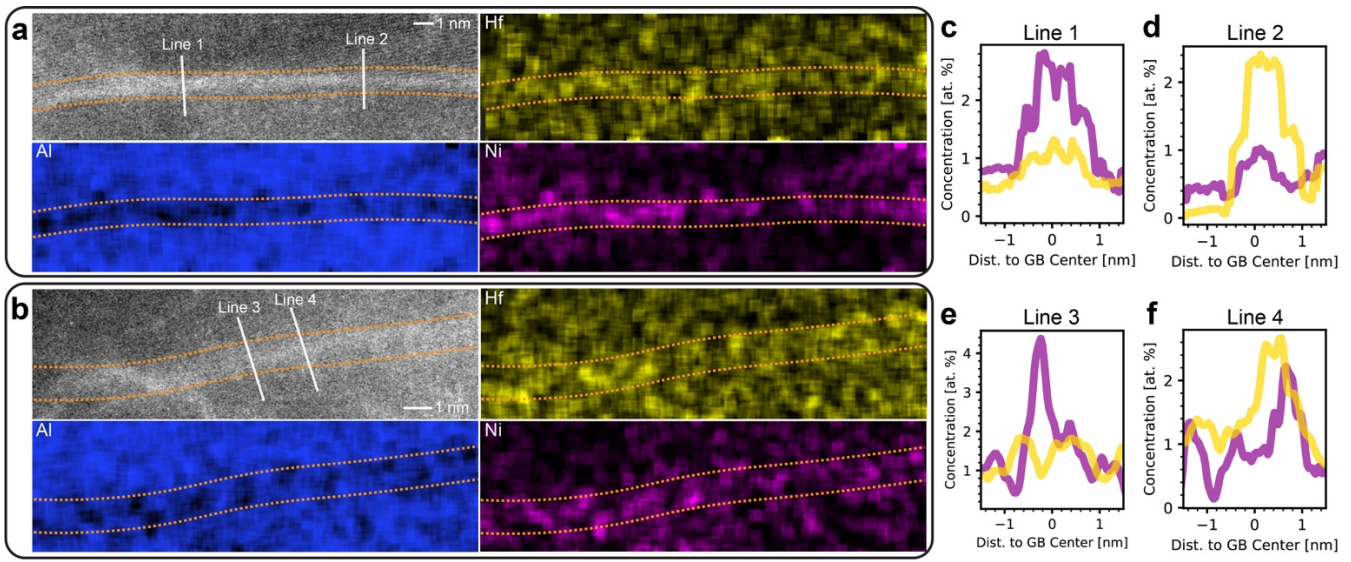

Fig. 4: Micrographs of observed enrichment for Ni and Hf at the grain boundaries in Al. In (a) and (b), we show STEM micrographs for two representative grain boundaries and EDS maps for the elements Al, Ni, and Hf. In (c)-(f), we show EDS line scans for Ni and Hf at the two boundaries. Three more representative boundaries are shown in Fig. S2.

To further probe segregation in the system, we took over 180 EDS line scans across 19 different grain boundaries. For each line scan, we measure the solute concentration at the grain boundary $X^{gb}$ as the maximum solute concentration in the grain boundary region (defined as ± 0.5 nm from the boundary center), and $X^{bulk}$ as the minimum solute concentration in the bulk region (defined as [−1.5, −0.5] nm, and [0.5, 1.5] nm from the center) of that same line scan. We denote the computed local enrichment values using this approach as $β^*$, to differentiate it from the global mean enrichment $β$. The results are shown in Fig. 5.

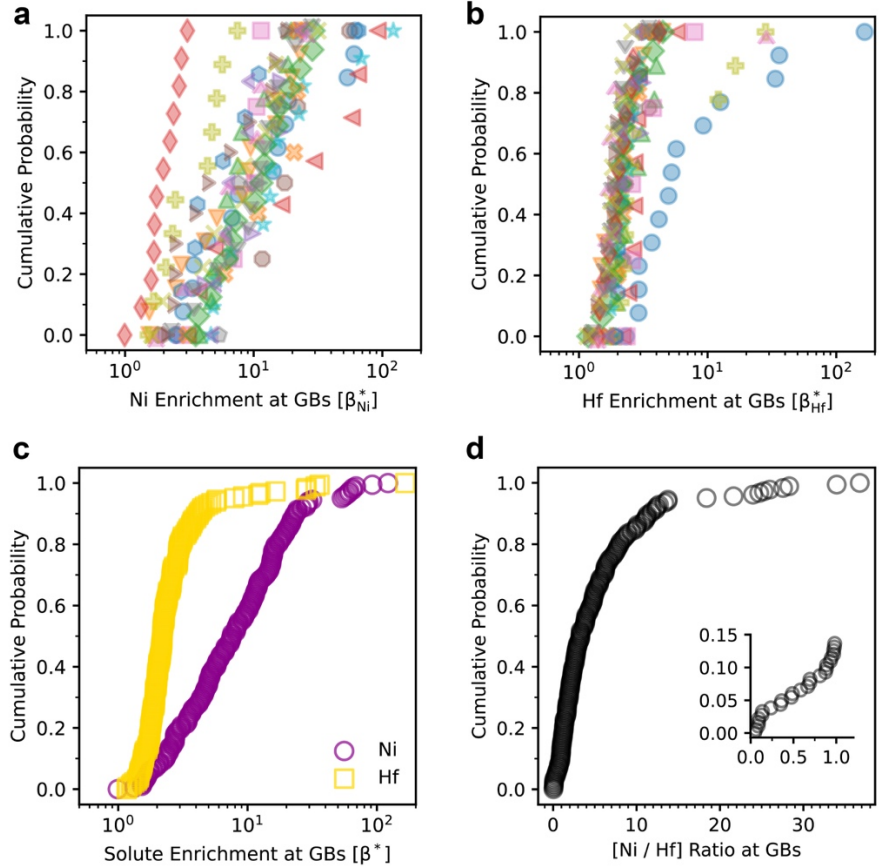

Fig. 5: Statistics of measured grain boundary enrichment for Ni and Hf in Al. In (a) and (b) we show the cumulative probability distribution for measured enrichment for Ni and Hf, respectively, using 184 line scans across 19 different grain boundaries in Al (shown in different symbols). In (c), we collapse all 184 line scans into a single distribution. (d) The distribution of observed ratio of Ni-to-Hf enrichment measured at each line scan.

There are four key experimental findings that validate our multi-solute design framework:

- First, *the spectral nature of segregation:* The observed enrichments for Ni and Hf, as shown in Fig. 5(a) and (b) for all 19 grain boundaries, exhibit a wide range of variation (*59*), not only across grain boundaries, but also within the same grain boundary itself.

- Second, *the relative magnitudes of segregation*: Both Ni and Hf co-segregate, as shown in Fig. 5(c), but Ni is the stronger segregator by a large margin—a factor of ~4x on average (i.e., at the median) and higher than 10x at the tail (Fig. 5(d)). This is consistent with the relative magnitudes of their spectra computed in Fig. 2(e), and Monte Carlo calculations in Fig. 3, which predict an average enrichment margin of a factor of ~8x.

- Third, *the wide range of co-segregation behavior:* In Fig. 5(d), we plot the ratio of Ni to Hf at the grain boundary regions. We observe a wide range of co-segregation behaviors; in some regions, the Hf content is almost an order of magnitude higher than Ni, whereas in others, the opposite is true. This observation is consistent with our prediction of a wide range of site-wise correlations across the grain boundary network (Fig. 2(e)) that range from complete favorability of Ni to the opposite.

- Fourth, *the fractions of grain boundary preference:* It is shown in Fig. 5(d) that ~15% of grain boundary regions are richer in Hf, while ~85% are richer in Ni. These percentages closely match the fraction of enthalpically favorable grain boundary sites for each solute, as shown earlier by the quadrants in Fig. 2(e).

The excellent qualitative and quantitative agreement between our computational predictions and experimental observations points directly to the site-wise nature of grain boundary segregation. It represents a key validation of the spectral segregation model, and demonstrates the power of spectra and their cross-correlations to design and control grain boundary segregation in multi-solute systems.

In conclusion, we have advanced and experimentally validated a new mode of cooperative co-segregation in multi-solute alloys. This segregation concept emerges from site-wise correlations in segregation energies that are physically intrinsic to true grain boundary environments, but have been tacitly neglected for decades in segregation theory. Our computational framework enabled efficient mapping of full polycrystalline and alloy spaces for ternary segregation with quantum accuracy, for both rapid screening and full design of multi-solute segregation in alloys, and is, in principle, easily extensible to other base metals and higher-order alloys. The experimental validation of all the important features of the model in the Al(Ni, Hf) system bolsters the quantitative value of the approach. The large database of ternary segregation correlations provided along with this article in the supplementary materials provides any researcher the same abilities in 780 ternary alloy families based on aluminum. We hope that this helps position the field to move beyond the frontier of optimizing multicomponent grain boundary chemistries for improved structural and functional properties of materials.

**Acknowledgments:**

The authors thank Dr. Yang Yu for his help with the FIB sample preparation.

**Funding:** This work was supported by the U.S. Department of Energy, Office of Basic Energy Sciences under grant number DE-SC0020180. M. Wagih acknowledges that this work was performed in part under the auspices of the U.S. Department of Energy by Lawrence Livermore National Laboratory under Contract DE-AC52-07NA27344. M. Wagih was supported by the Lawrence Fellowship at Lawrence Livermore National Laboratory, and the Laboratory Directed Research and Development (LDRD) program under project tracking code 24-ERD-049. The microscopy work was performed using MIT.nano Characterization facilities.

**Author contributions:** M.W. and C.A.S. conceived the research idea and designed the workflow. M.W. wrote all code, performed all calculations, and designed the experiments. Y.N. and M.W. fabricated the samples. T.L. performed the STEM characterization. C.A.S. provided overall guidance. All the authors analyzed the data, discussed the results, and contributed to writing and reviewing the manuscript.

**Competing interests:** The authors declare no competing interests.

**Data and materials availability:** The full segregation correlation database and all data supporting the findings of this study are available within this article and its Supplementary Materials. Additional data and code related to this work are available from the authors upon request.

**Supplementary Materials**

Materials and Methods

Supplementary Text

Figs. S1 to S2

References (*60–63*)

Data S1

# Designing for Cooperative Grain Boundary Segregation in Multicomponent Alloys

Malik Wagih[1,2], Yannick Naunheim[1], Tianjiao Lei[1], and Christopher A. Schuh[1,3]*

[1] Department of Materials Science and Engineering, Massachusetts Institute of Technology, 77 Massachusetts Avenue, Cambridge, MA 02139, USA

[2] Materials Science Division, Lawrence Livermore National Laboratory, 7000 East Avenue, Livermore, CA 94550, USA.

[3] Department of Materials Science and Engineering, Northwestern University, 2145 Sheridan Road, Evanston, IL 60208, USA

*Corresponding author: schuh@northwestern.edu

**The PDF file includes:**

Materials and Methods
Supplementary Text
Figs. S1 to S2
References (*60–63*)

**Other Supplementary Materials for this manuscript include the following:**

Data S1

## Materials and Methods

### Monte Carlo Simulations

For a given polycrystal with solute atoms, we assign bulk sites an energy of zero for solute occupation, and assign grain boundary sites the prescribed/calculated segregation energies for the solute species. A total of 100 Monte Carlo cycles are conducted for equilibration. For a ternary A(B, C) system, a Monte Carlo cycle consists of one swap per solute atom of (A,B) swaps, 1 swap per solute atom of (A,C) swaps, and 1 swap per solute atom of (B,C) swaps. Swap attempts proceed using the Metropolis criterion, with an acceptance probability that is the minimum of $[1, \exp(-\Delta E/kT)]$, where $\Delta E$ is the change in energy for the system, k is Boltzmann's constant and T is the temperature. The prescribed swap rates lead to a total of $\sim 2.5 \times 10^7$ swap attempts for simulations in Fig. 3 (main text) and Fig. S1. For simulations with solute-solute interactions (Supplementary Text S2), we use the same Monte Carlo procedure outlined, but with tracking of Ni–Hf bonds in the system. To accomplish this, we use the polycrystal in Fig. 2(a) as the input to the simulations; for each atomic site, we count the number of Ni–Hf bonds formed with its twelve nearest neighbor sites, and assign an energy change for the formation/deletion of a Ni–Hf bond (see Supplementary Text S2 for the magnitude of this interaction energy). We assume Ni–Hf bonds to have the same magnitude at the bulk and grain boundaries.

### Sample fabrication

Aluminum powder (Fisher Scientific, < 44 μm, 99.97%), Hafnium powder (Fisher Scientific, < 44 μm, 99.6%) and Nickel powder (Thomas Scientific, 3-7 μm, 99.9%) were ball milled using a high-energy SPEX 8000D mixer/mill for 20 h. Hardened steel was used for grinding jar and milling media with a ball-to-powder ratio of 10:1 and addition of 0.1 ml/g of dry ethanol as a process control agent to balance fracturing and welding of the powders during milling. The as-milled powder composition was measured to be Al – 3.2 ($\pm$ 0.3) at. % Ni – 0.8($\pm$ 0.2) at. % Hf using wavelength-dispersive X-Ray diffraction (WDS) on a JEOL JXA-8200 Superprobe EPMA (voltage of 15 kV and beam current of 10 nA). The composition and the Fe contamination of <0.1 at. % were determined on 20 spots. After ball milling, the as-milled powder was annealed at 300 °C for 1 h in a protective atmosphere containing high-purity Ar and 4% $H_2$ (with heating and cooling ramps of 10 °C/min).

### Microstructural characterization

High-annular angle dark-field (HAADF) scanning transmission electron microscopy (STEM) paired with energy dispersive X-ray spectroscopy (EDS) were used to identify location of grain boundaries and measure elemental concentrations at the grain boundaries, respectively, using a Thermo Fisher Scientific Themis Z probe aberration-corrected STEM. The microscope is operated at 200 kV and equipped with super X-4 quadrant EDS detector which offers atomic resolution elemental mapping. To prepare thin lamellae for STEM, focused ion beam (FIB) lift-out method (*60*) was used with a final polish of 5 kV and 15 pA to minimize damage from the Ga ion beam.

**Supplementary Text**

S1. Impact of site-wise correlations

In Fig. 3 in the main text, we compare the results of Monte Carlo simulations for two cases: (i) the 20 nm$^3$ polycrystal presented in Fig. 2(a), which uses the actual segregation energies for Ni and Hf, including site-wise correlations; and (ii) a hypothetical polycrystal of similar grain size, in which the grain boundary sites are naively assigned Ni and Hf segregation energies in order of increasing magnitude (i.e., assuming a positive correlation) as shown in Fig. 3. For both sets of simulations, we fix the total concentration of solute atoms to be equal to the volume fraction of grain boundary sites, and we use a Ni-to-Hf ratio of 1:1 in the system. For both sets, we plot the change in enrichment of Ni and Hf at the grain boundaries as a function of temperature; we define enrichment as $\beta = X^{gb}/X^{bulk}$, where $X^{gb}$, $X^{bulk}$ are the solute concentrations at the grain boundaries, and bulk (intra-grain) lattice, respectively.

The two simulation sets produce completely opposing results for Hf, both in magnitude and sign of segregation. In the true, correlated polycrystal, Hf co-segregates with Ni, which has a higher enrichment magnitude. The co-segregation is predictable from the site-wise correlations in Fig. 2(e), as quadrants II and IV show that the grain boundaries can accommodate both solute species without competing. And, the stronger enrichment of Ni is also predictable given the asymmetry between quadrants II and IV, as well as Ni's generally lower segregation energies that favor it to win the competition with Hf in quadrant III. The hypothetical control polycrystal, however, shows no co-segregation whatsoever—to the degree of predicting Hf to be depleted at the grain boundaries. Because Ni is the stronger segregator, the assumption of site competition leaves no room for Hf at the grain boundaries, displacing it with Ni even below its average concentration.

Leaving out the true site-wise correlations leads not only to incorrect predictions of segregation, but also to incorrect predictions of temperature-dependent physical behavior, as shown in Fig. 3, for the concentration of Hf at the grain boundaries; in the real polycrystal, it decreases with temperature, whereas in the hypothetical control polycrystal, it increases. This can be explained by the contribution of configurational entropy: as the temperature increases, entropy increases the randomization and drives the solution toward an enrichment factor of $\beta = 1$, which represents a completely random mixture. Thus, for a system that is enriched at the grain boundaries, ($\beta > 1$), this is reflected as a decrease in solute segregation with increasing temperature, i.e., entropic desegregation, which we see in both simulation sets for Ni, and only in the true, correlated polycrystal for Hf, as shown in Fig. 3. And, similarly, if a solute is depleted at the grain boundaries, ($\beta < 1$), its concentration increases at the grain boundaries with increasing temperature, as observed for Hf in the hypothetical case.

S2. Impact of solute-solute interactions in non-dilute systems

We note here, that the proposed mode of "cooperative" co-segregation differs from the classical "synergetic co-segregation" (*12, 25–27*), whereby a strongly segregating solute species enhances the segregation of another species by means of their attractive solute-solute interactions. In the classical approach, as briefly noted in the manuscript, attractive interactions are the primary mechanism for the co-existence of two species at the grain boundaries. However, although solute-solute interactions may enhance or depress the co-segregation of competing/cooperative species in non-dilute systems, it is not necessarily the main driving force for co-segregation. To illustrate, we conduct more computationally expensive Monte Carlo simulations (see Methods) that approximate the expected solute-solute interactions for the polycrystal in Fig. 2(a), by explicitly tracking Al–Hf, Al–Ni, and Ni–Hf bonds in simulations of Al(Ni,Hf). The bond (interaction) energy, $\omega$, is calculated from $\Delta H^{mix} = z\omega X(1 - X)$, where $\Delta H^{mix}$ is the mixing enthalpy, z is the coordination number (z = 12), and X is the alloy concentration. For X=0.5, the reported liquid mixing enthalpies for Al–Hf (*61*), Al–Ni (*62*), and Ni–Hf (*63*), are on the order of –30 kJ/mol, –50 kJ/mol, and –50 kJ/mol, respectively, which results in interaction energies on the order of –10 kJ/mol, –15 kJ/mol, and –15 kJ/mol, respectively. In the simulations, we assign Al–Hf and Al–Ni bonds an energy of zero, and Ni–Hf bonds an energy of –5 kJ/mol, giving it a net favorability for formation (i.e., attractive solute-solute interactions). We show the simulation results in Fig. S1(a).

Even for this attractive pair of solute species, the effect of site-wise correlations is higher by almost an order of magnitude (note the log-scale of the y-axis in Fig. S1) than the interactions alone (which elevate Hf to be slightly enriched at $\beta \approx 2$). The effect of correlation is even more pronounced in systems with repulsive solute-solute interactions, because correlations can change, not only the magnitude, but also the sign of segregation. For example, in Fig. S1(b), we re-calculate the results for the Al(Ni, Hf) system but with a hypothetical repulsive (i.e., mimicking a solute species with positive mixing enthalpies) Ni-Hf bond of the same magnitude as in Fig. S1(a), +5 kJ/mol per bond. In this case, site-wise correlations are the difference between Hf depletion and enrichment at the boundaries. The interplay between solute correlations and interactions provides a rich avenue for future exploration of segregation design in concentrated multicomponent alloys.

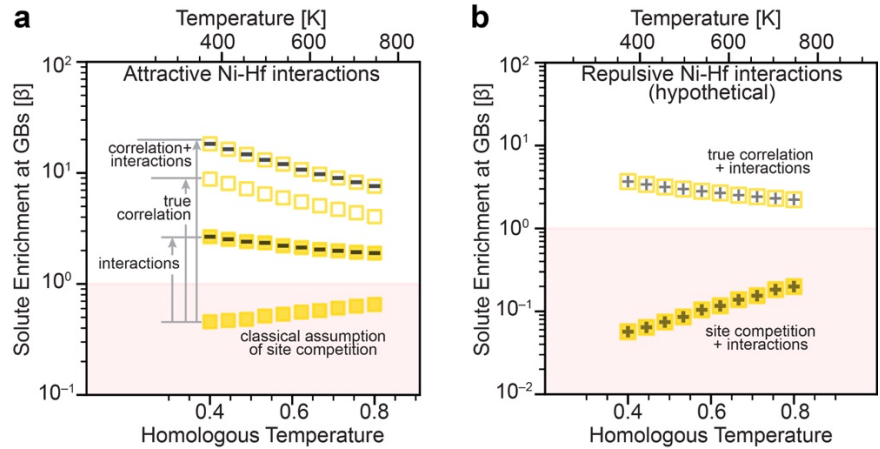

Fig. S1: The impact of solute-solute interactions and site-wise correlations on the equilibrium co-segregation state. We show the results of Monte Carlo simulations for (a) the polycrystal in Fig. 2(a) that delineate the contributions of (i) expected attractive solute-solute interactions for Ni and Hf, and (ii) the site-wise correlations at the grain boundaries for the two solutes. In (b) we show the computed equilibrium segregation state, accounting for both site-wise correlations and solute-interactions, for a hypothetical case in which Ni and Hf have repulsive interactions (of the same magnitude as (a) but with a different sign). For all simulations, we fix the total concentration of solute atoms to be equal to the volume fraction of grain boundary sites, and use a Ni-to-Hf ratio of 1:1 in the system.

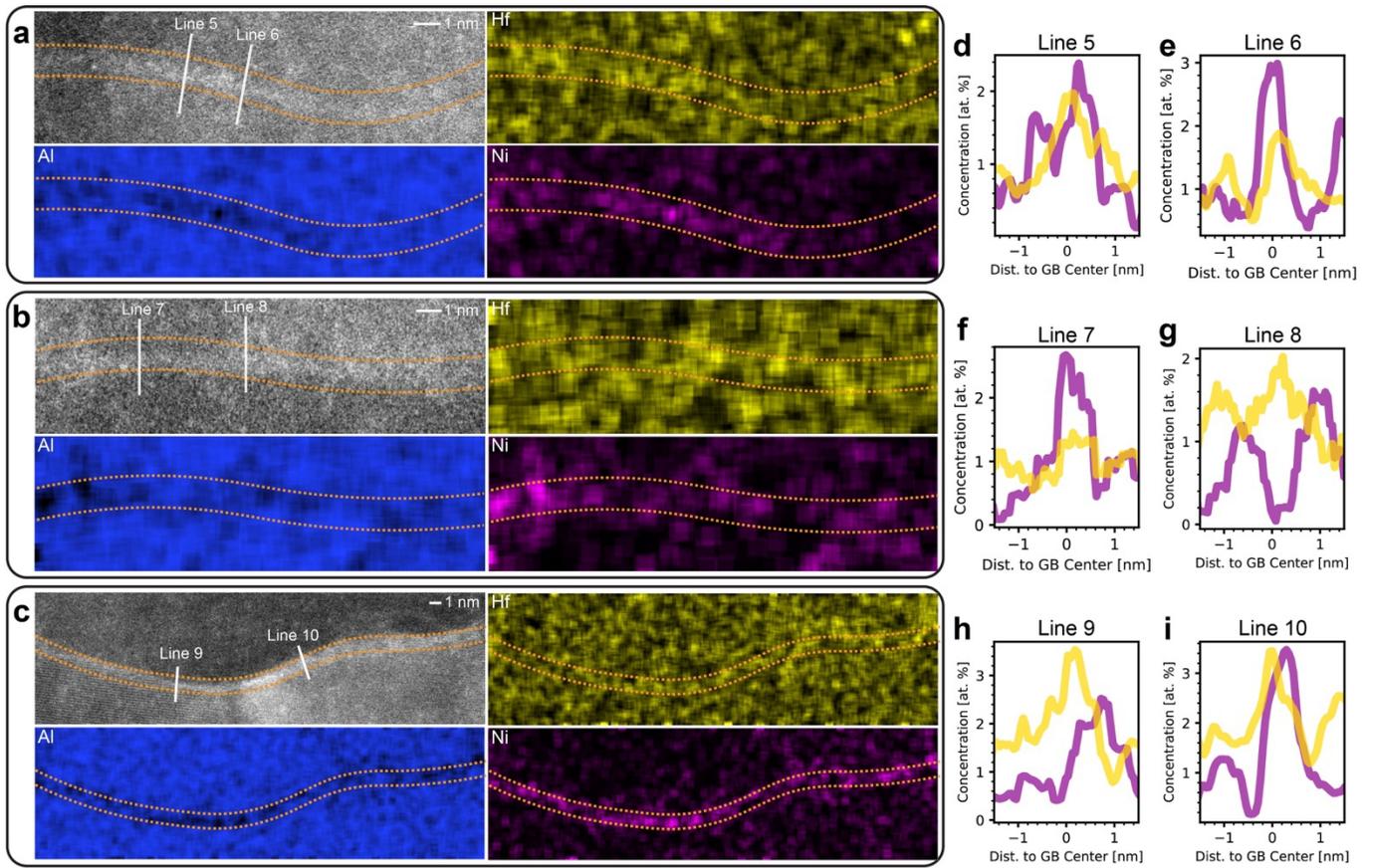

Fig. S2: Micrographs of observed enrichment for Ni and Hf at the grain boundaries in Al. In (a), (b), and (c), we show STEM micrographs for three additional representative grain boundaries and EDS maps for the elements Al, Ni, and Hf. In (d)-(i), we show EDS line scans for Ni and Hf at the three boundaries.

# Designing for Cooperative Grain Boundary Segregation in Multicomponent Alloys

Malik Wagih[1,2], Yannick Naunheim[1], Tianjiao Lei[1], and Christopher A. Schuh[1,3]*

[1] Department of Materials Science and Engineering, Massachusetts Institute of Technology, 77 Massachusetts Avenue, Cambridge, MA 02139, USA

[2] Materials Science Division, Lawrence Livermore National Laboratory, 7000 East Avenue, Livermore, CA 94550, USA.

[3] Department of Materials Science and Engineering, Northwestern University, 2145 Sheridan Road, Evanston, IL 60208, USA

*Corresponding author: schuh@northwestern.edu

**The PDF file includes:**

      Data S1

**Other Supplementary Materials for this manuscript include the following:**

      Materials and Methods
      Supplementary Text
      Figs. S1 to S2
      References

**Data S1. Site-wise correlations for multi-solute segregation in Al-based ternary alloys**

For 40 chemical elements, we present their site-wise correlations for segregation at the grain boundaries in ternary Al–based alloys. The 780 ternary combinations are presented in alphabetical order, and are not repeated, i.e., Al (Zr, Ag) is found under Al (Ag, Zr).

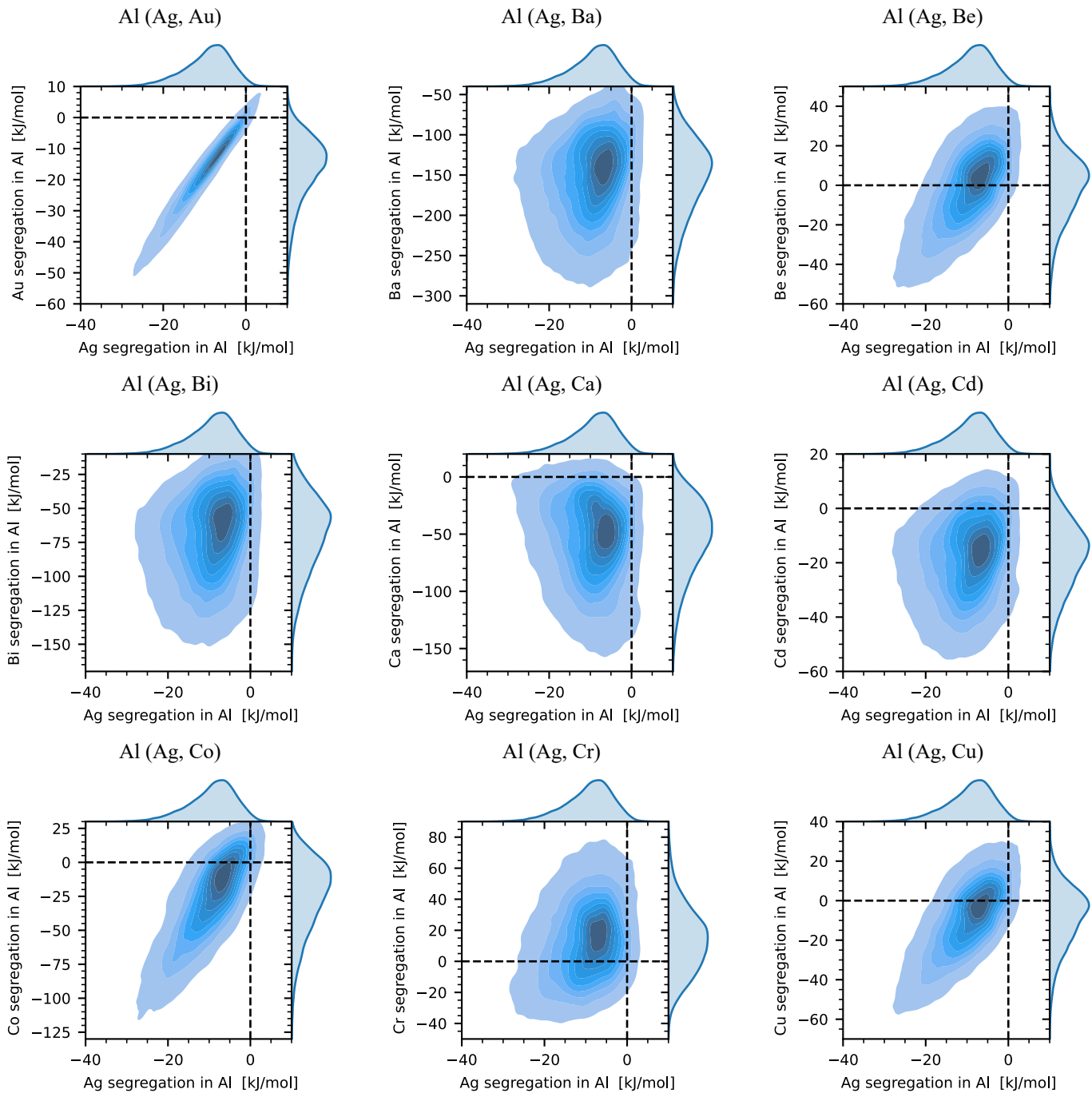

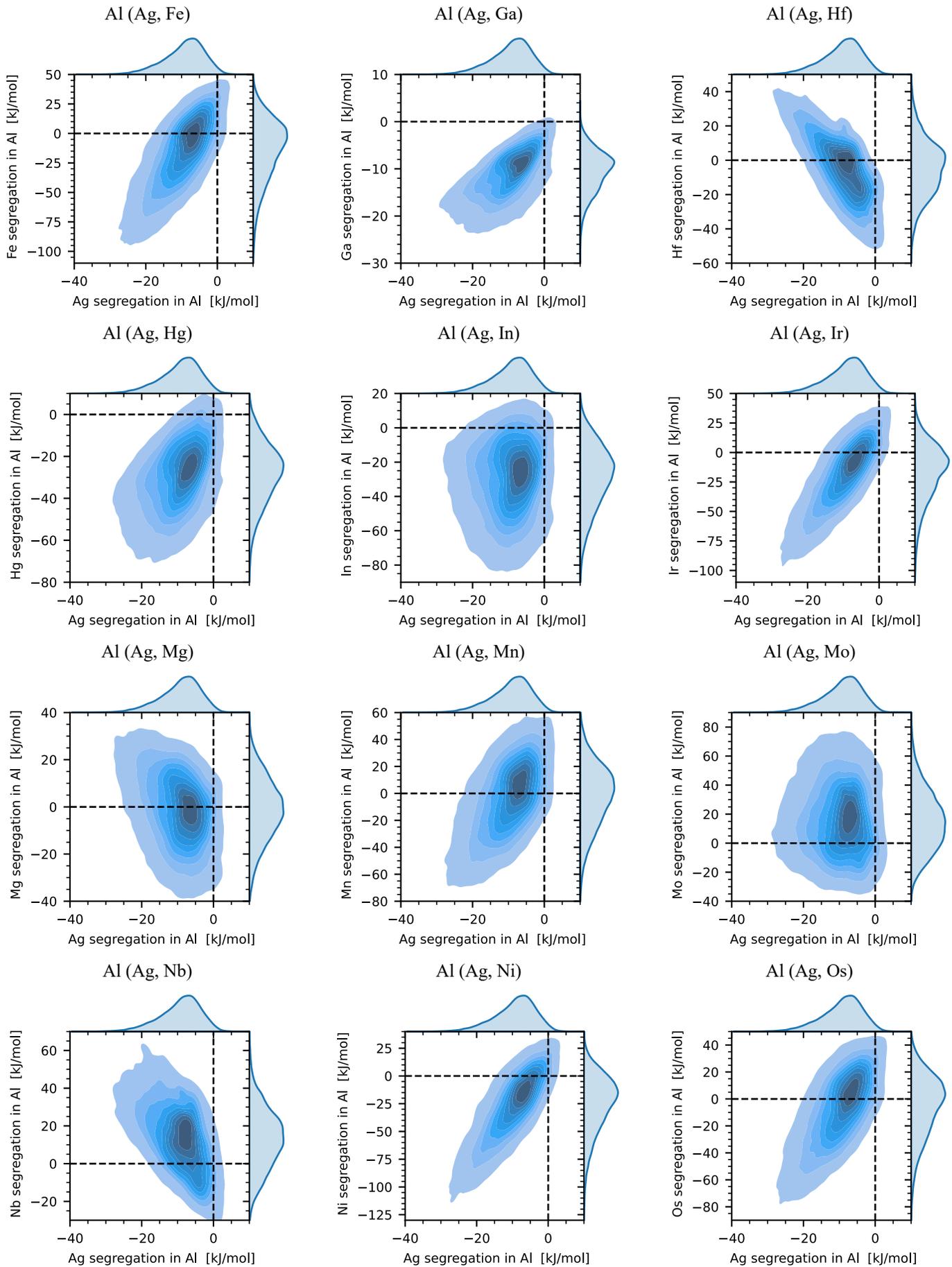

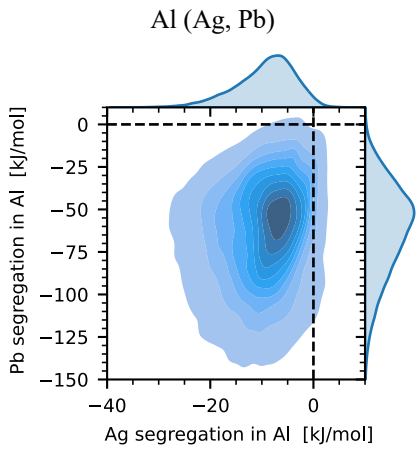

Al (Ag, Pb)

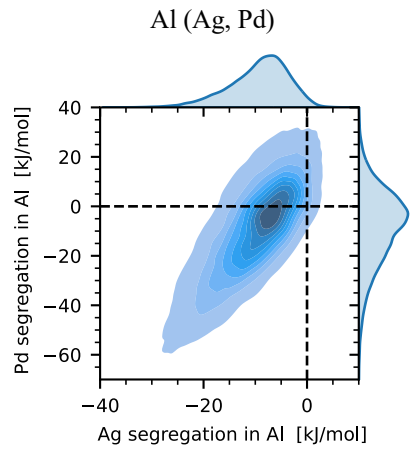

Al (Ag, Pd)

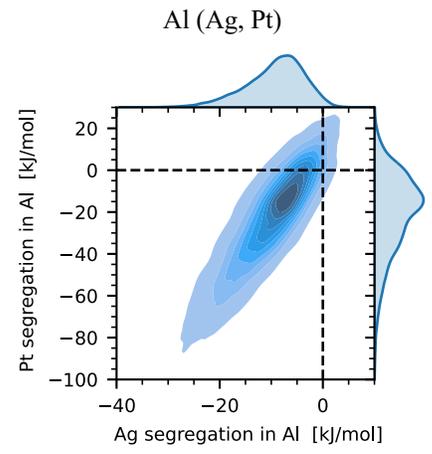

Al (Ag, Pt)

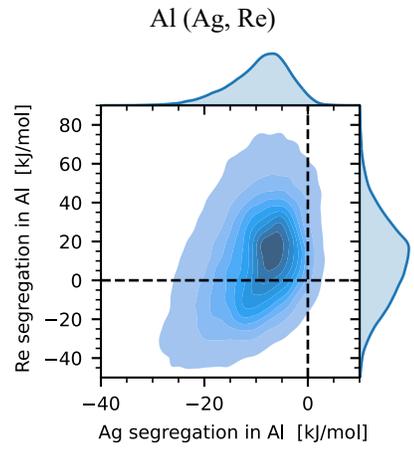

Al (Ag, Re)

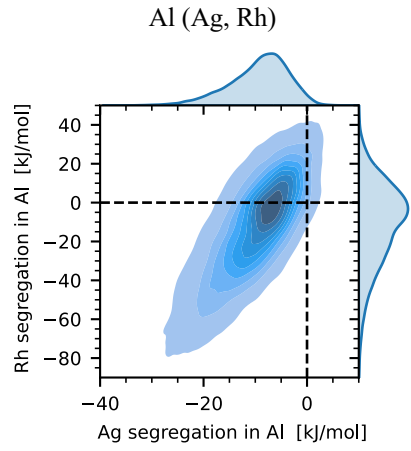

Al (Ag, Rh)

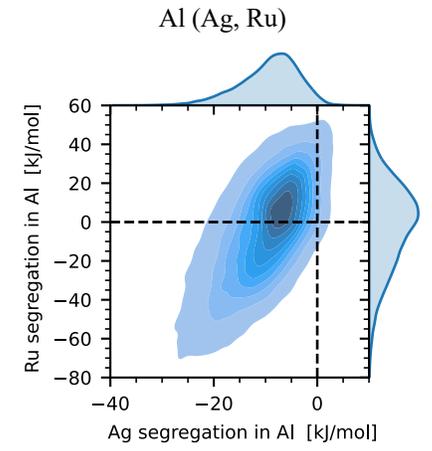

Al (Ag, Ru)

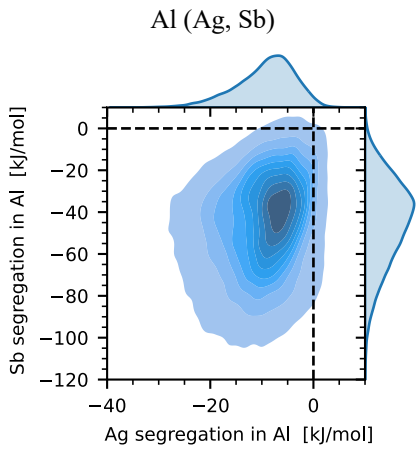

Al (Ag, Sb)

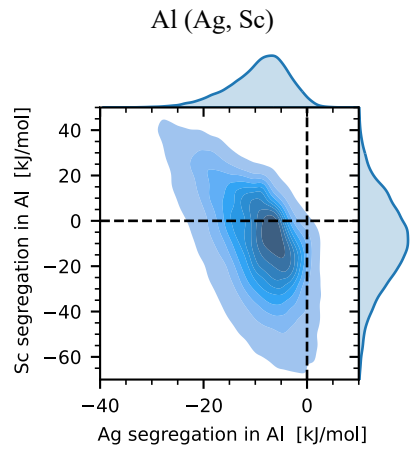

Al (Ag, Sc)

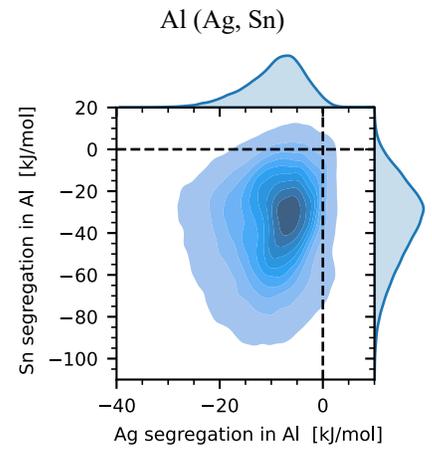

Al (Ag, Sn)

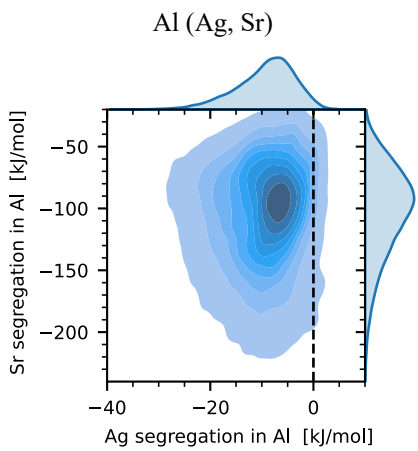

Al (Ag, Sr)

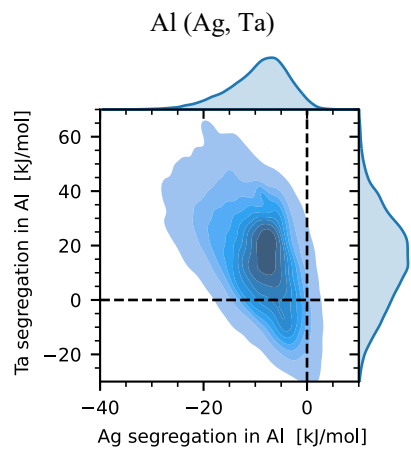

Al (Ag, Ta)

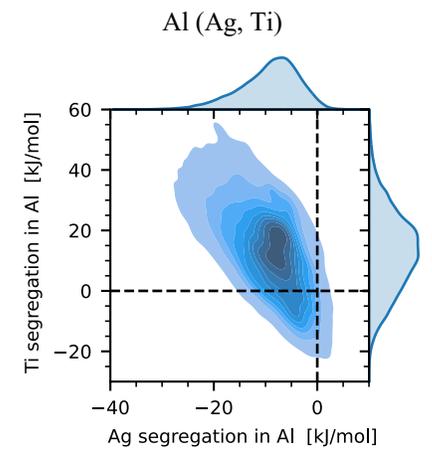

Al (Ag, Ti)

Page 4 of 67

## Al (Ag, Tl)

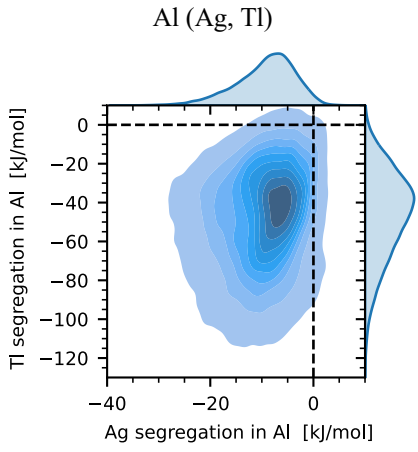

## Al (Ag, V)

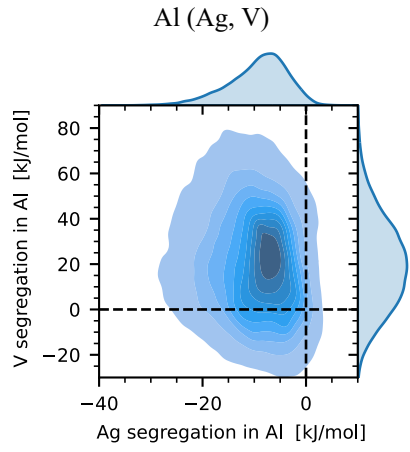

## Al (Ag, W)

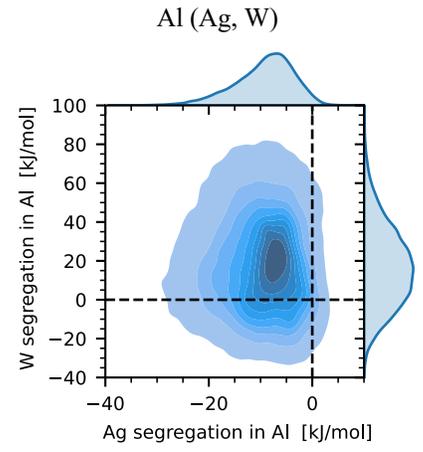

## Al (Ag, Y)

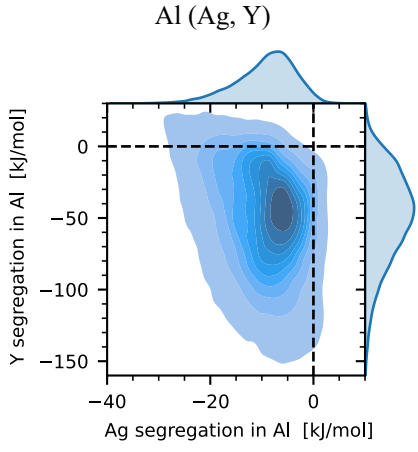

## Al (Ag, Zn)

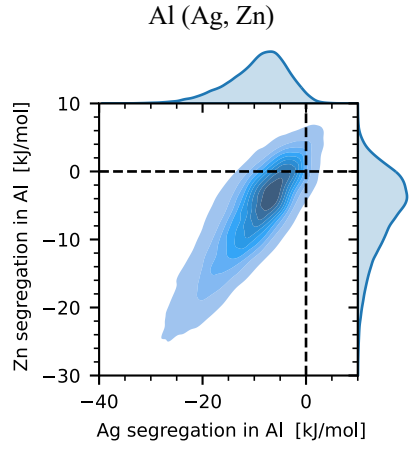

## Al (Ag, Zr)

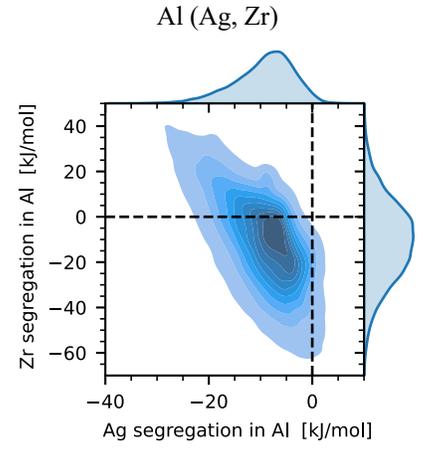

## Al (Au, Ba)

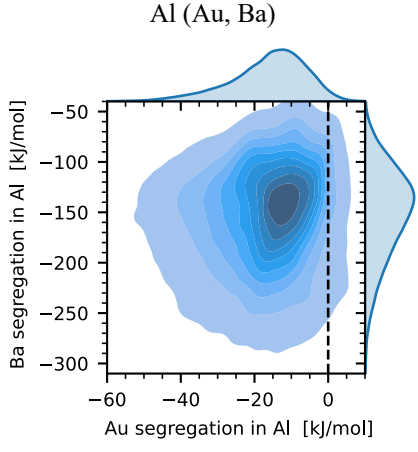

## Al (Au, Be)

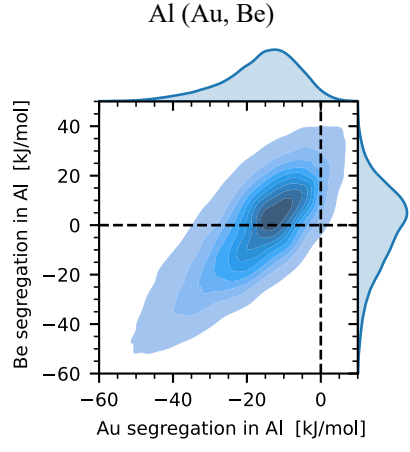

## Al (Au, Bi)

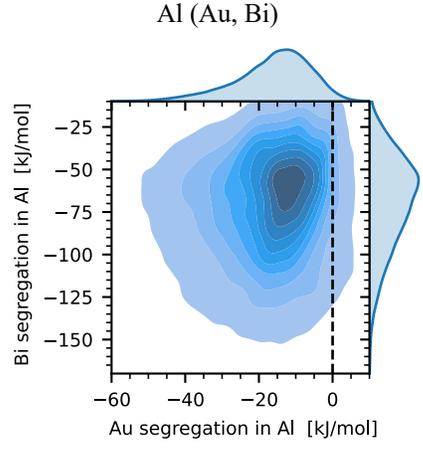

## Al (Au, Ca)

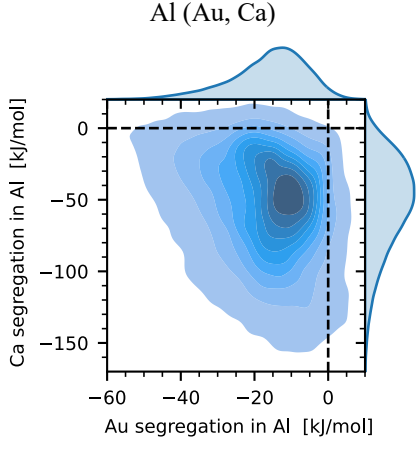

## Al (Au, Cd)

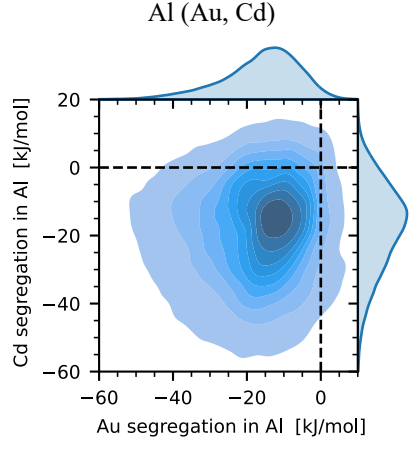

## Al (Au, Co)

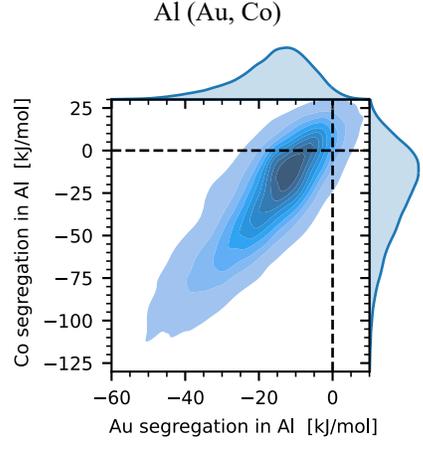

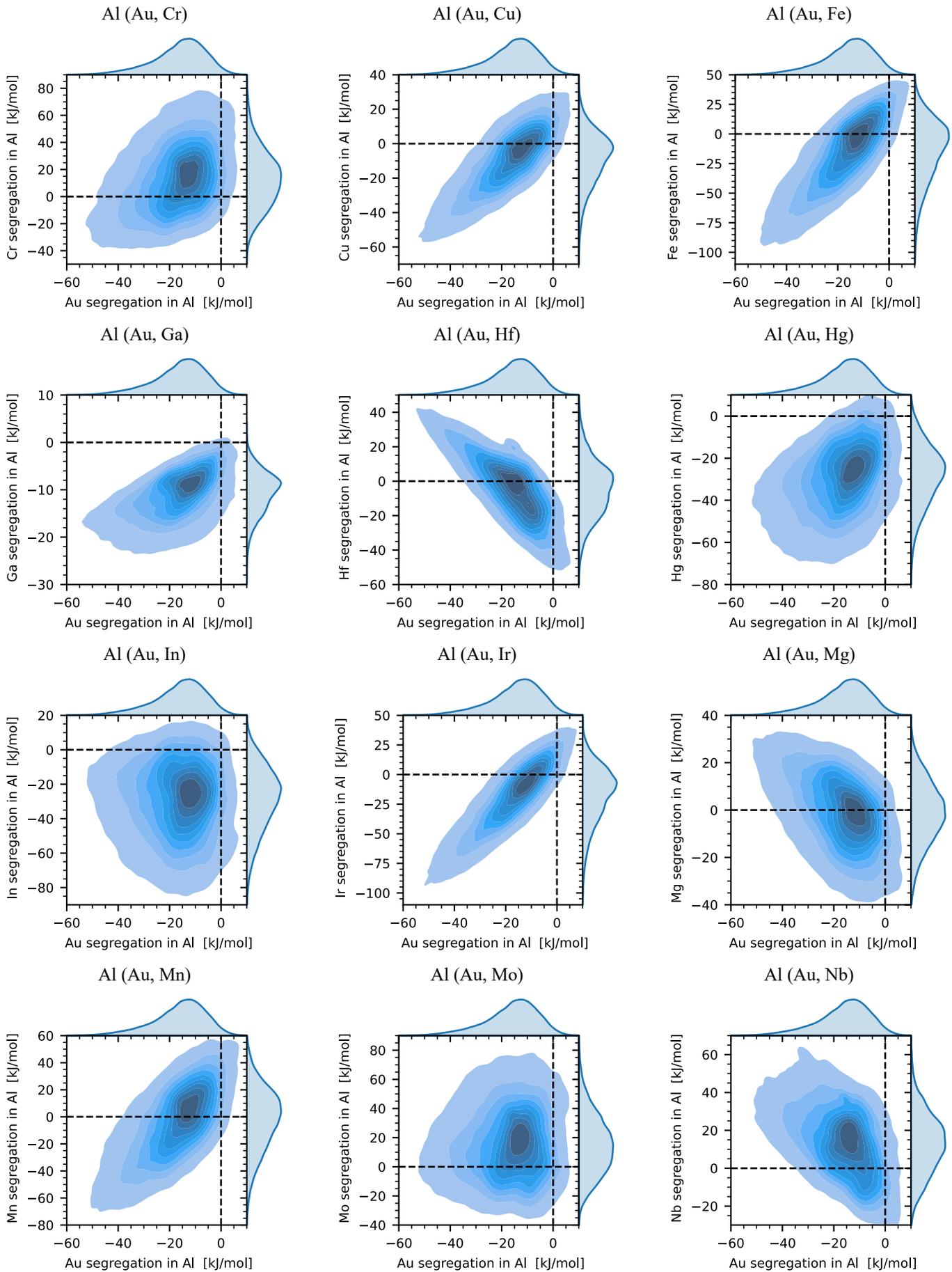

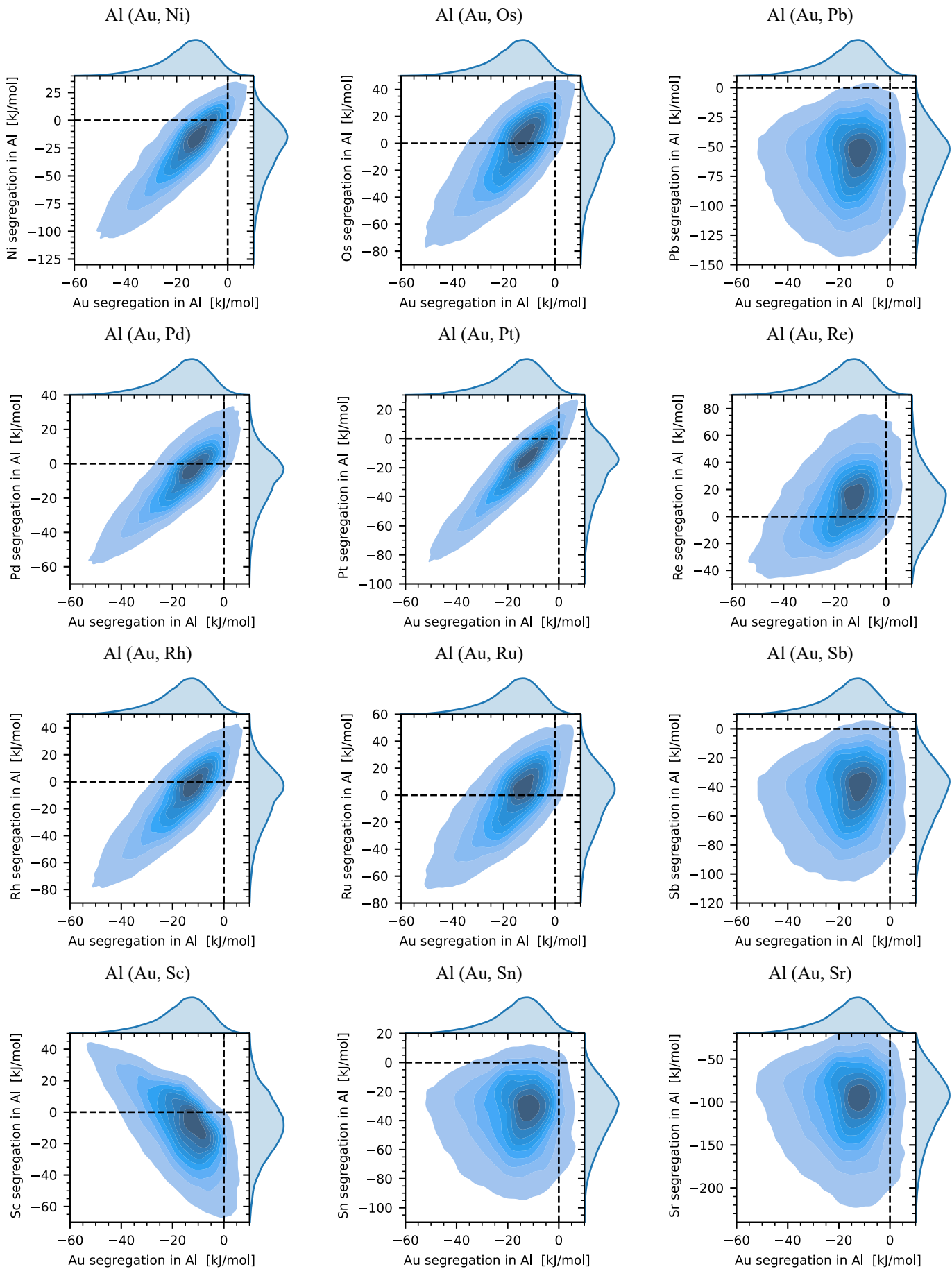

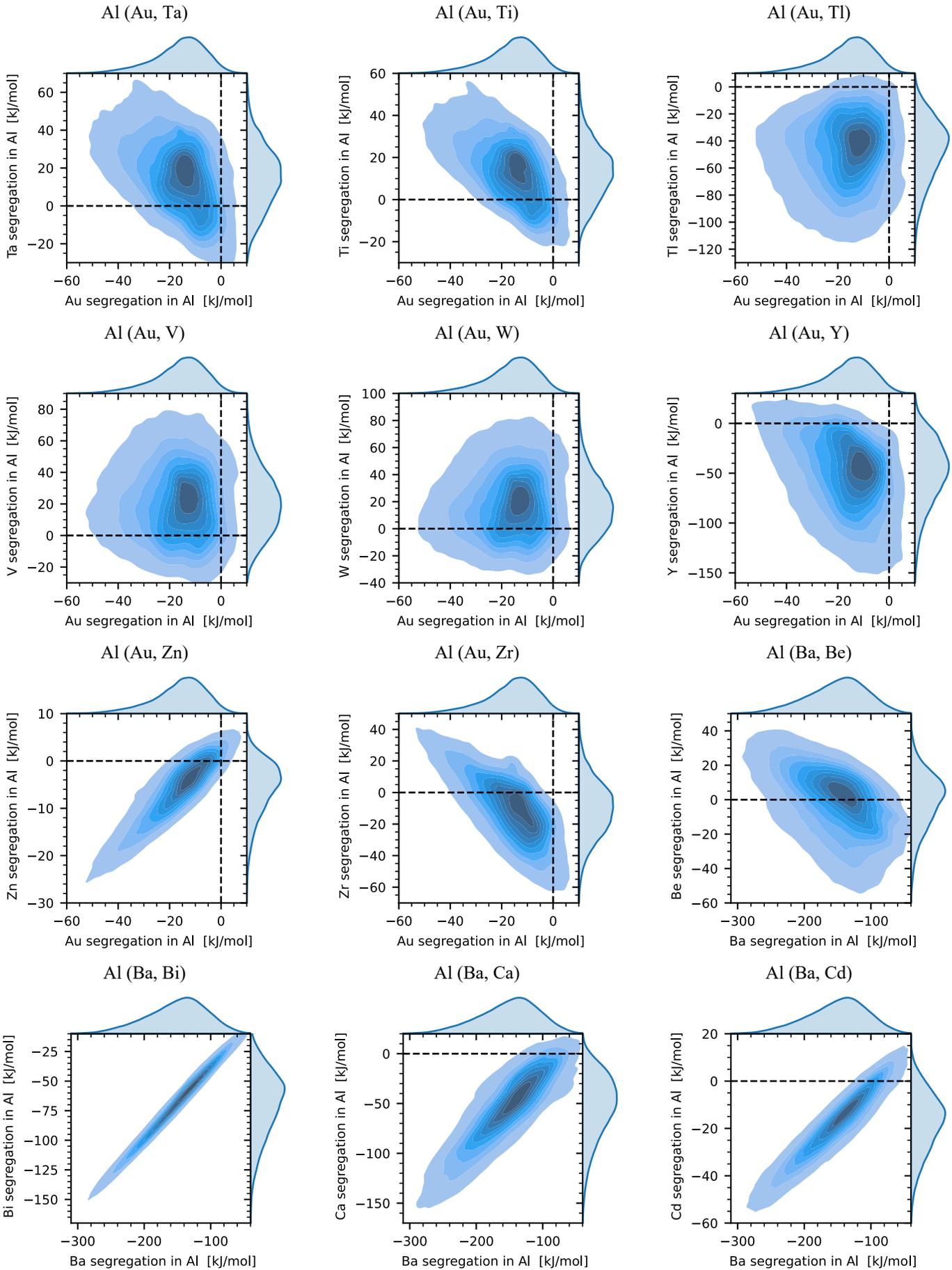

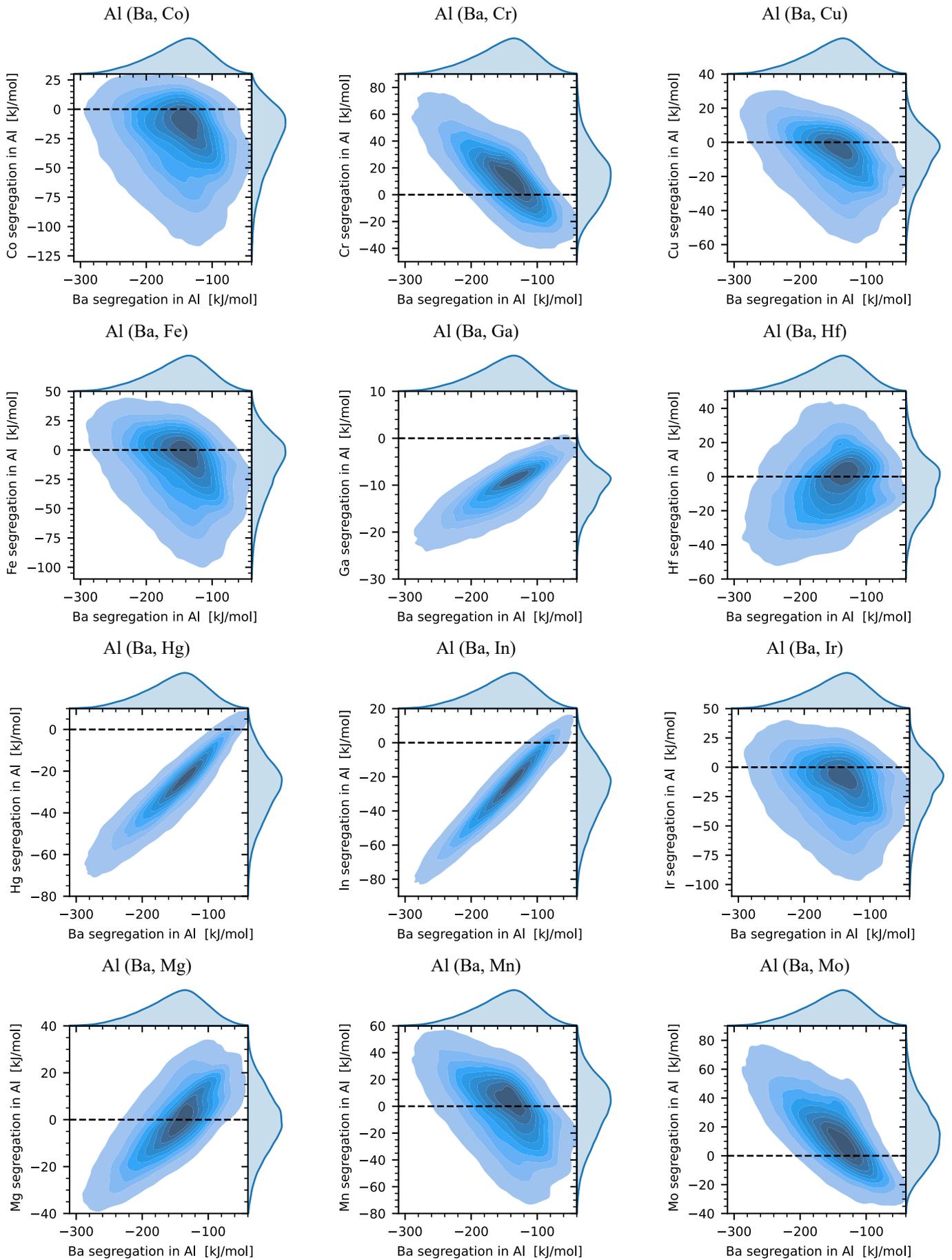

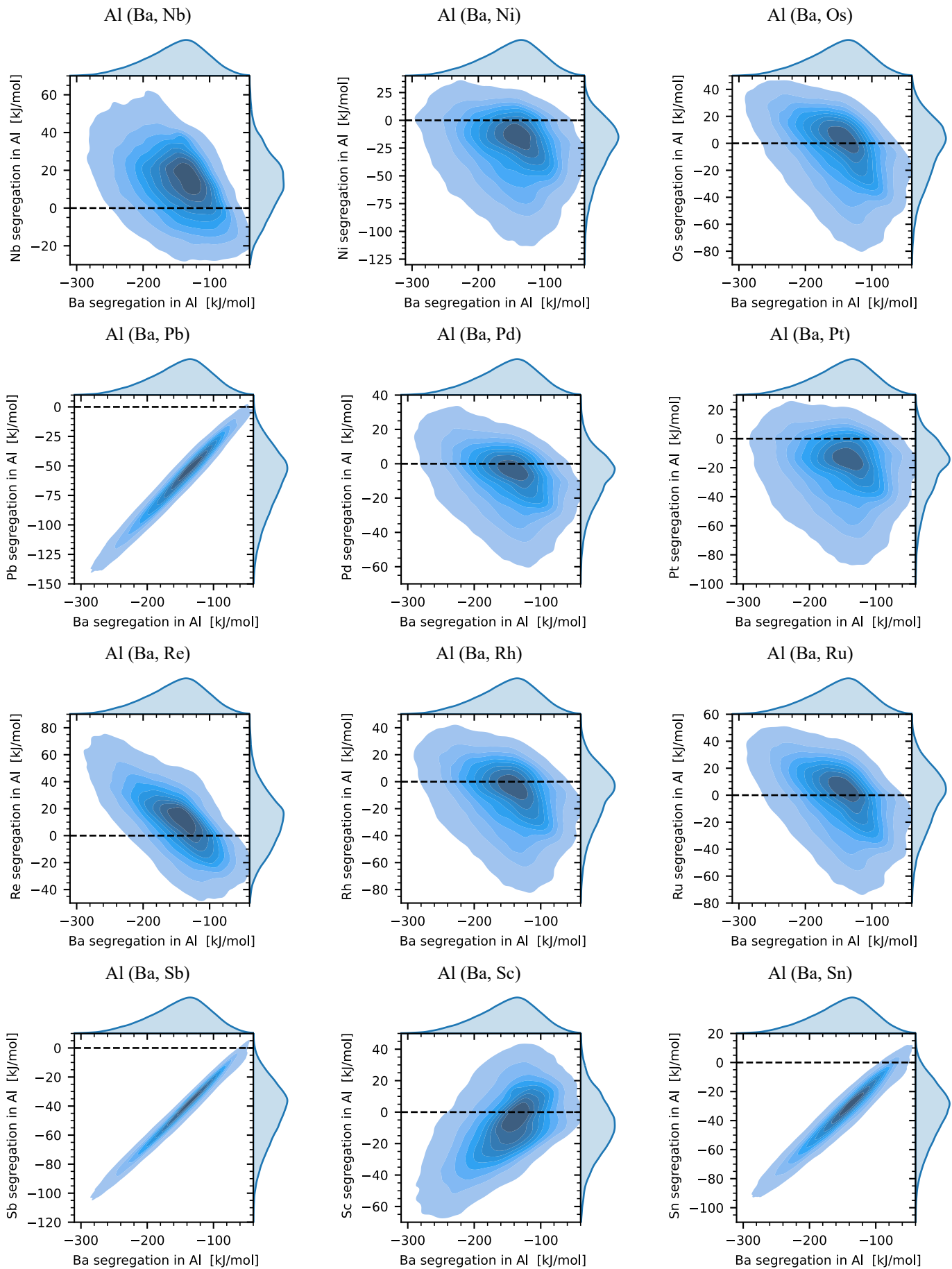

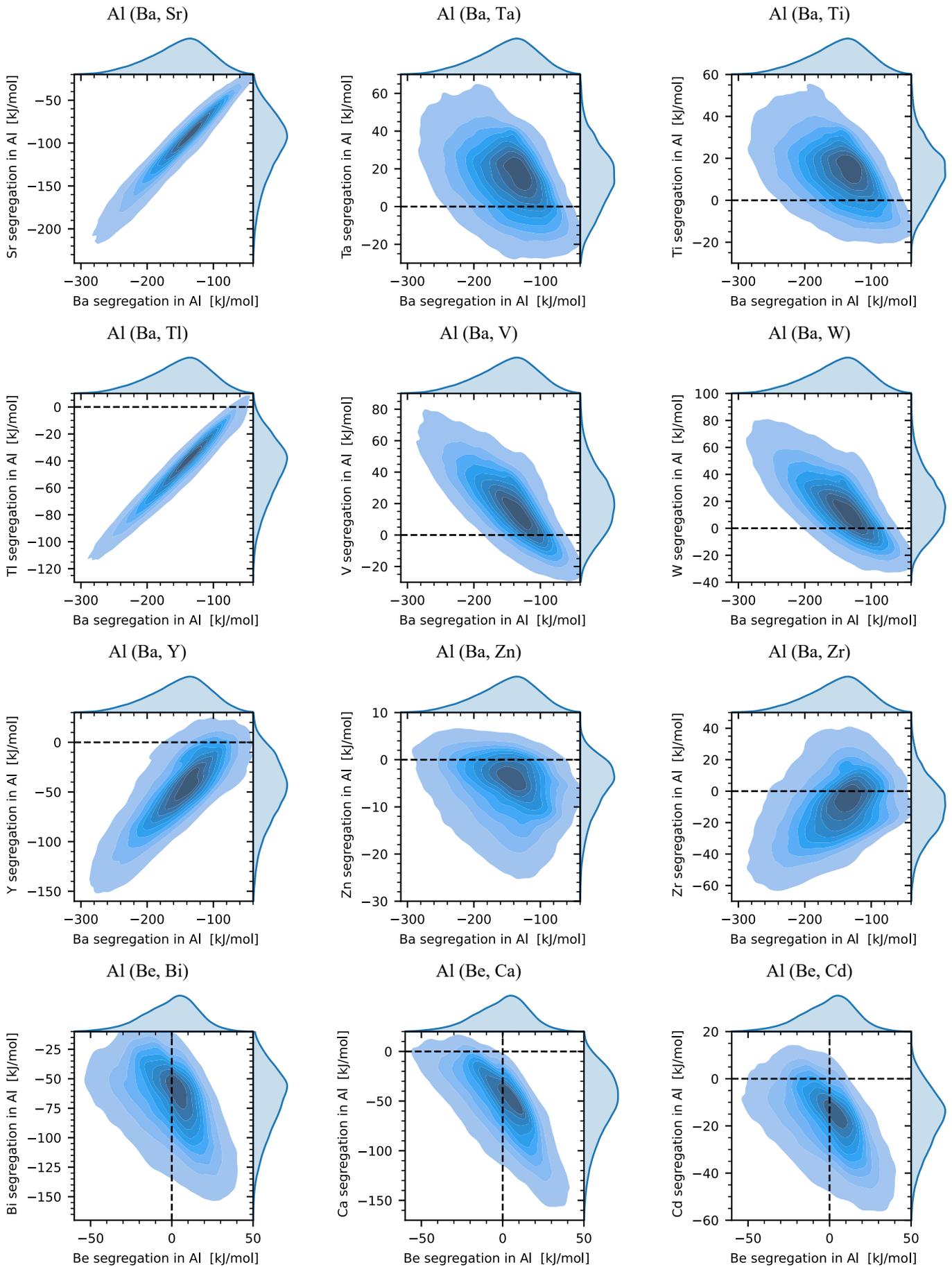

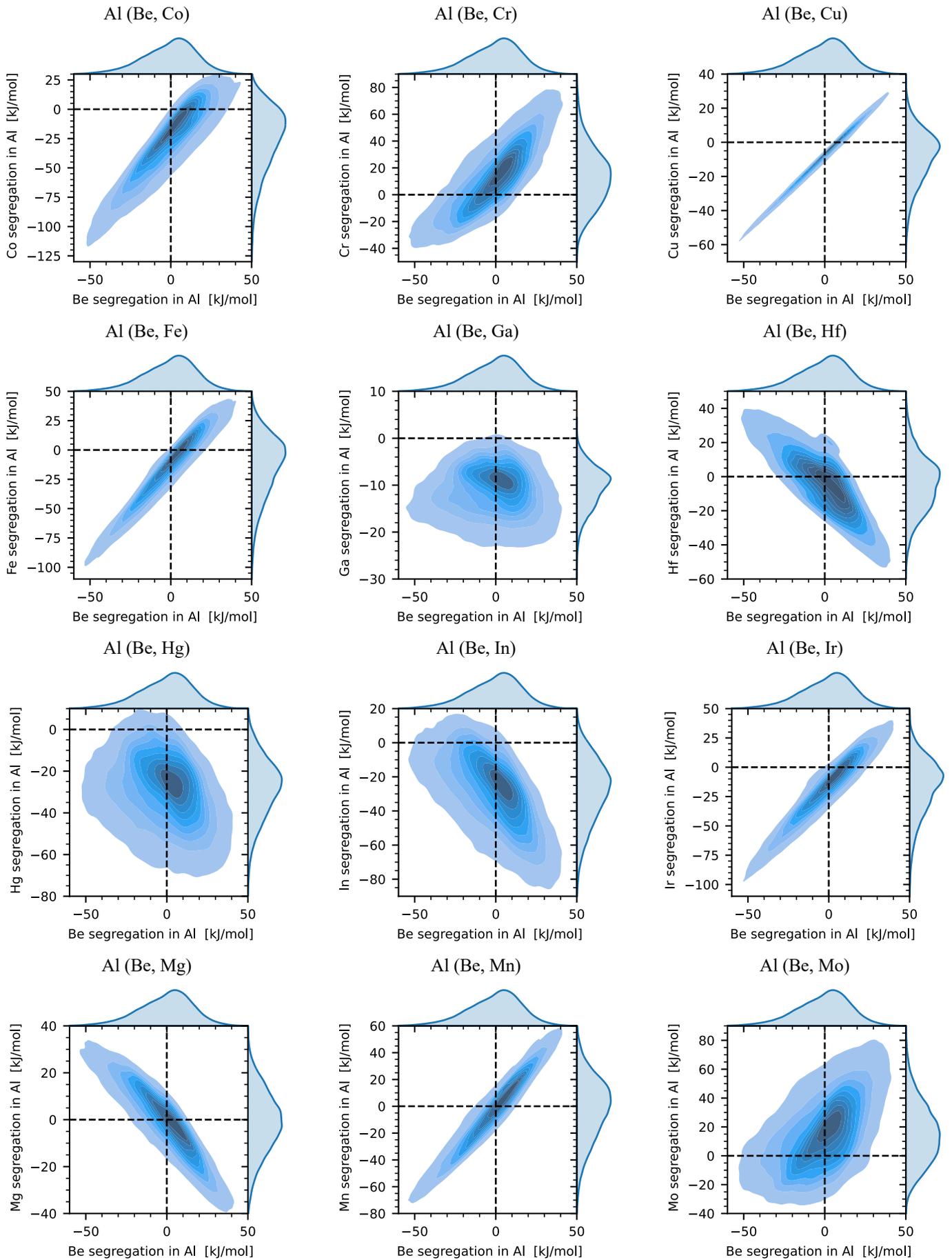

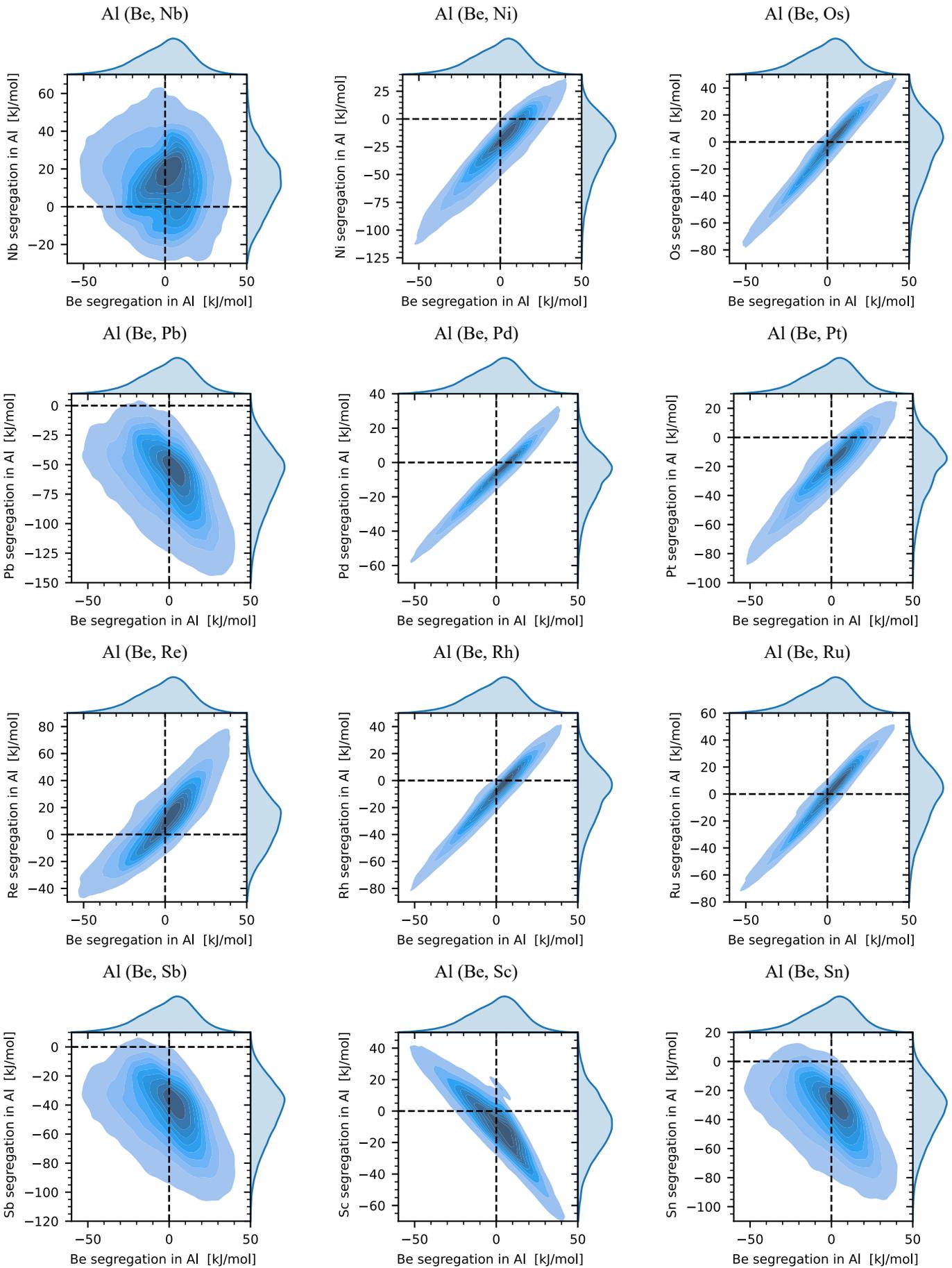

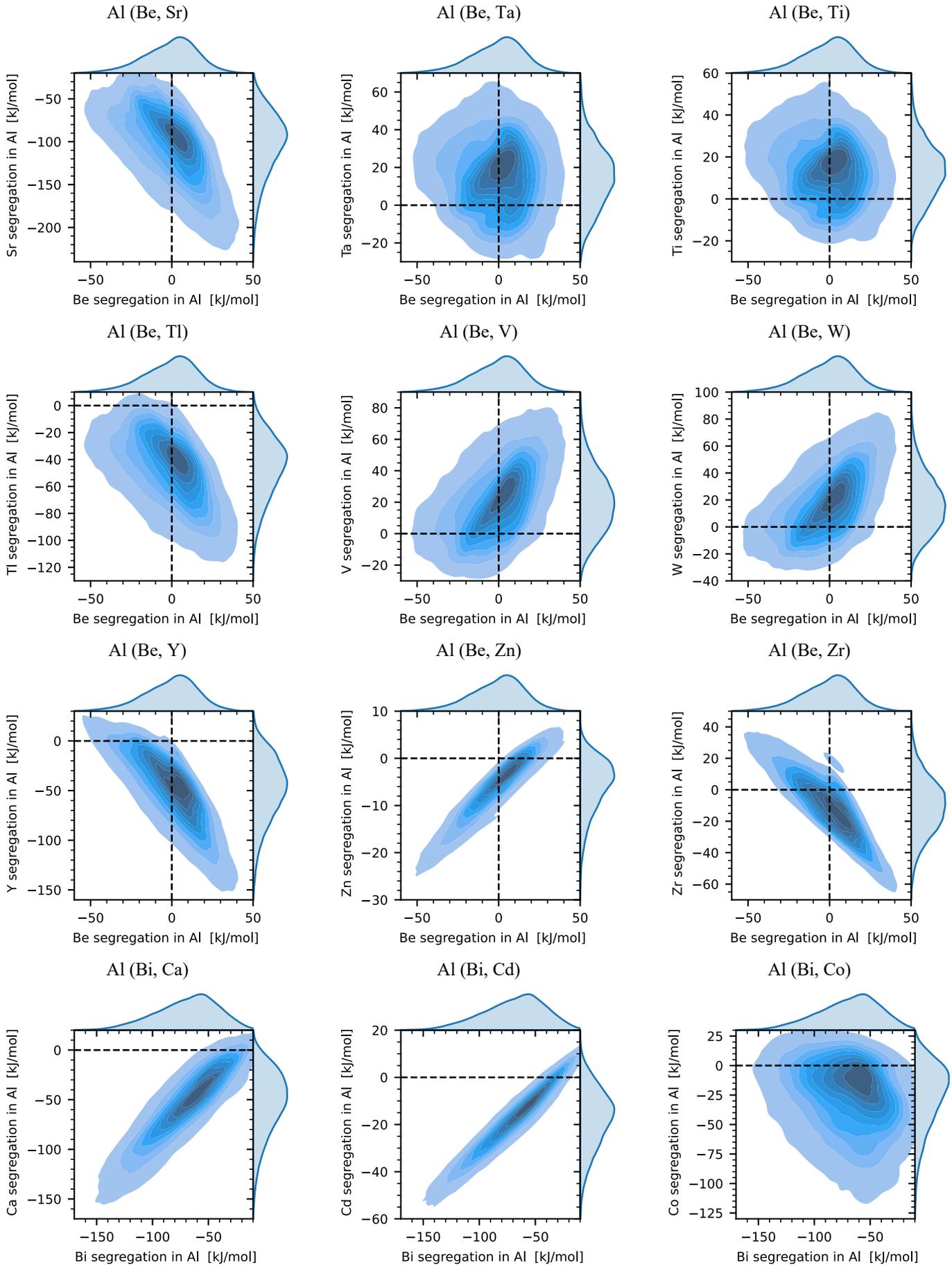

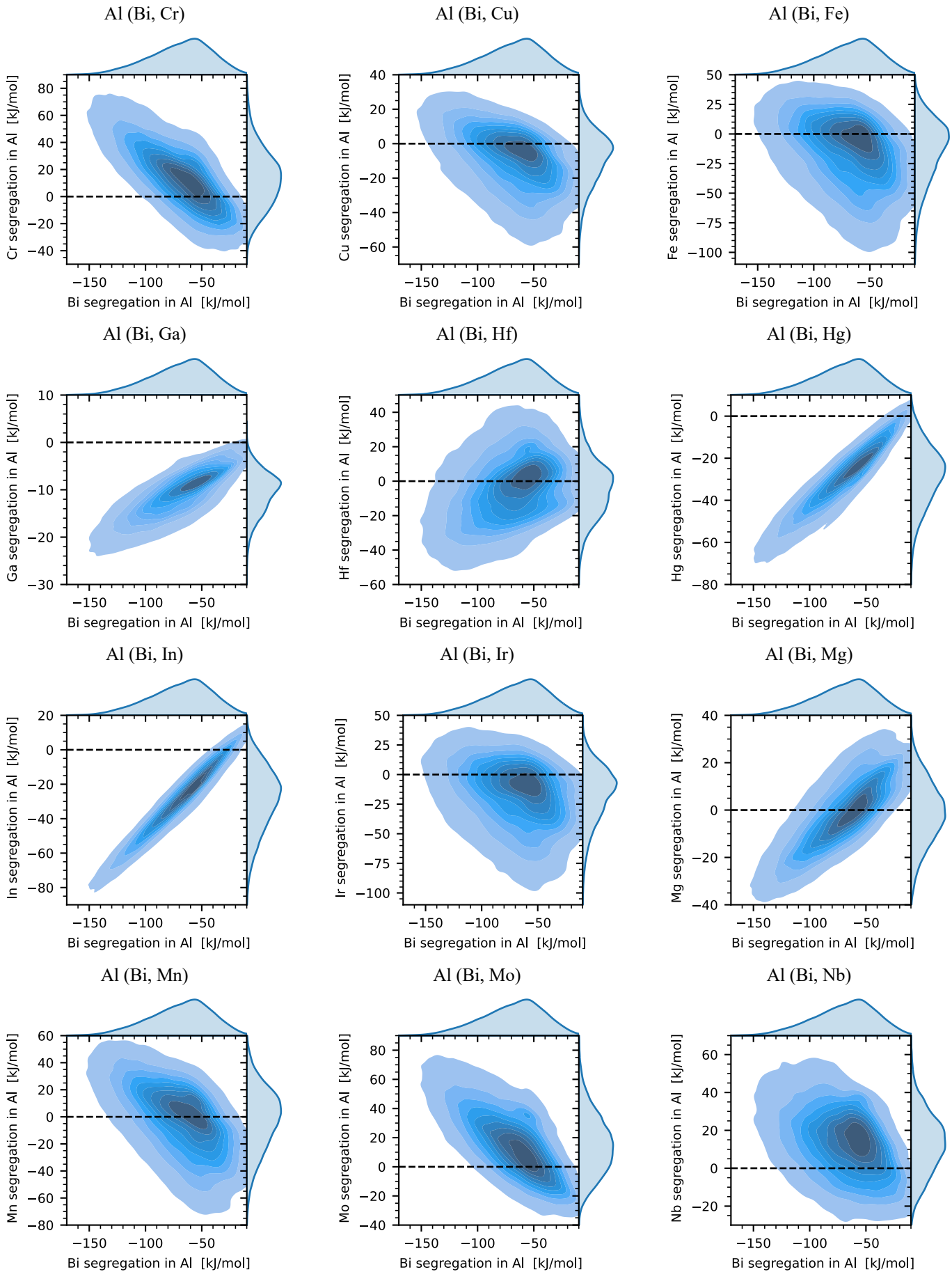

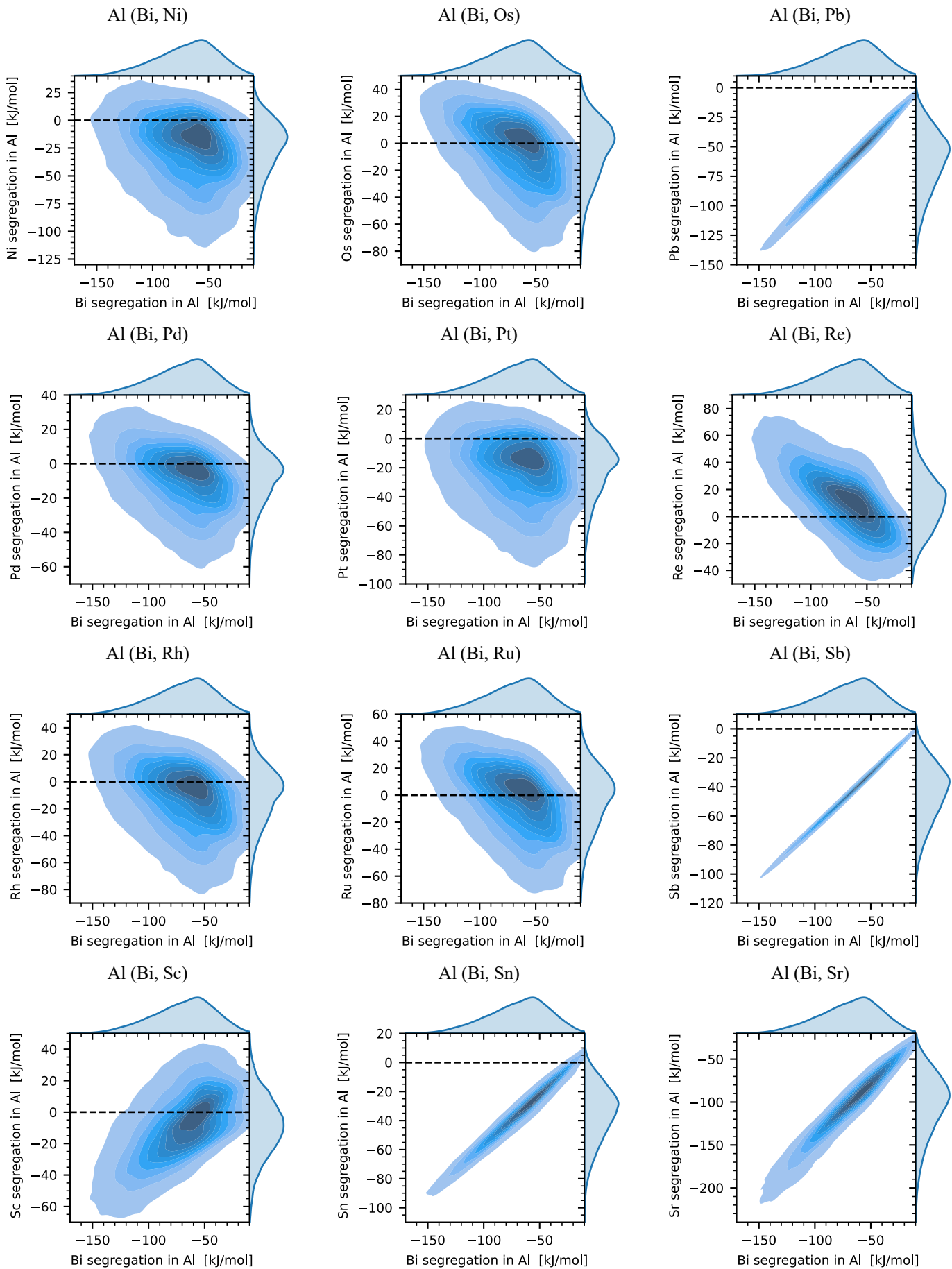

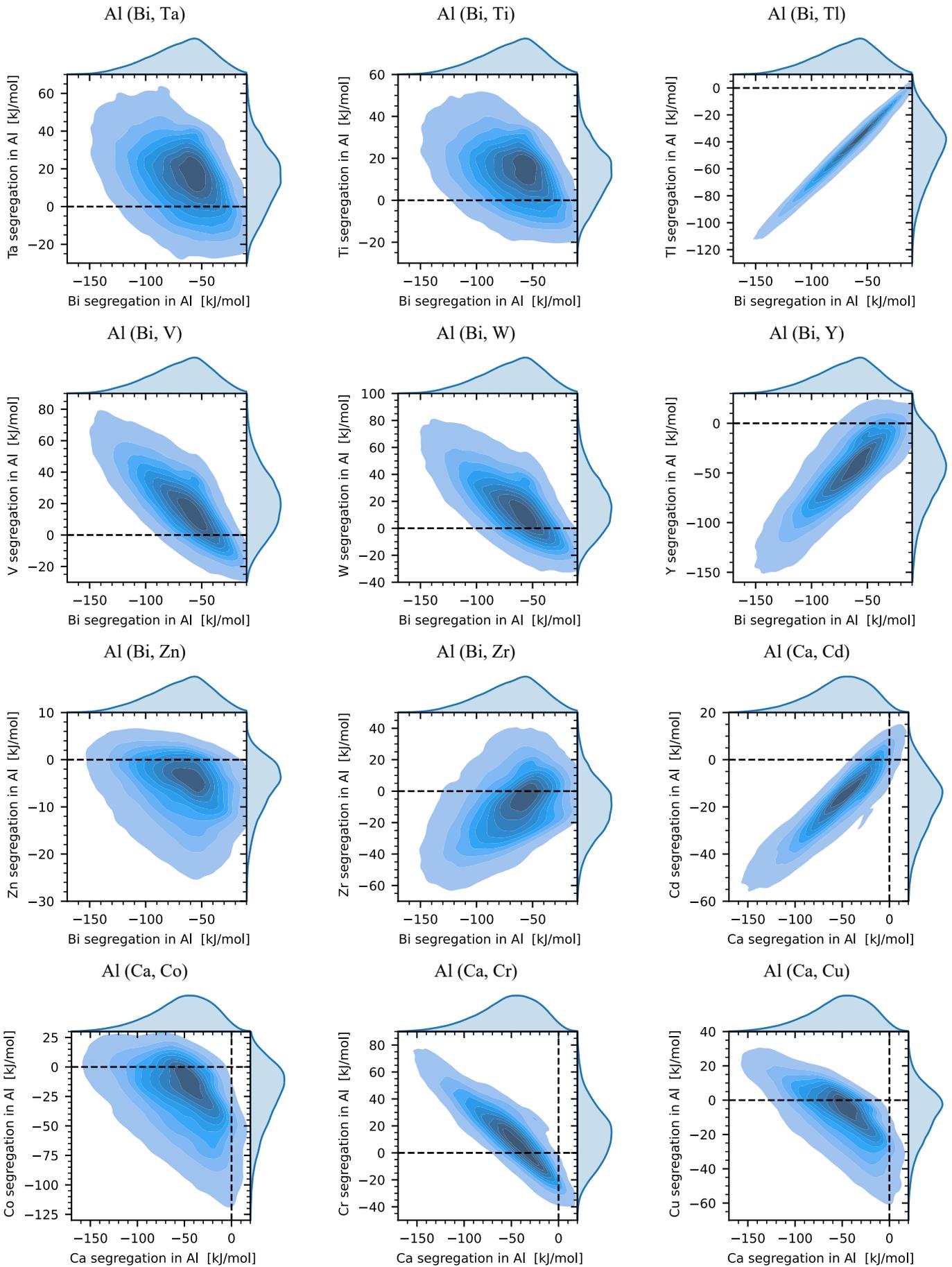

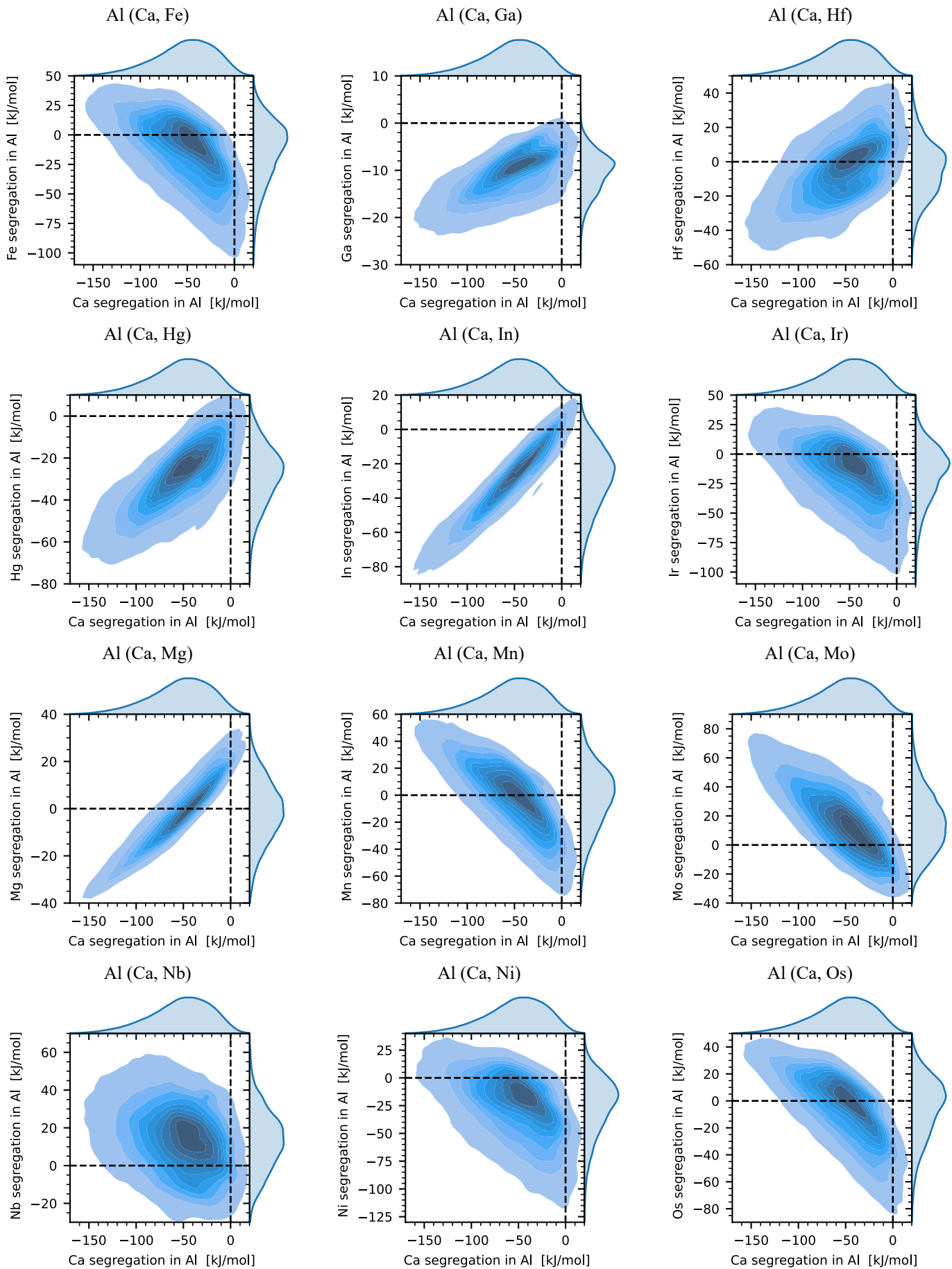

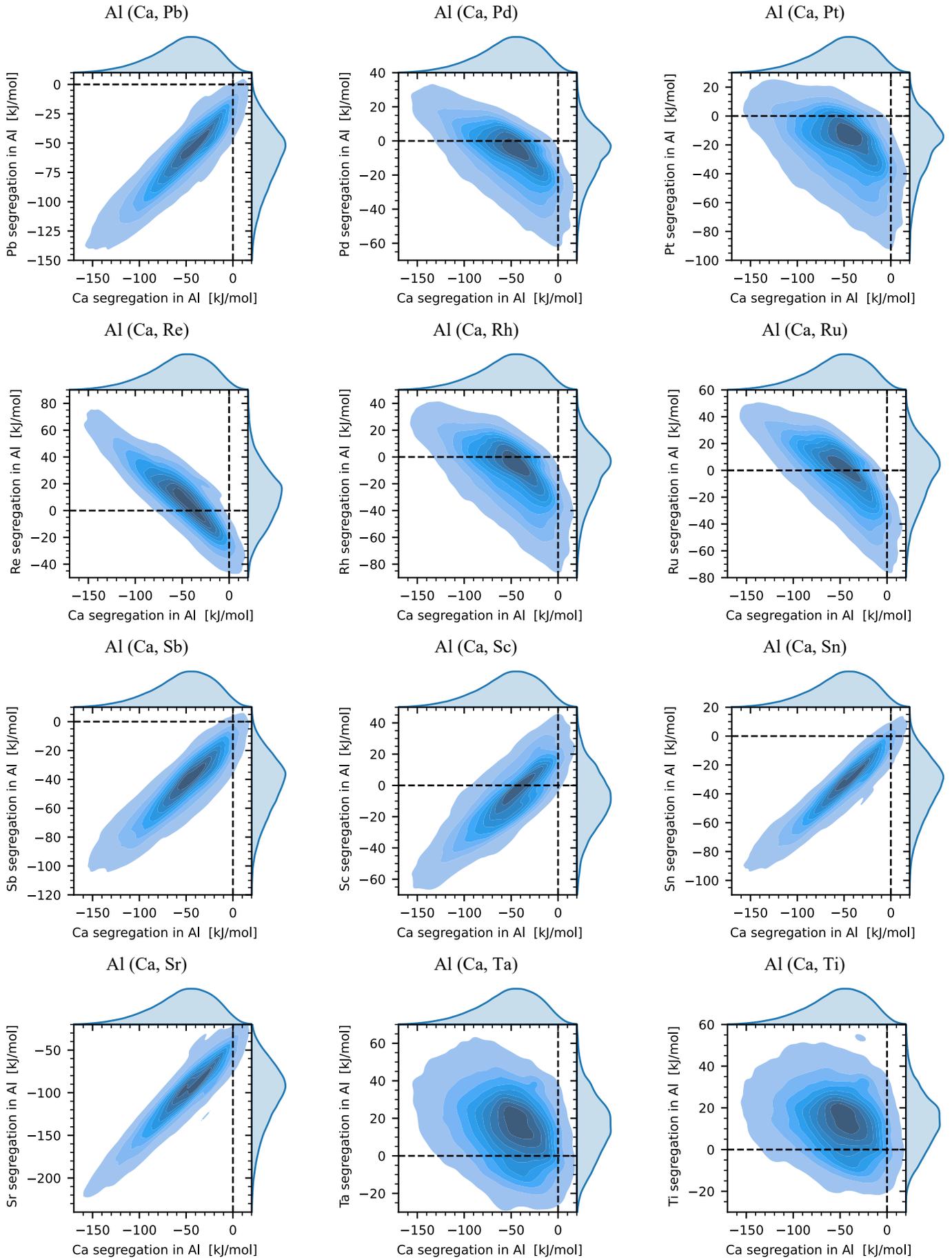

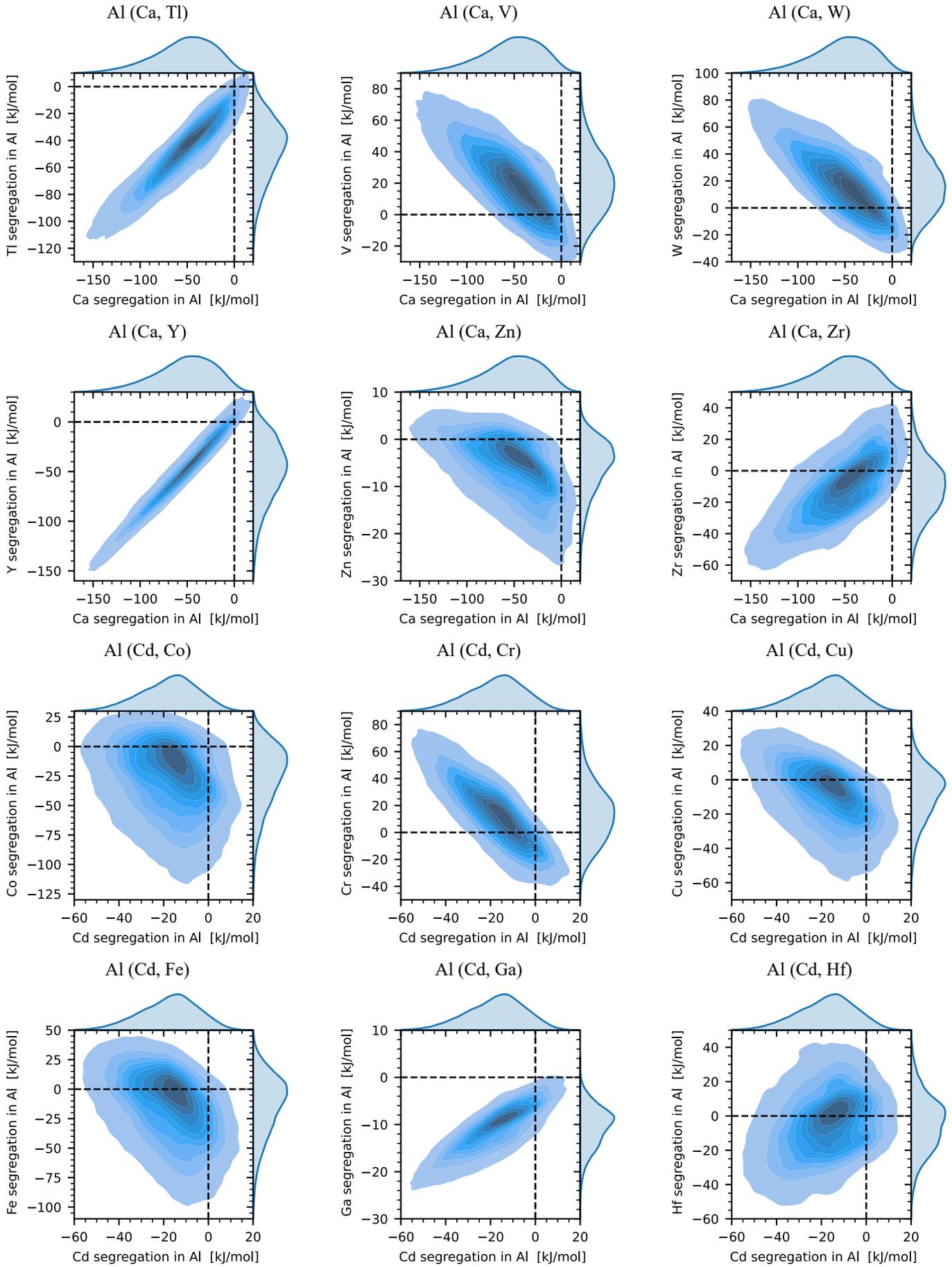

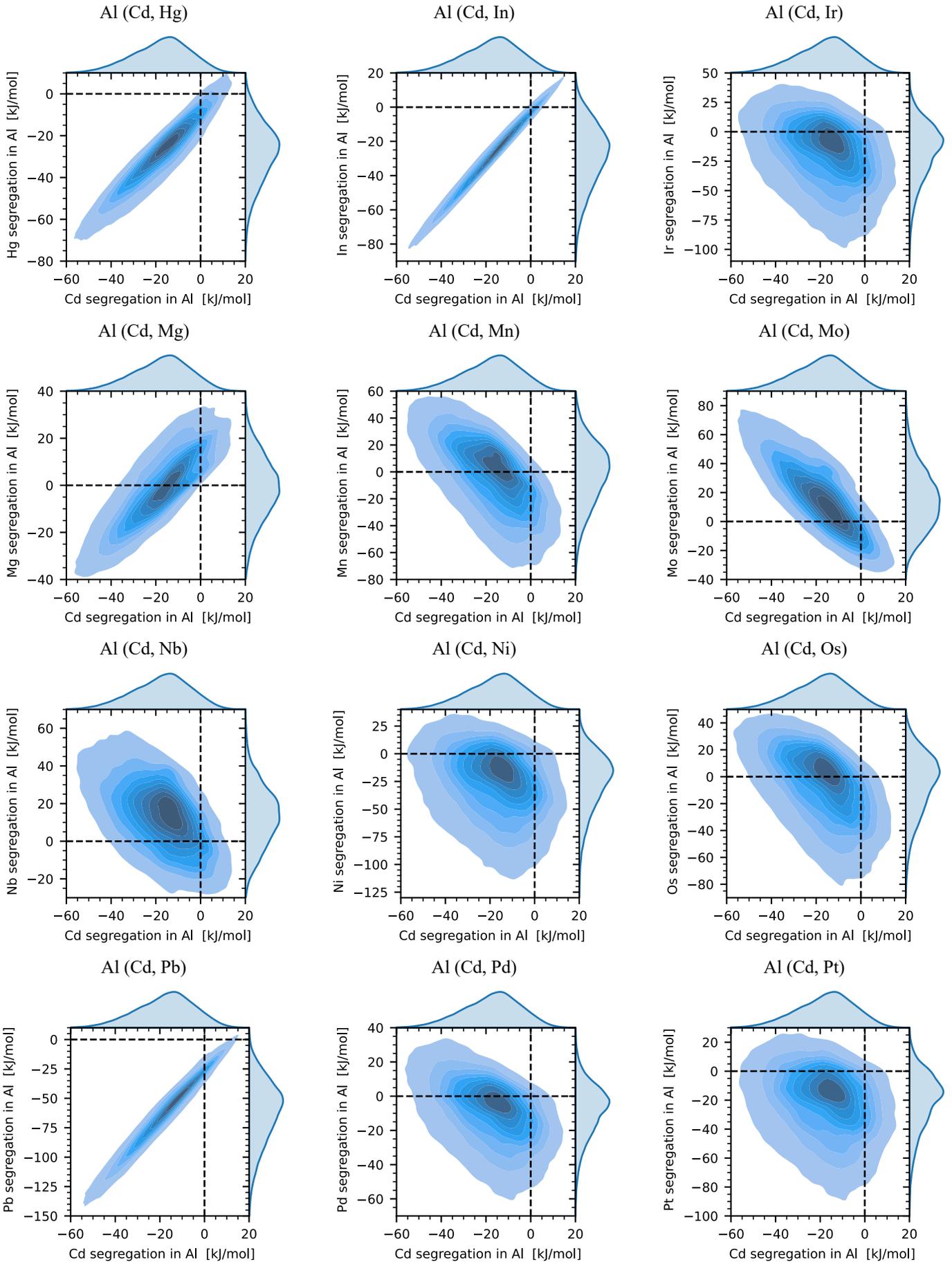

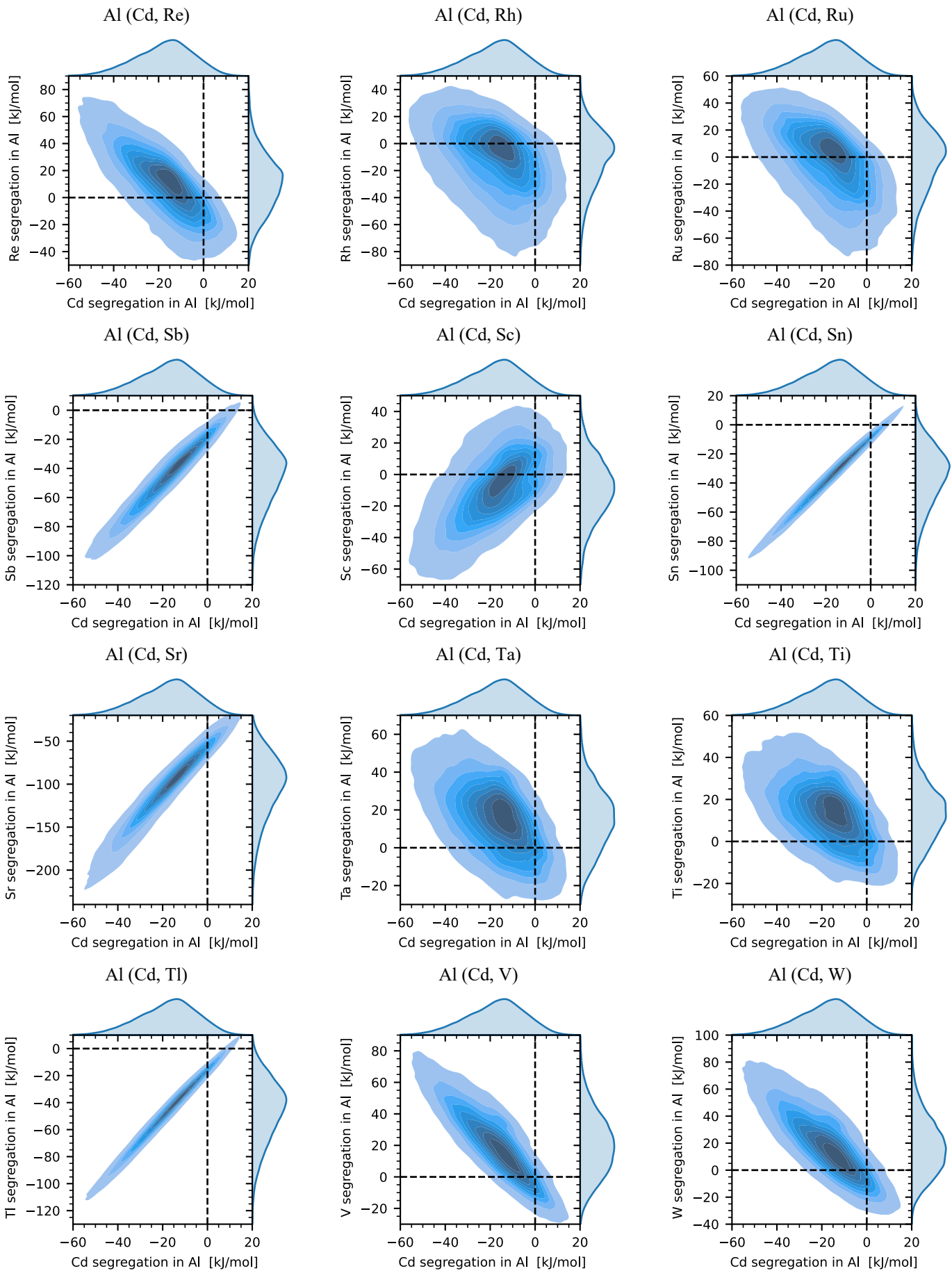

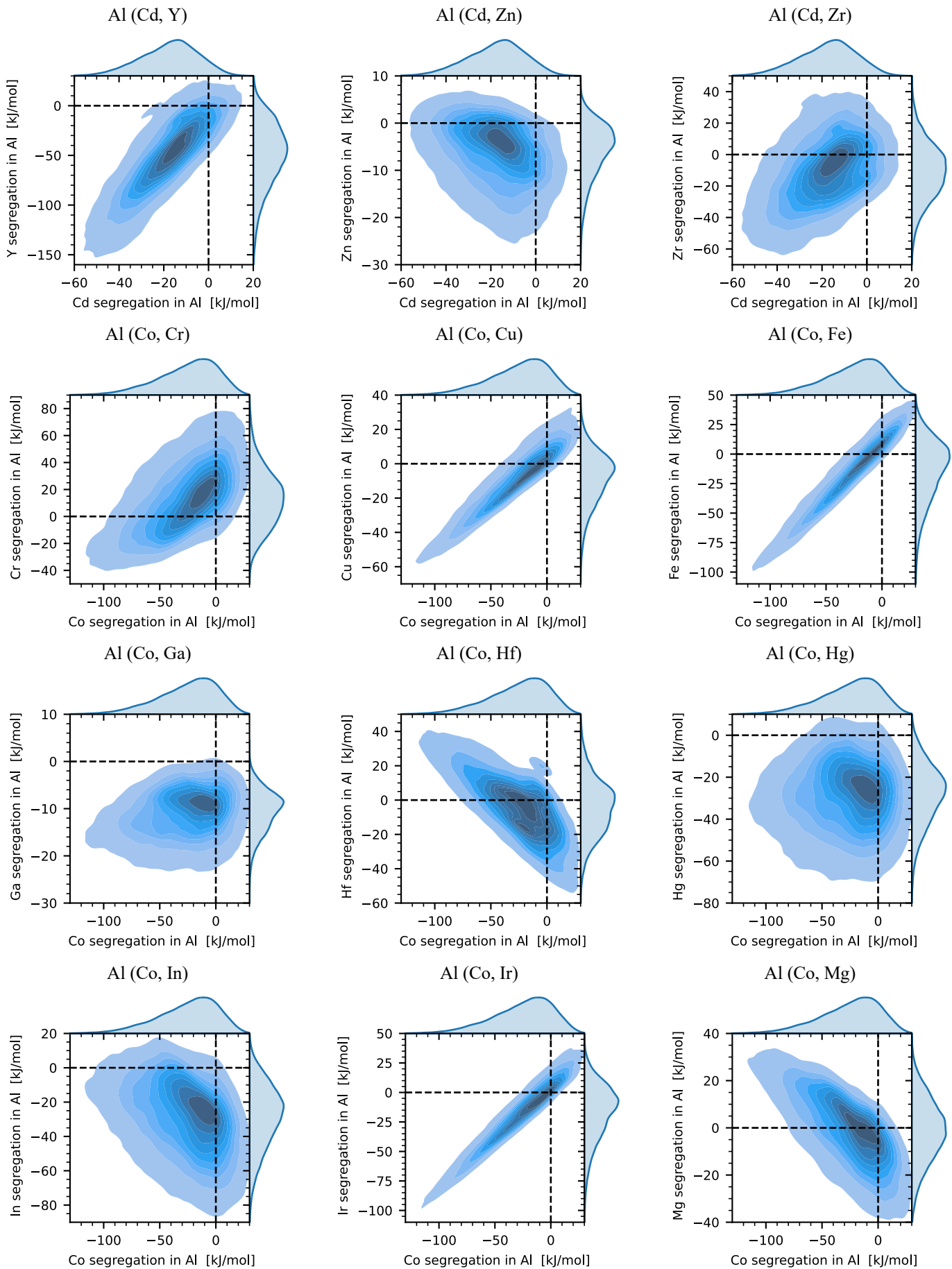

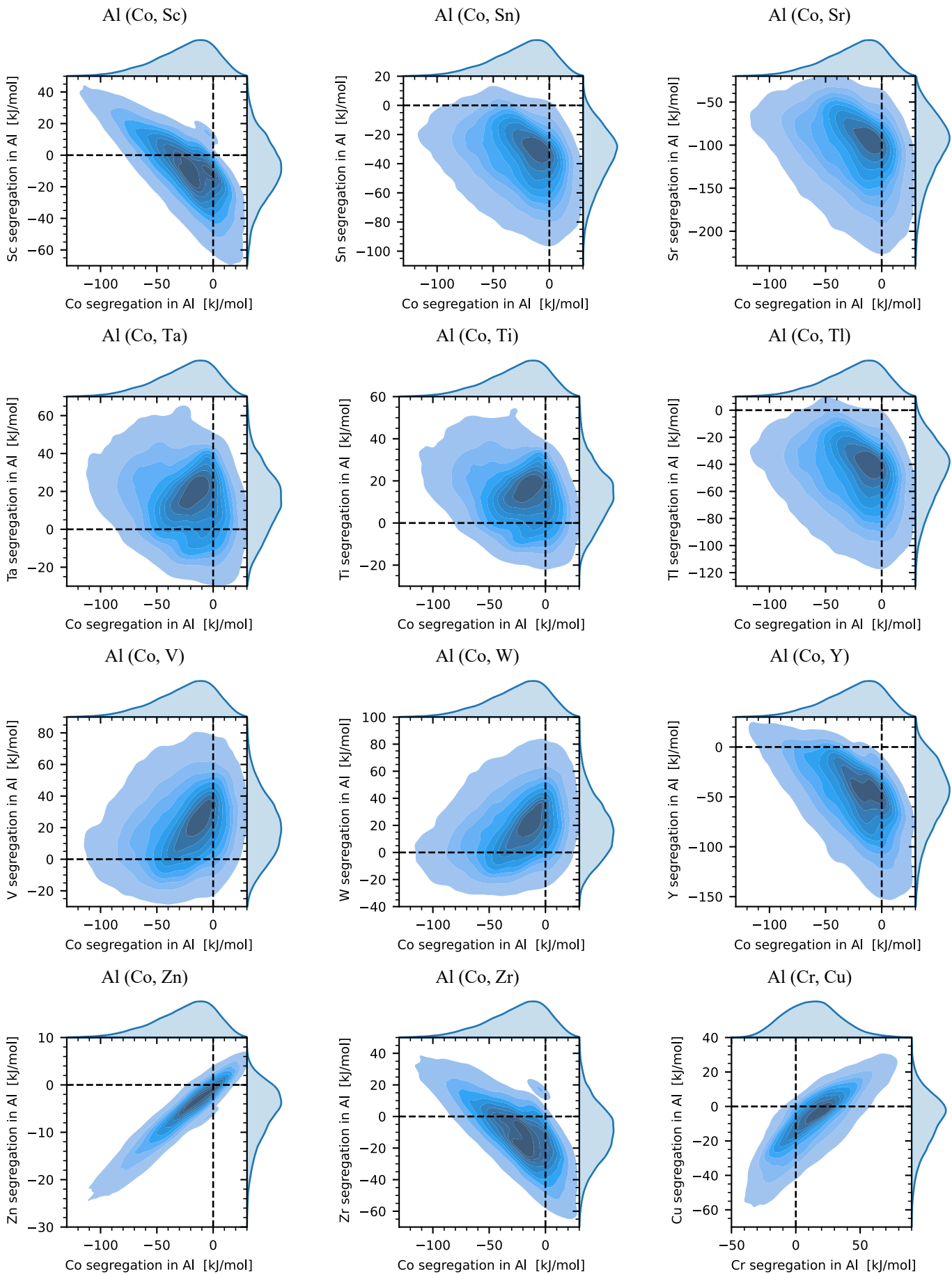

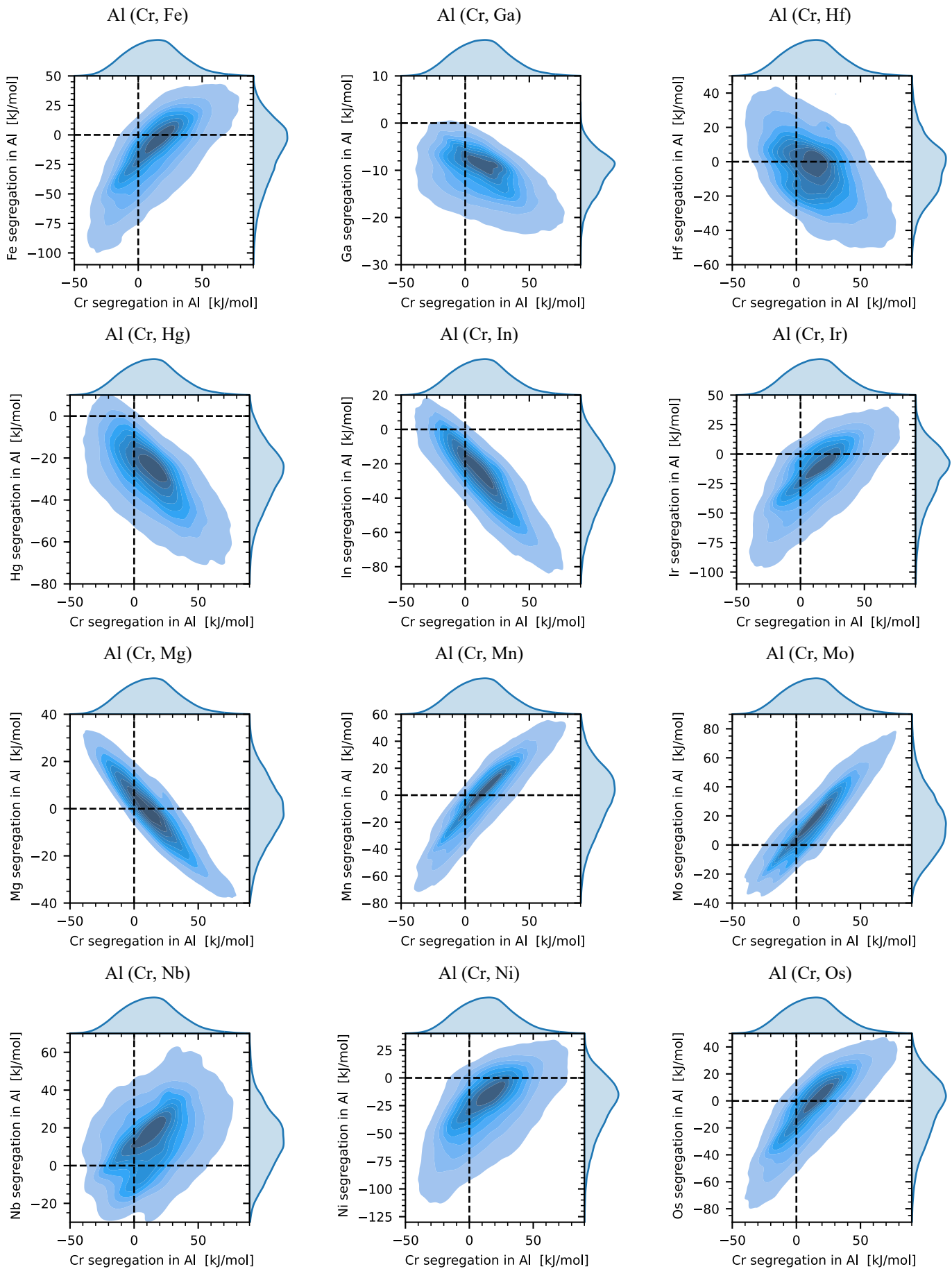

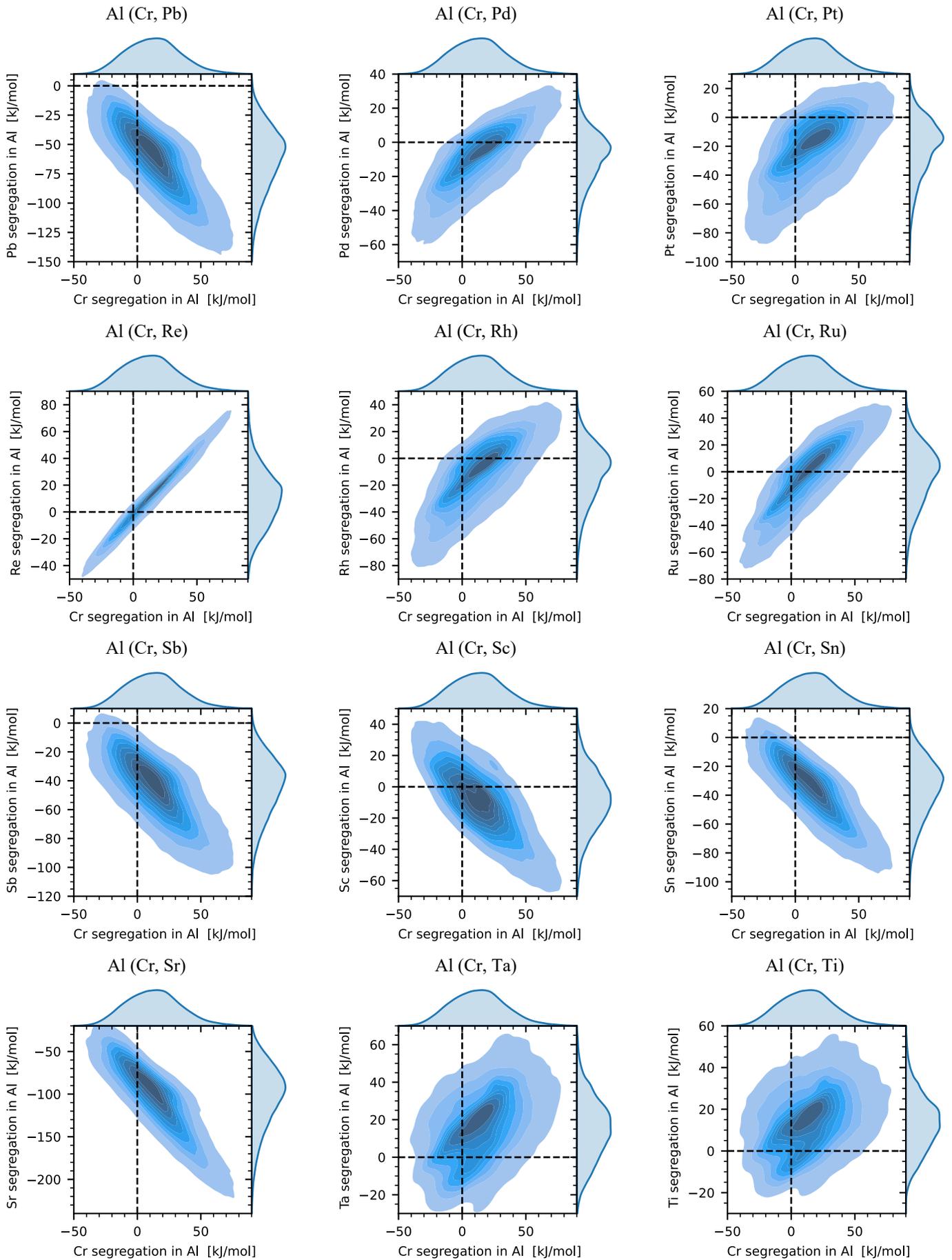

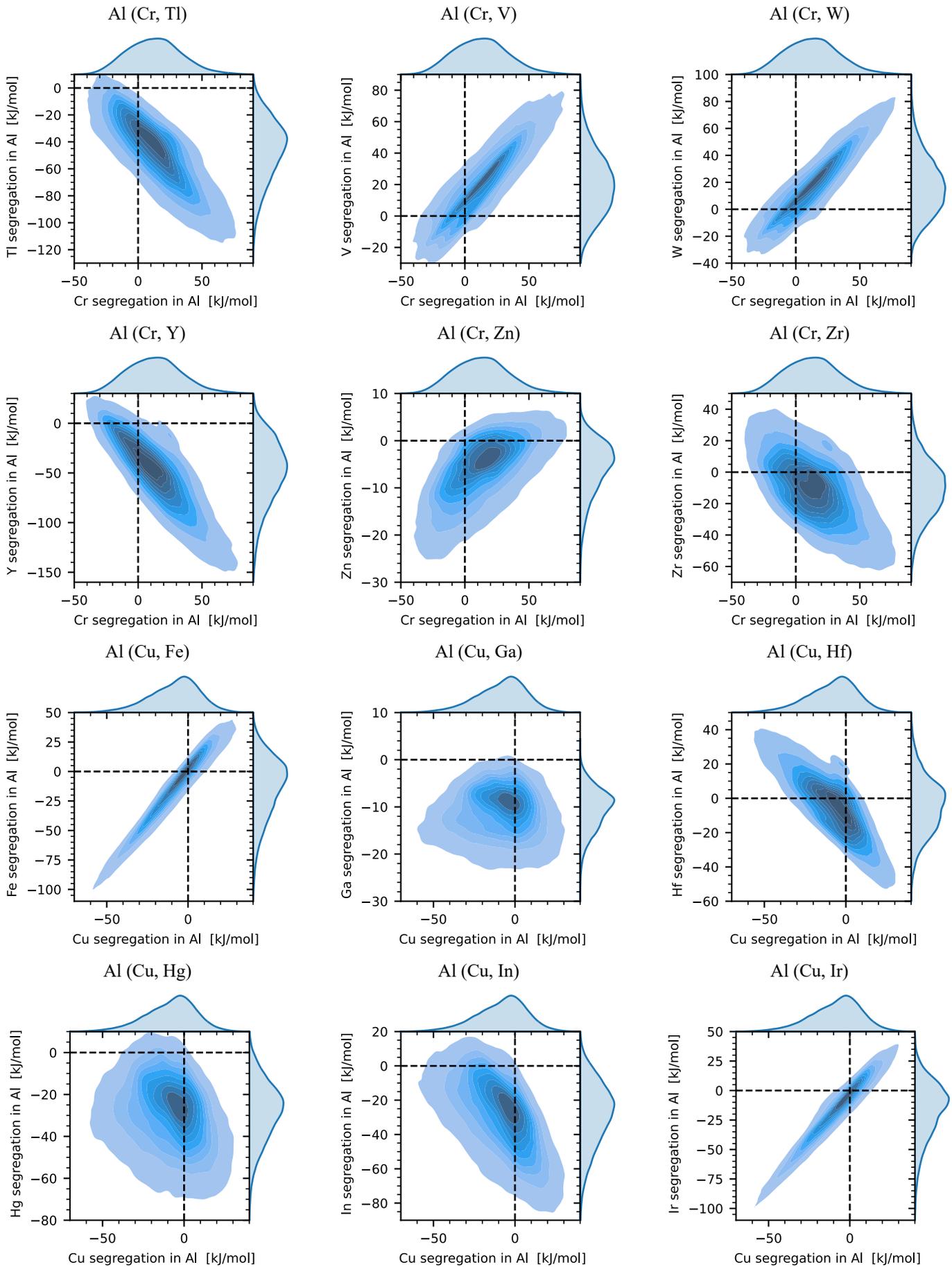

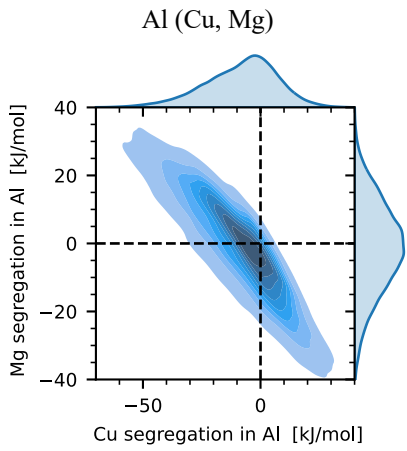

Al (Cu, Mg)

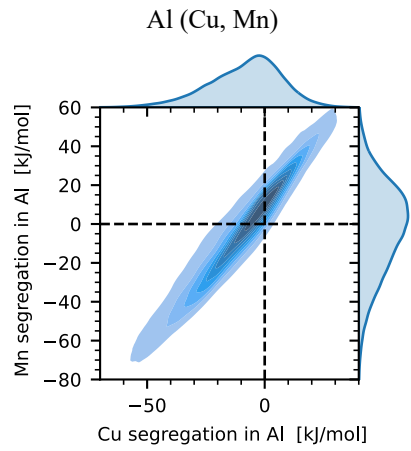

Al (Cu, Mn)

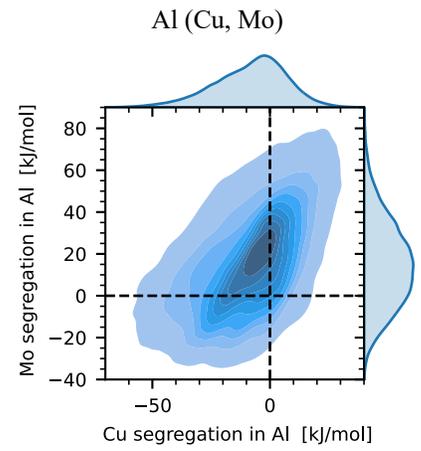

Al (Cu, Mo)

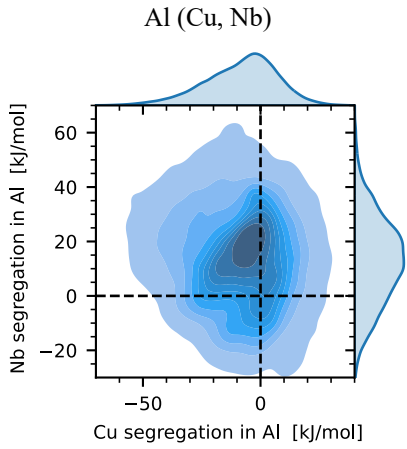

Al (Cu, Nb)

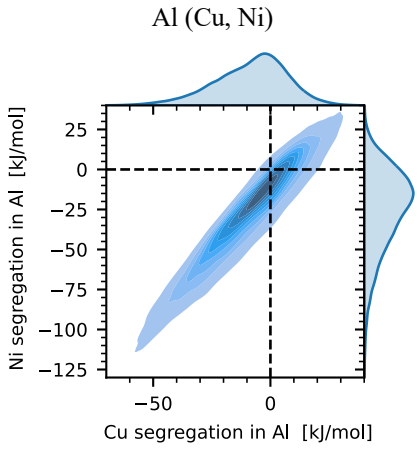

Al (Cu, Ni)

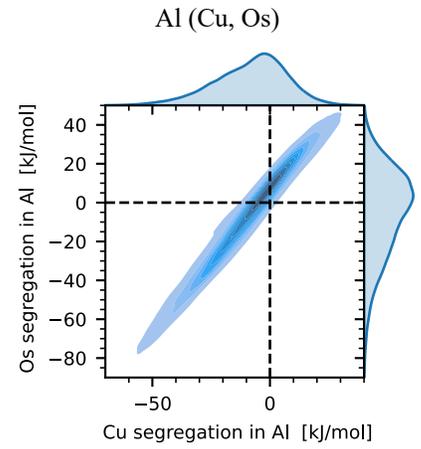

Al (Cu, Os)

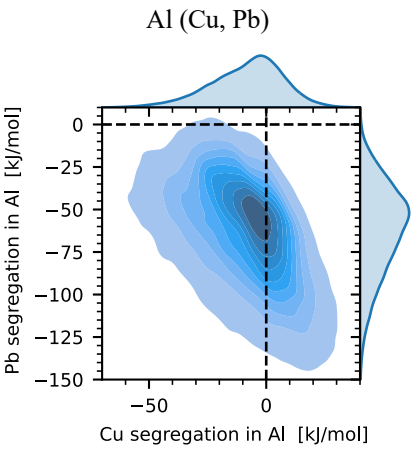

Al (Cu, Pb)

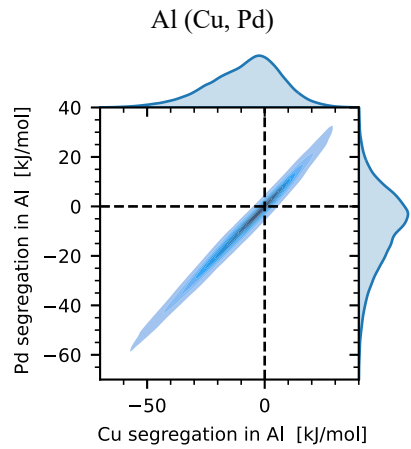

Al (Cu, Pd)

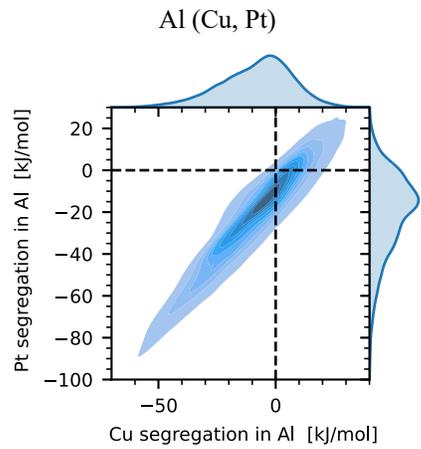

Al (Cu, Pt)

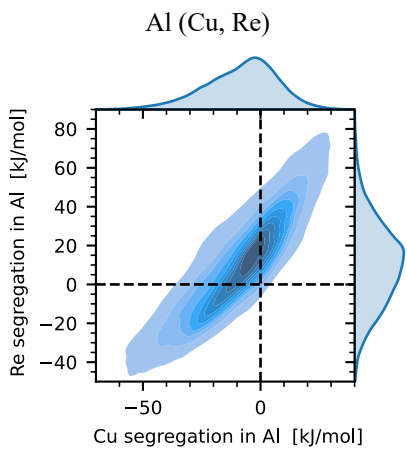

Al (Cu, Re)

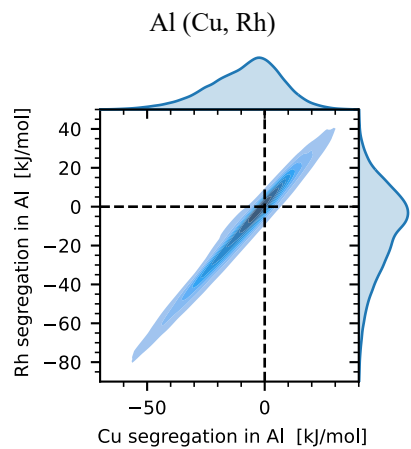

Al (Cu, Rh)

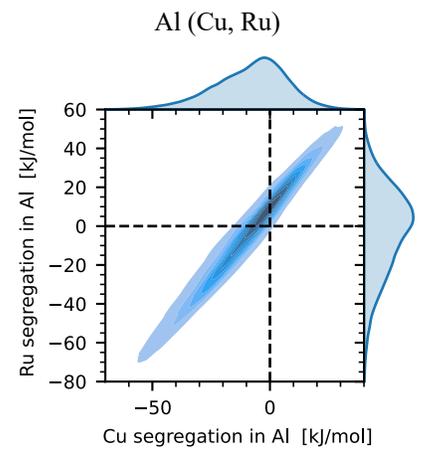

Al (Cu, Ru)

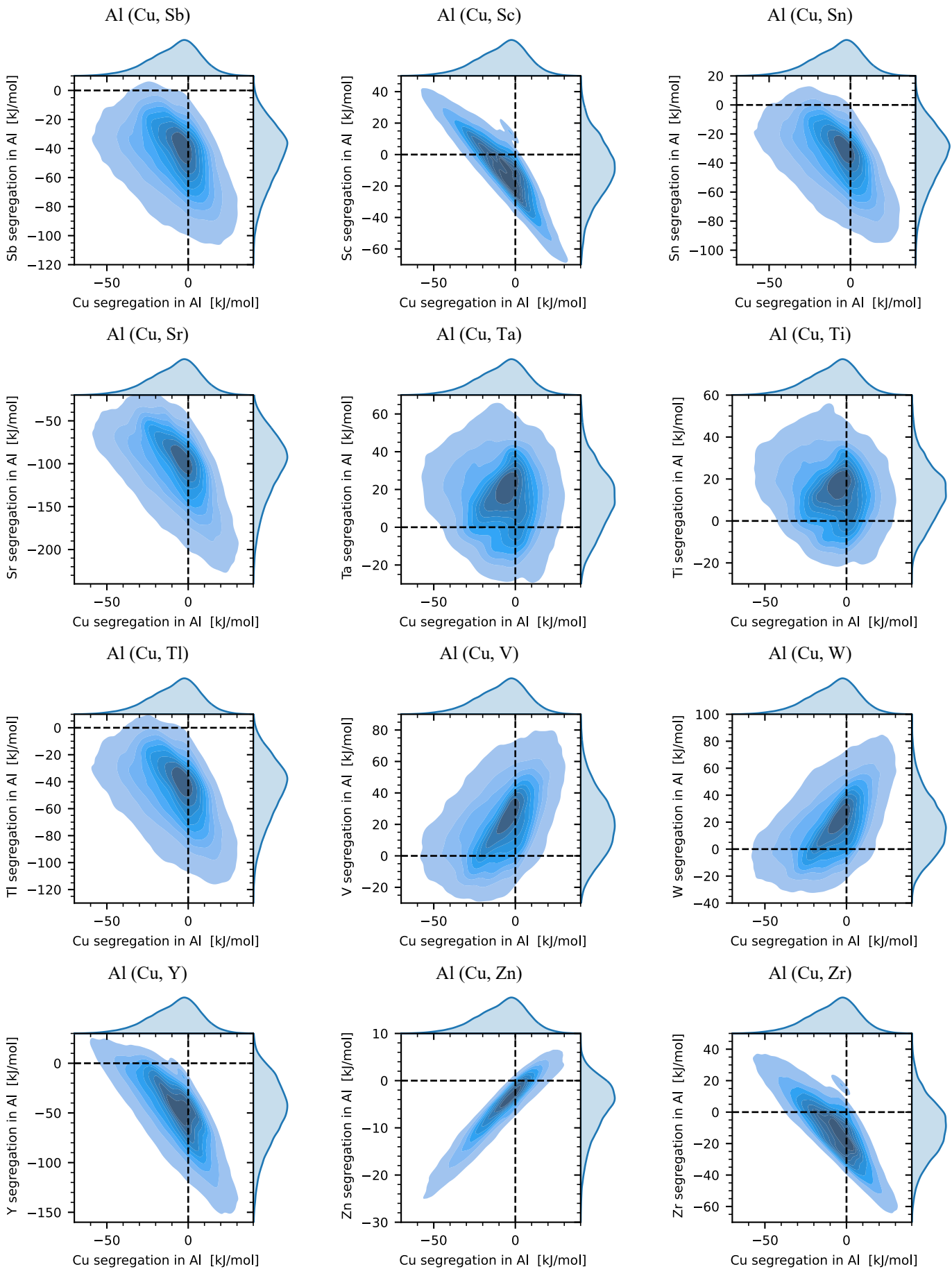

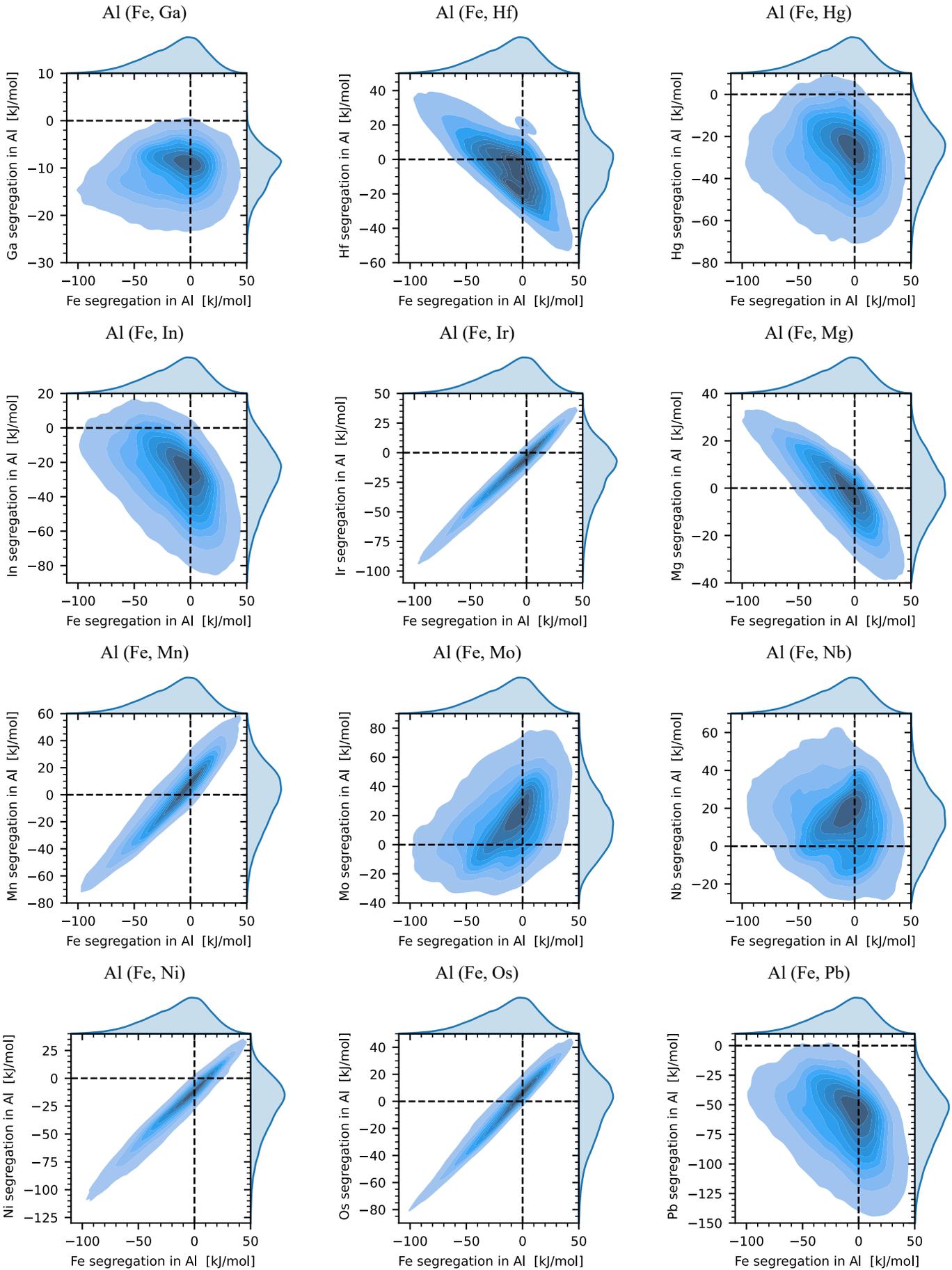

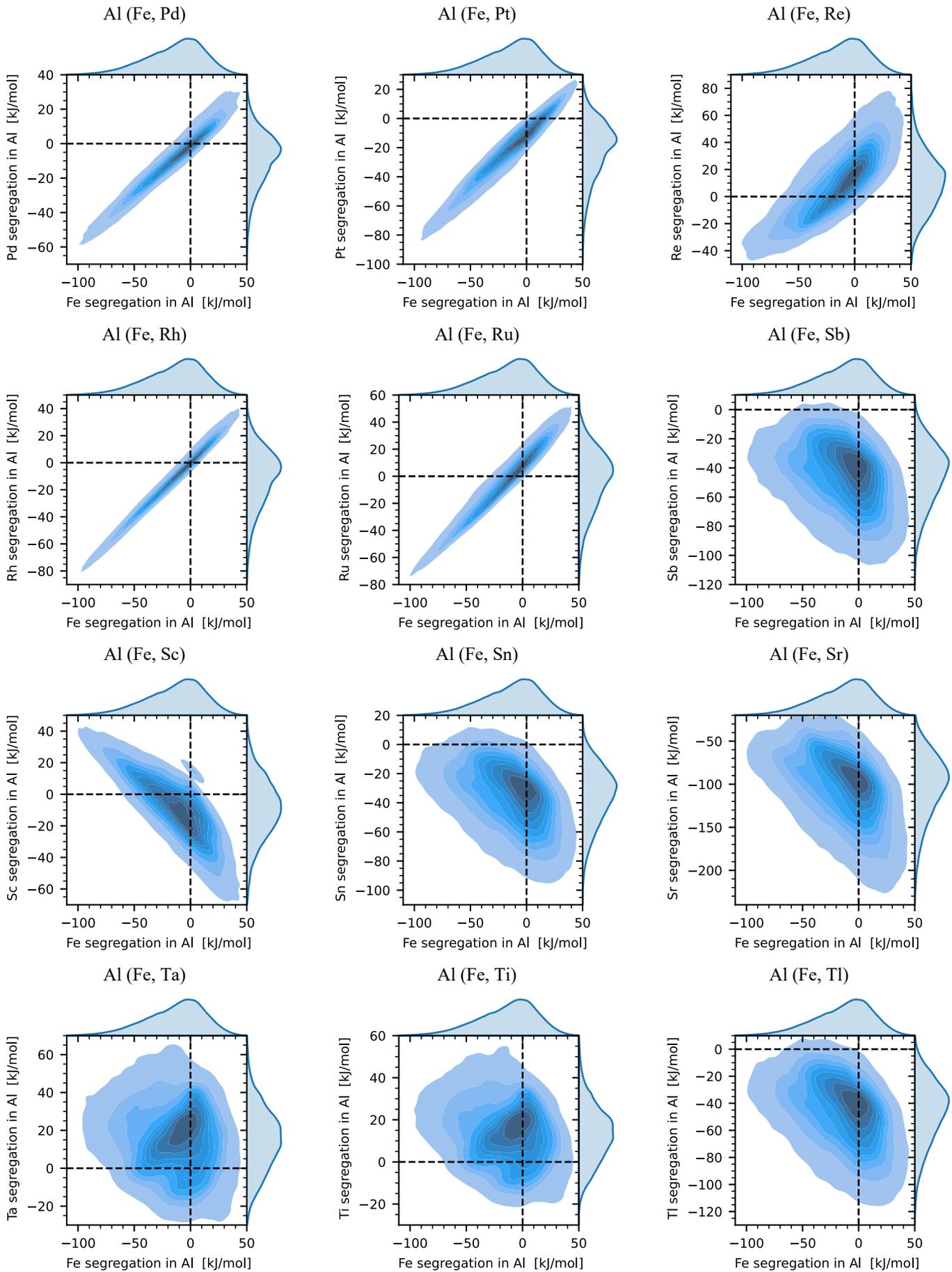

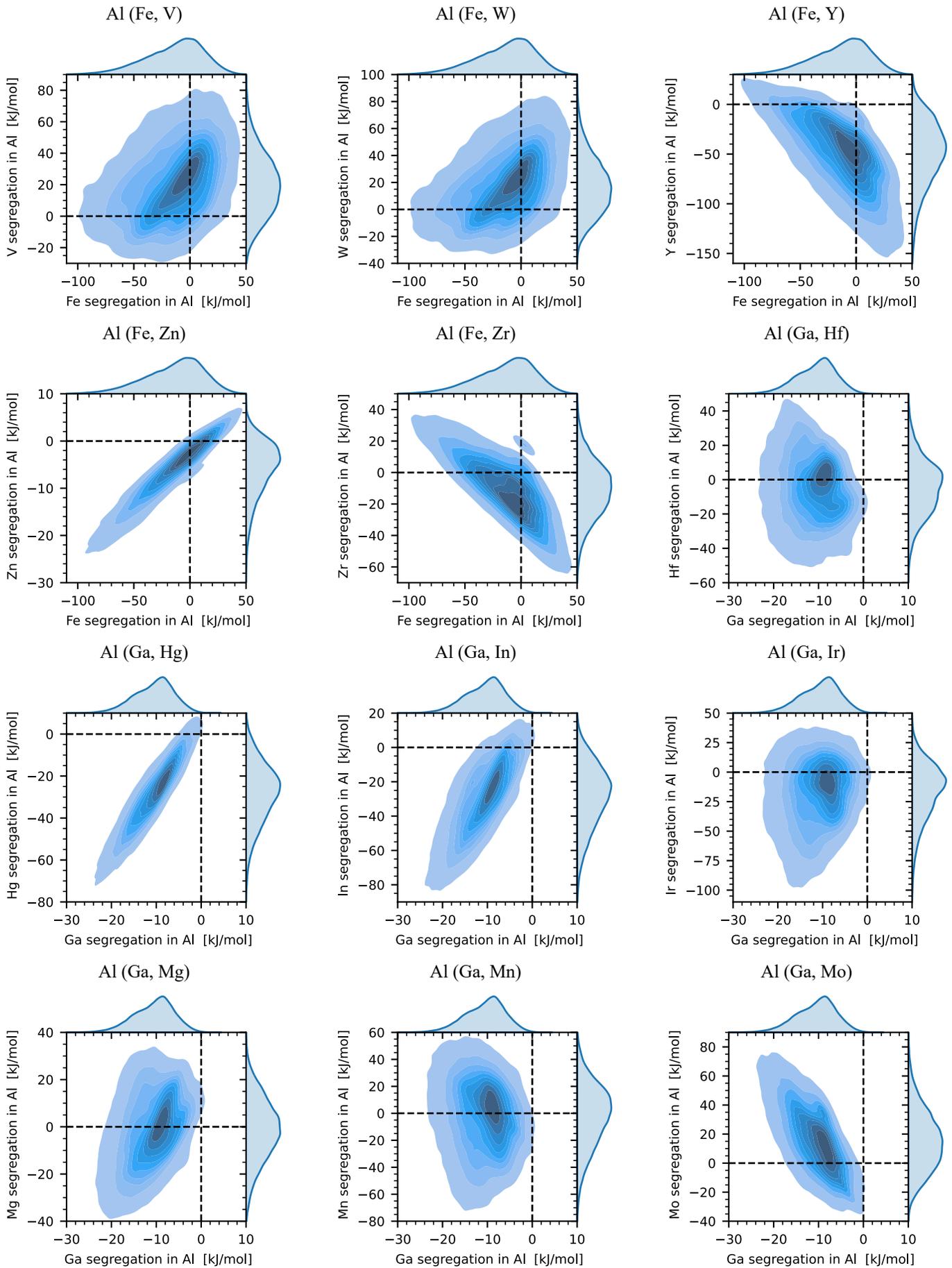

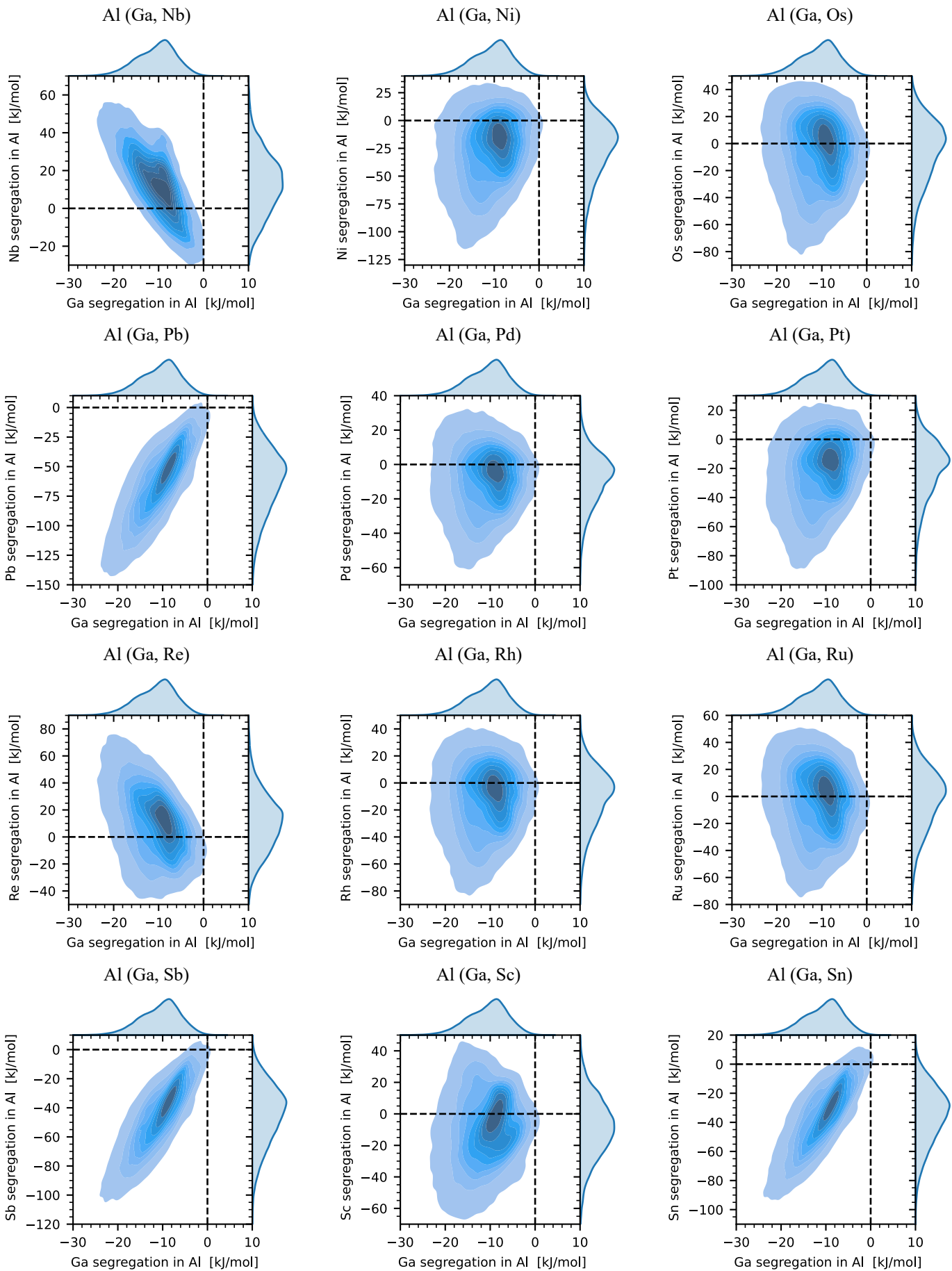

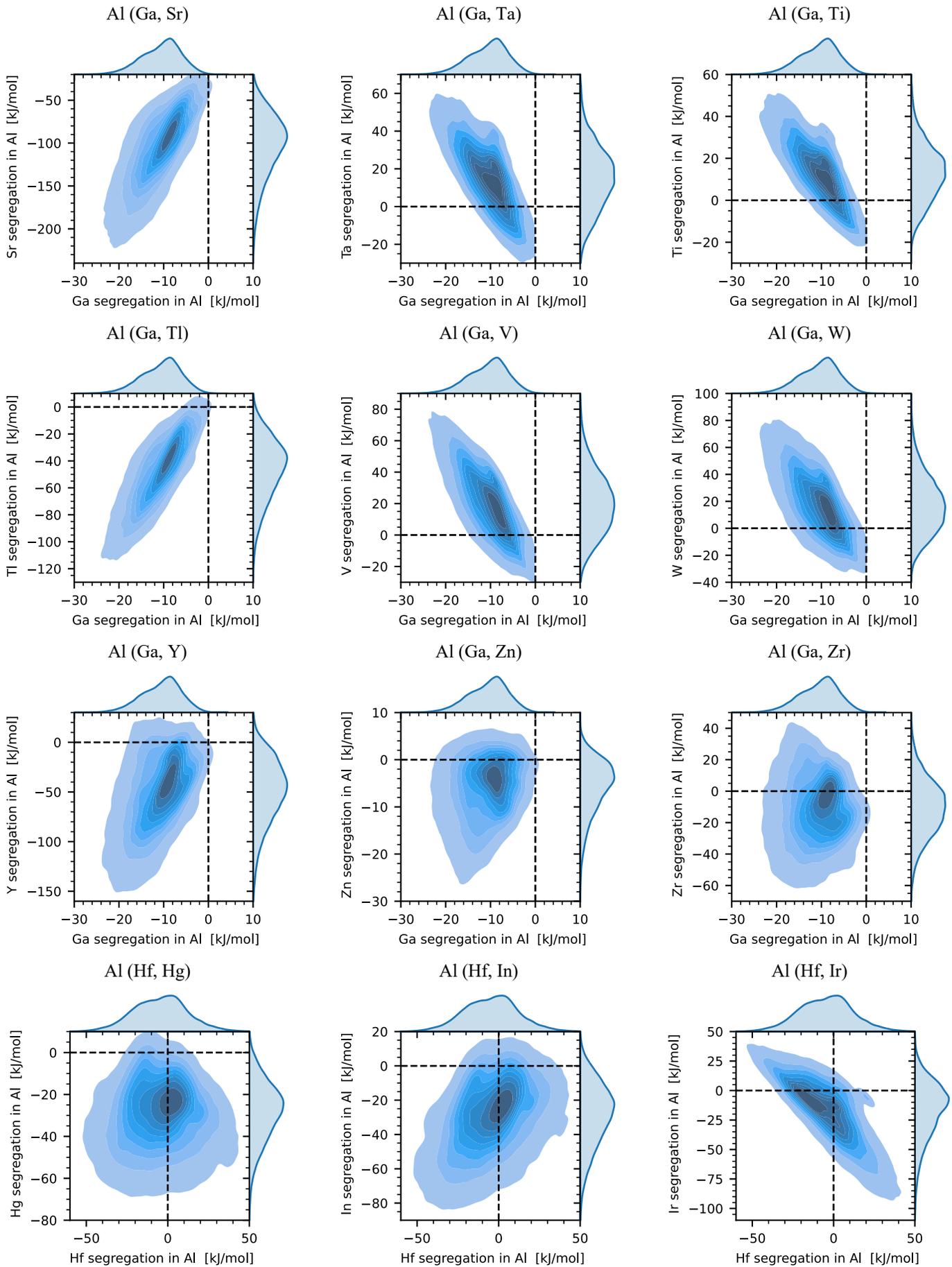

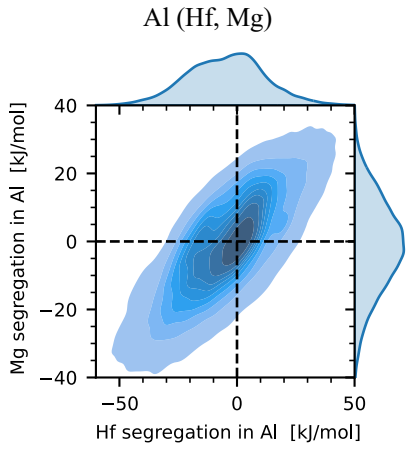

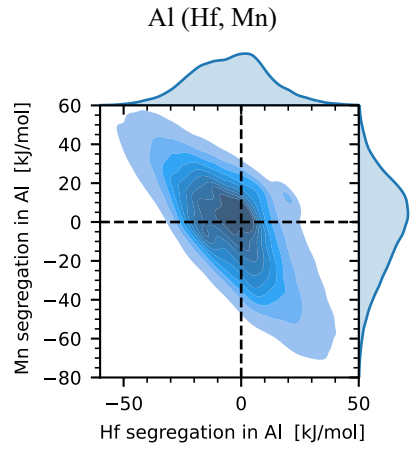

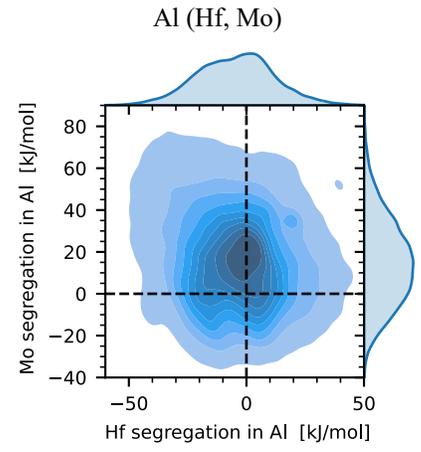

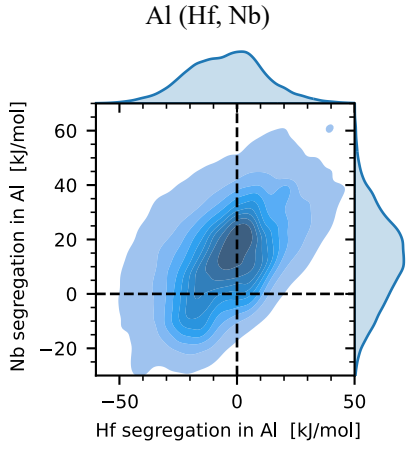

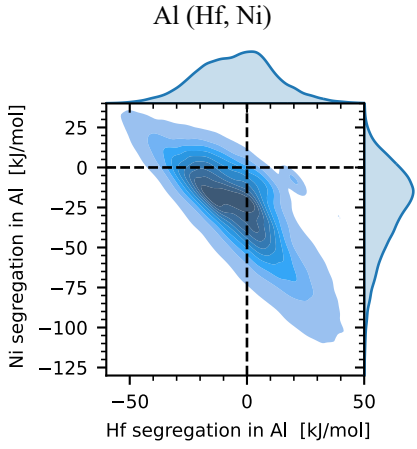

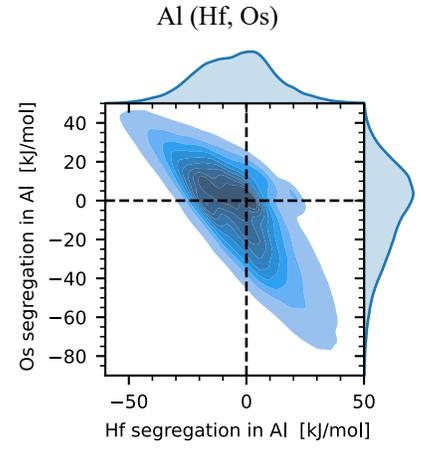

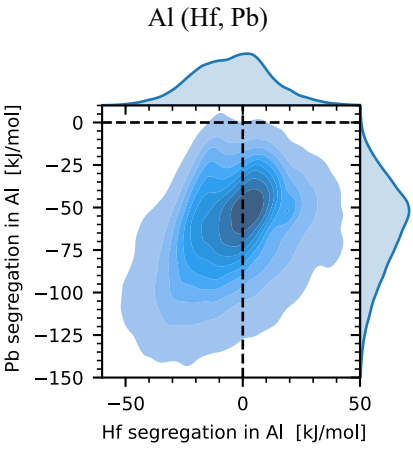

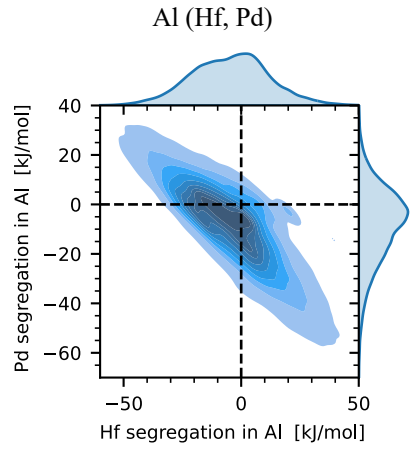

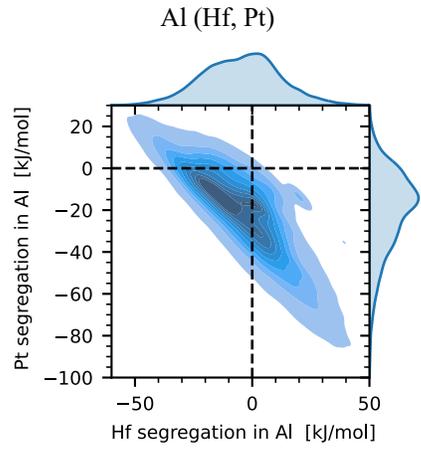

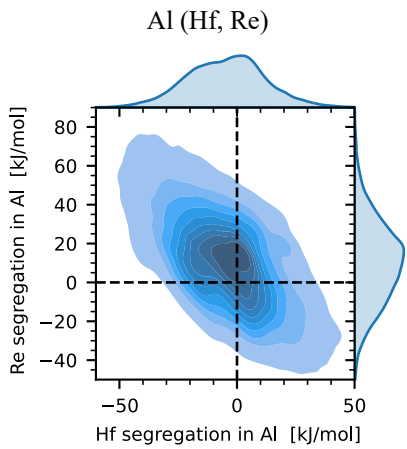

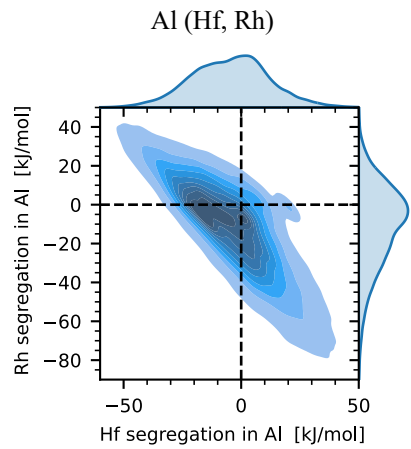

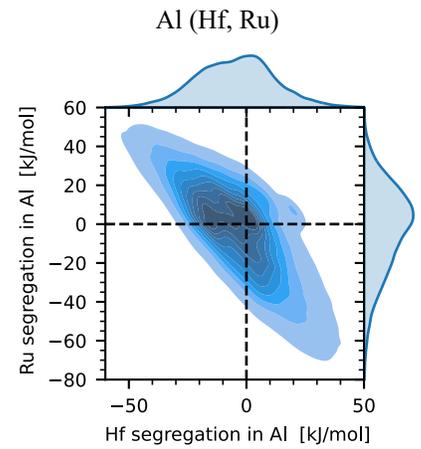

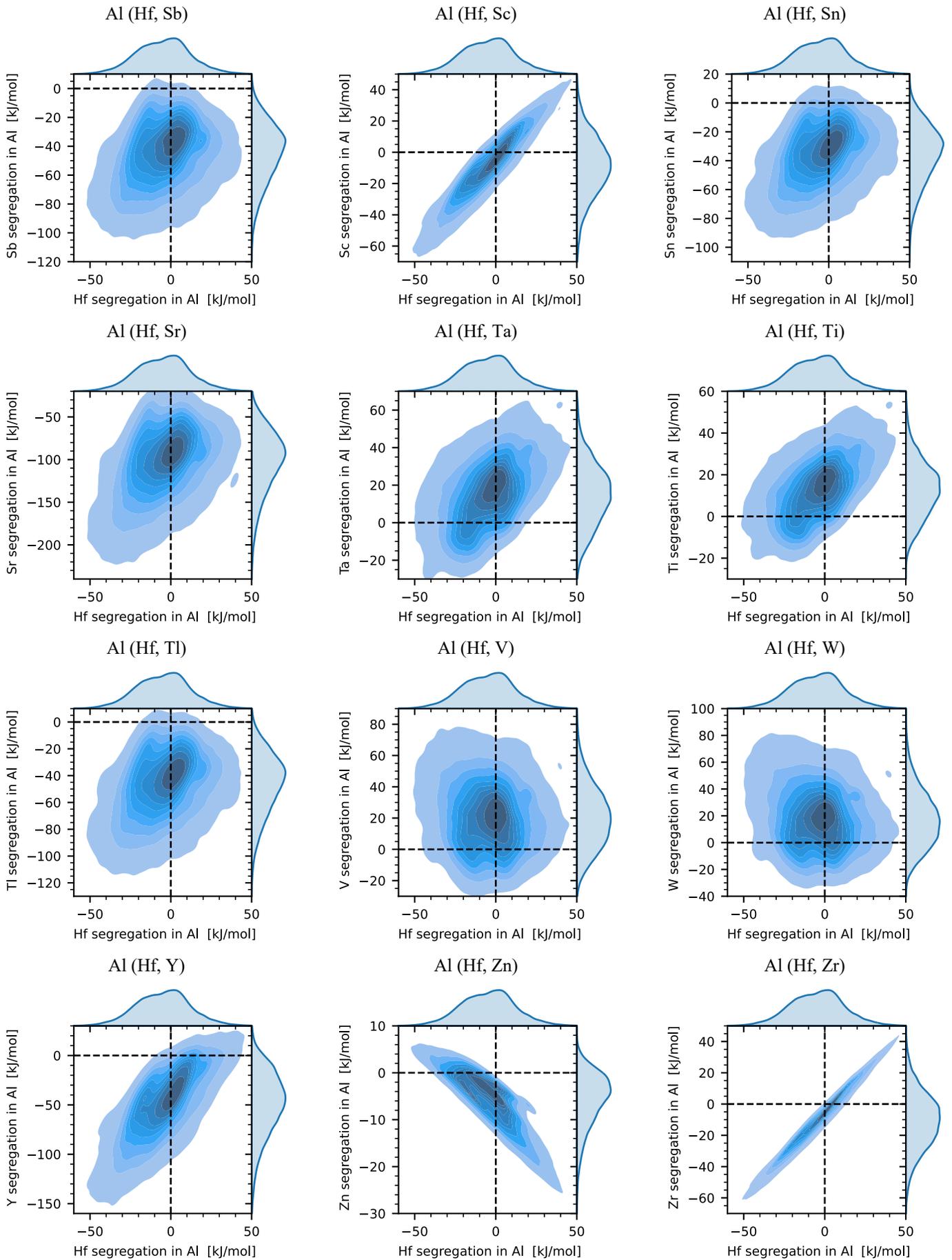

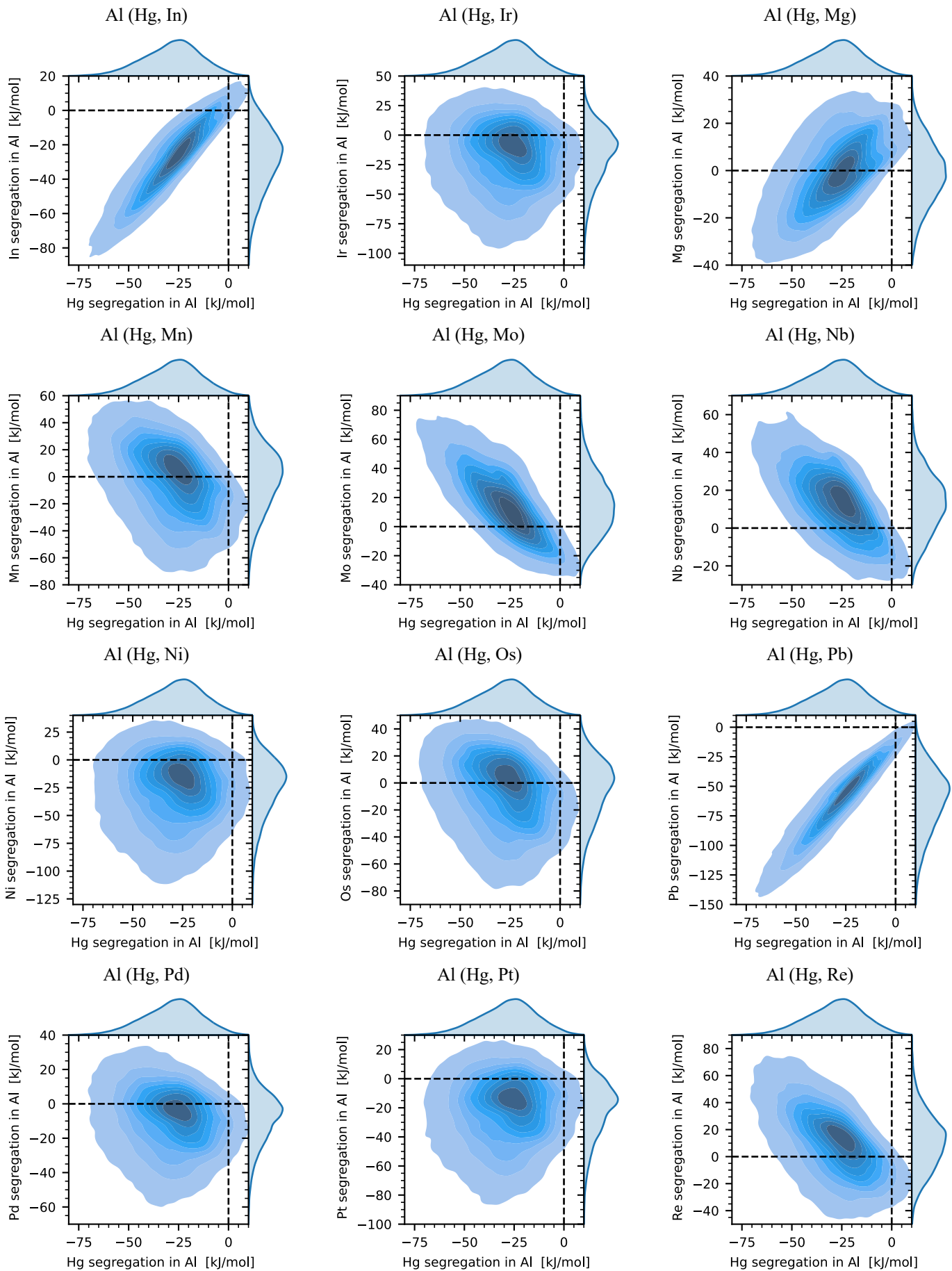

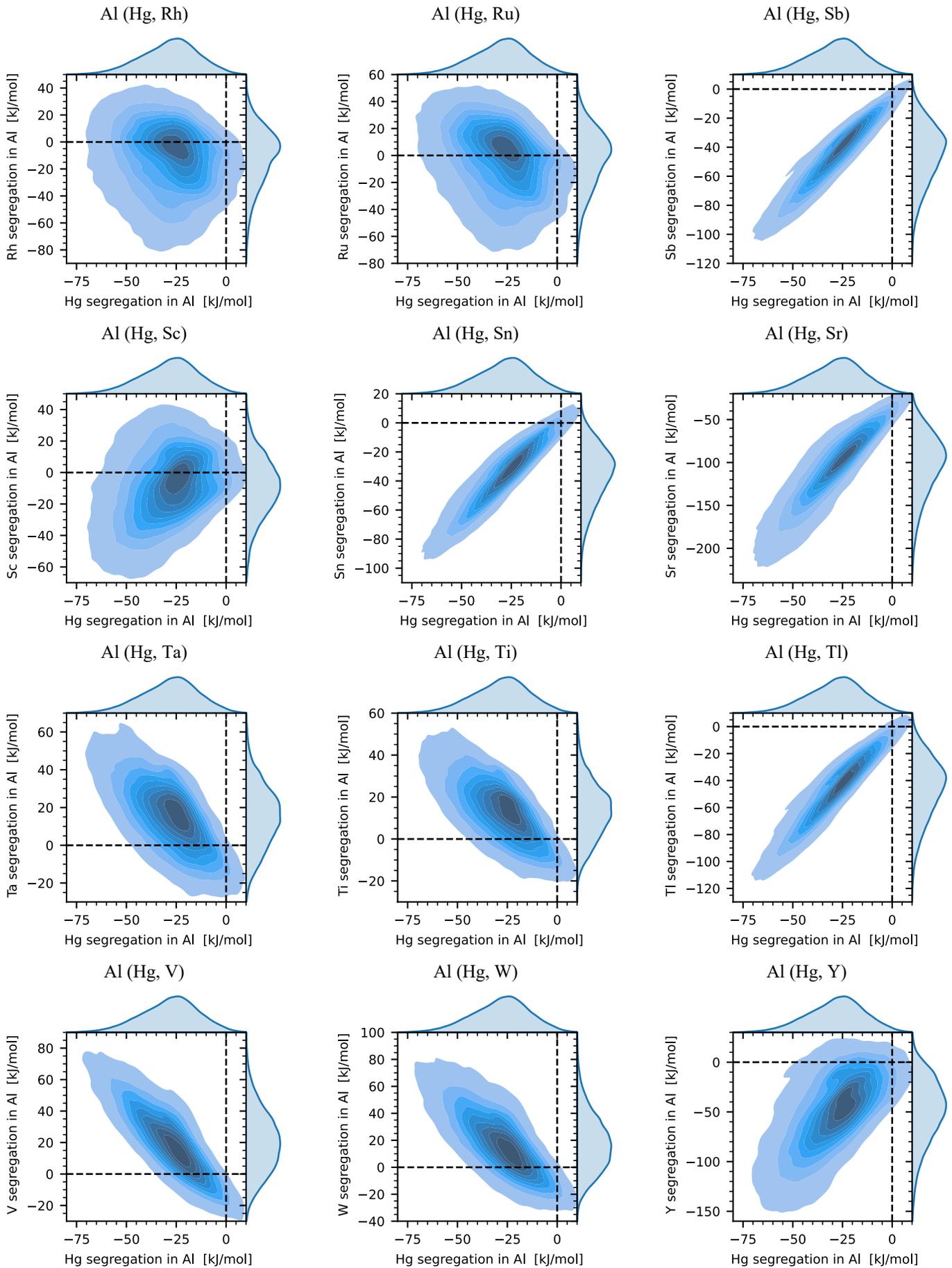

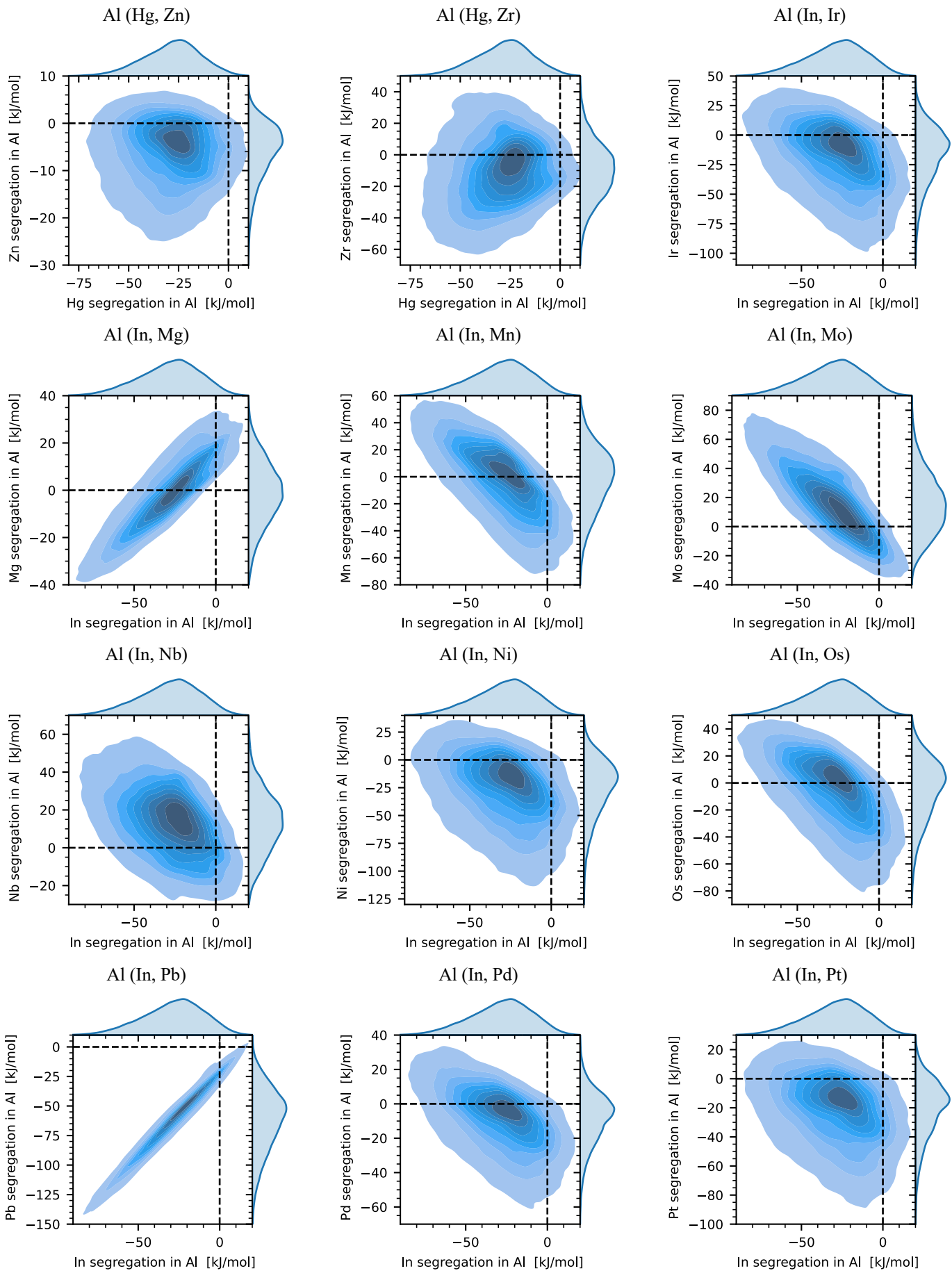

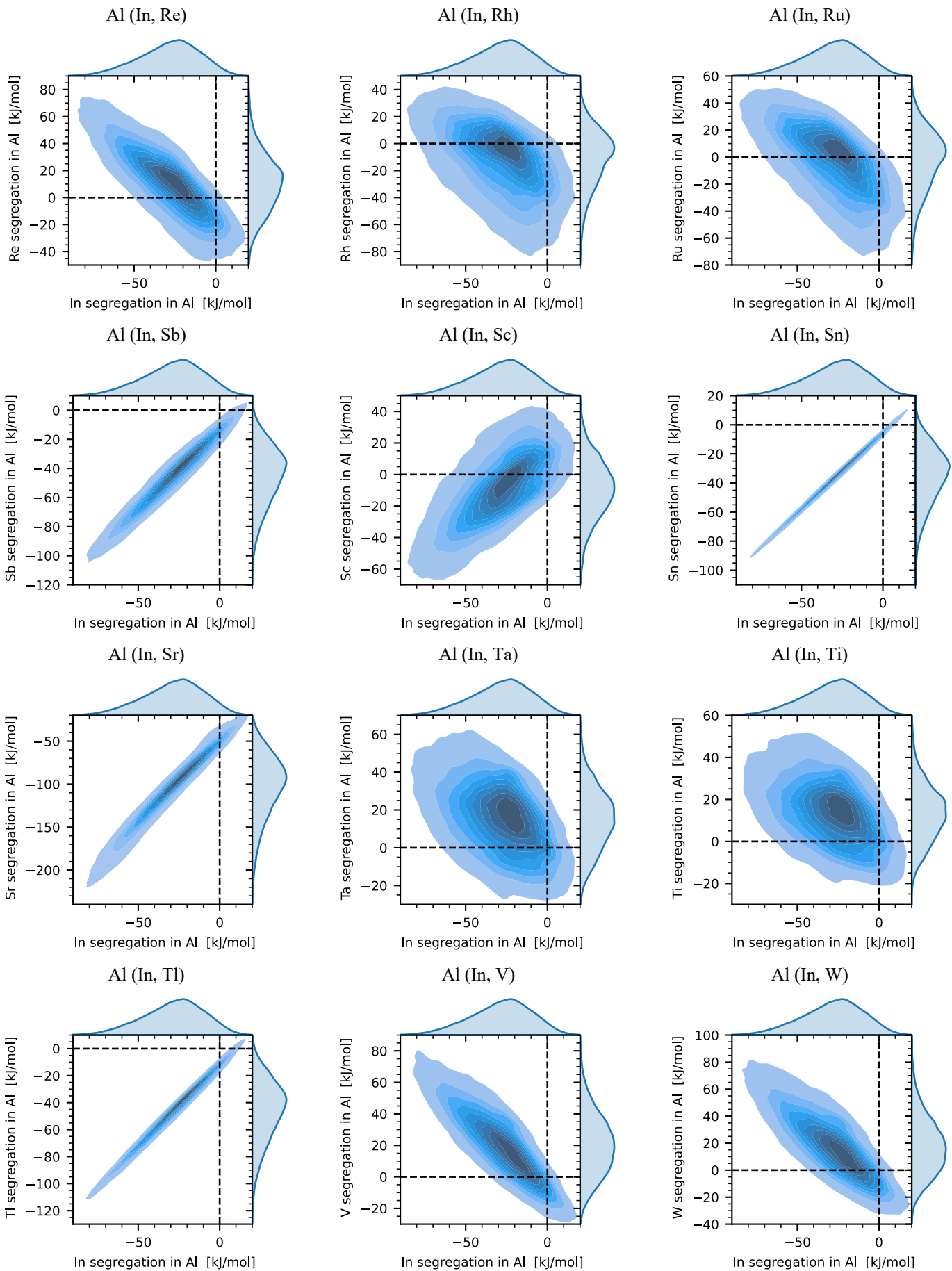

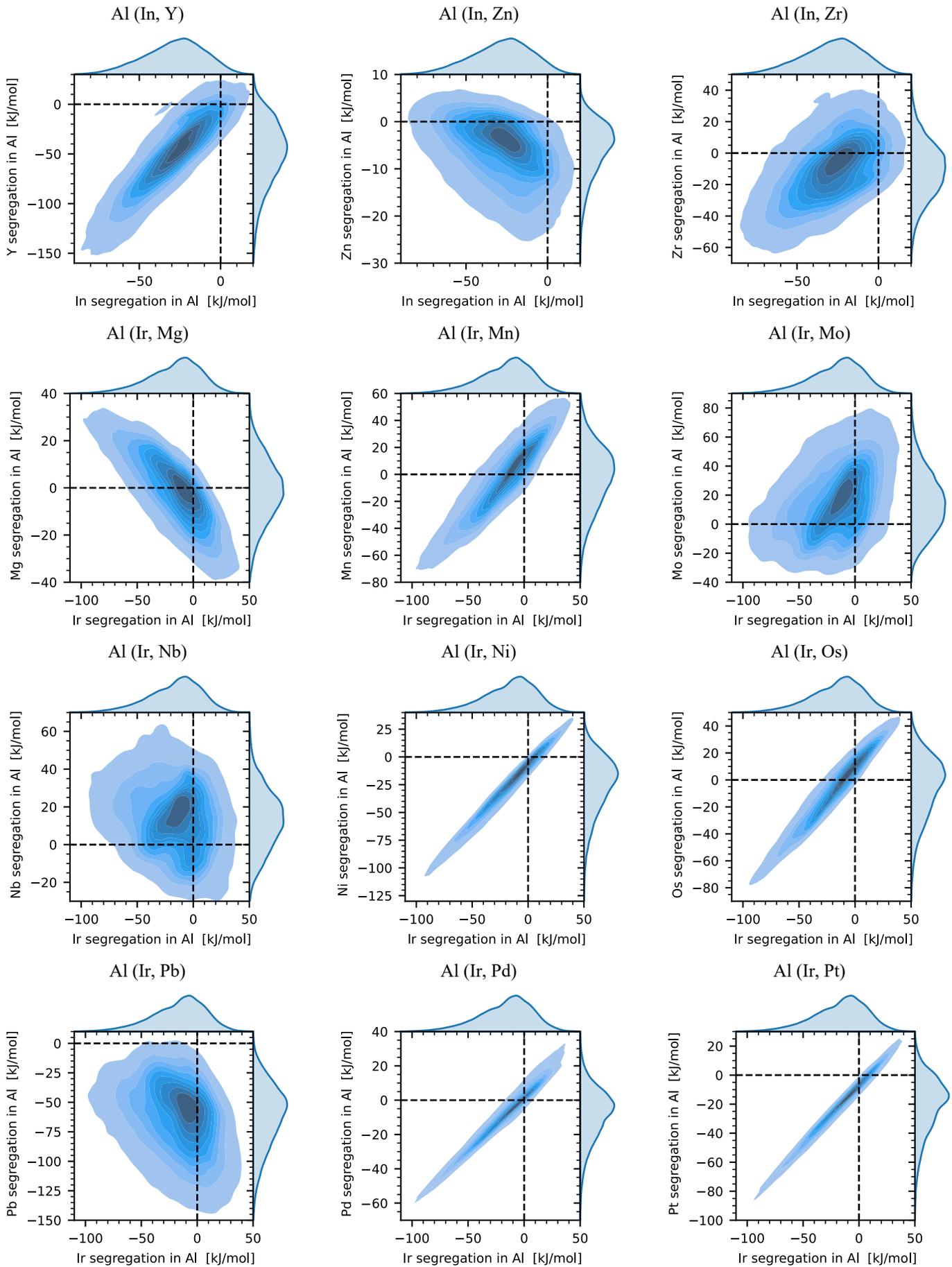

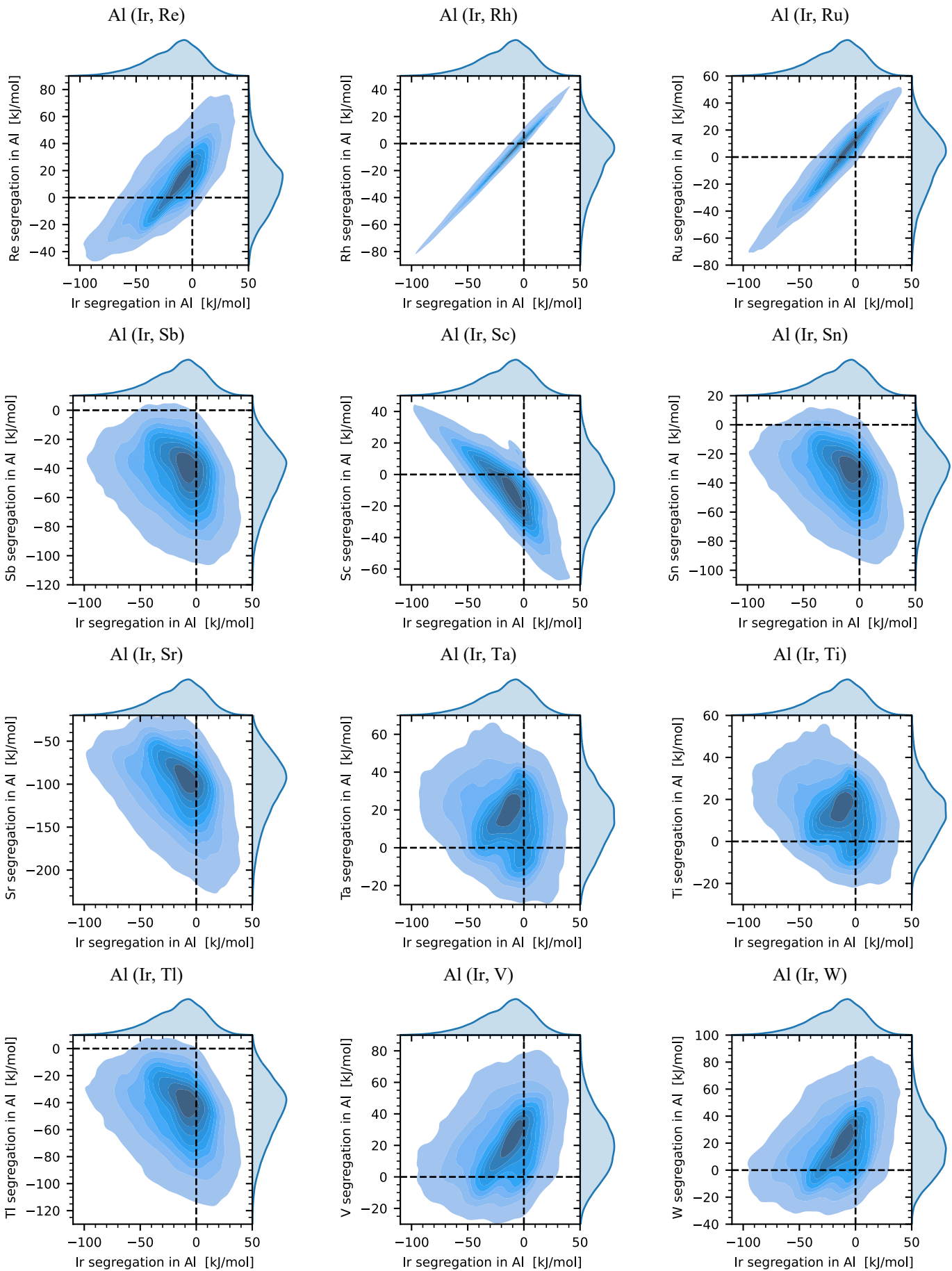

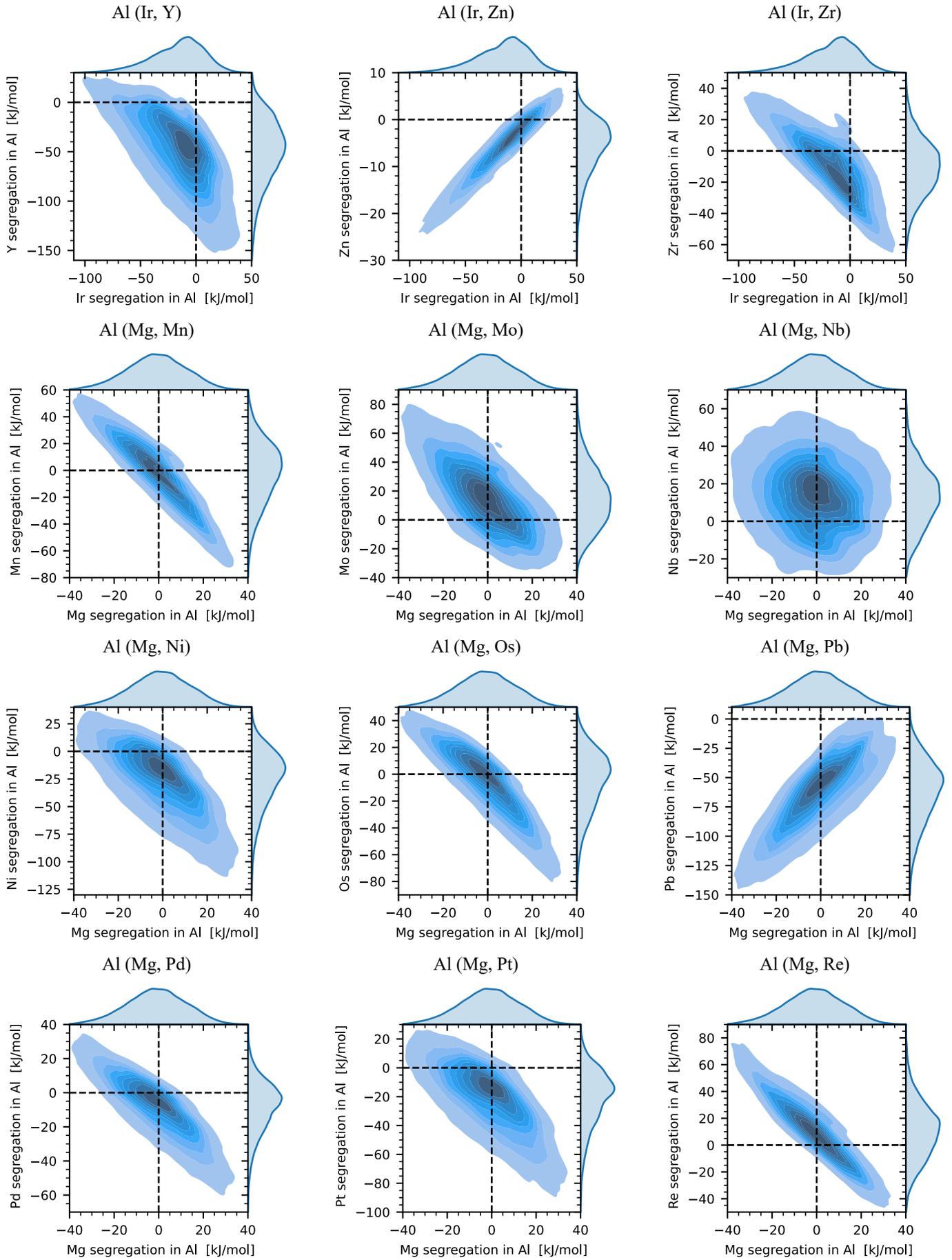

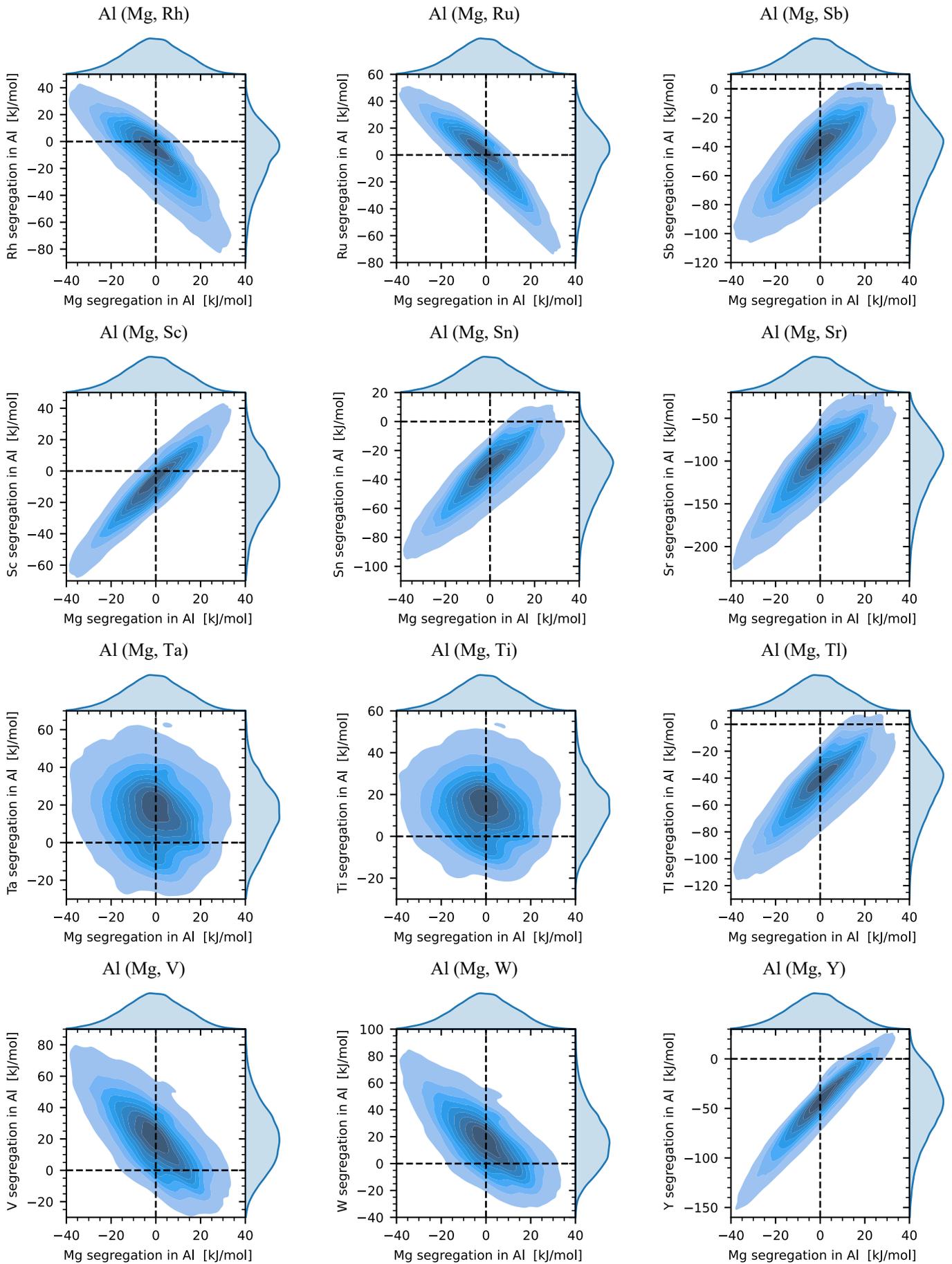

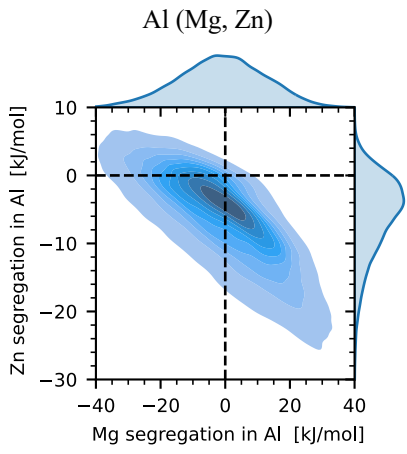

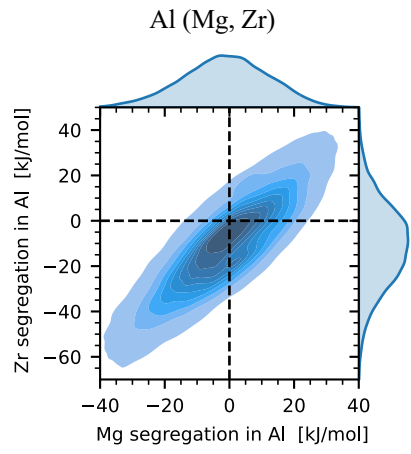

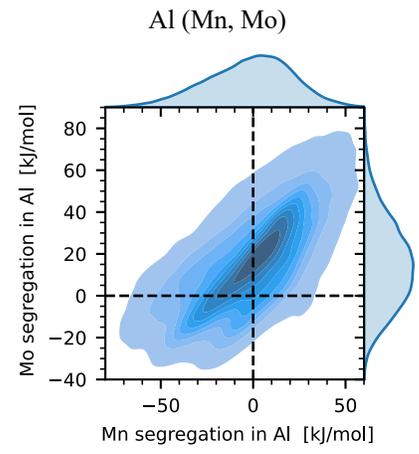

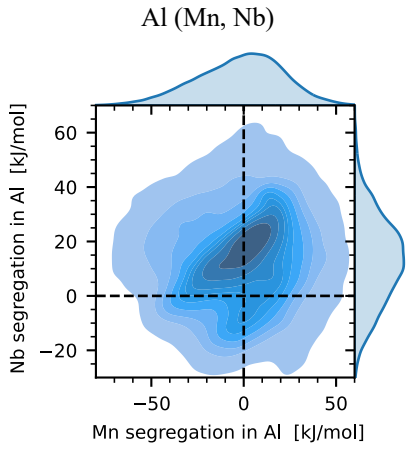

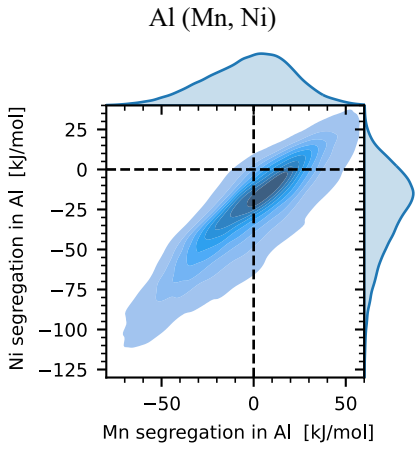

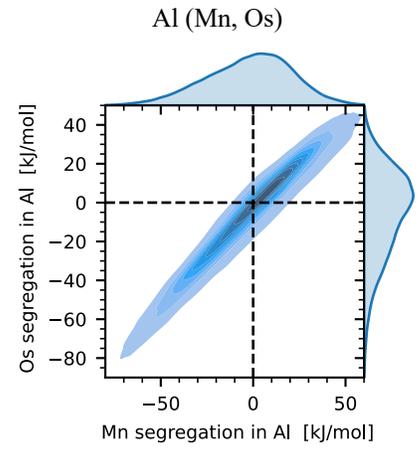

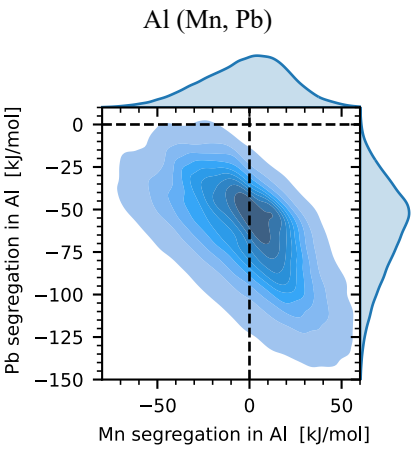

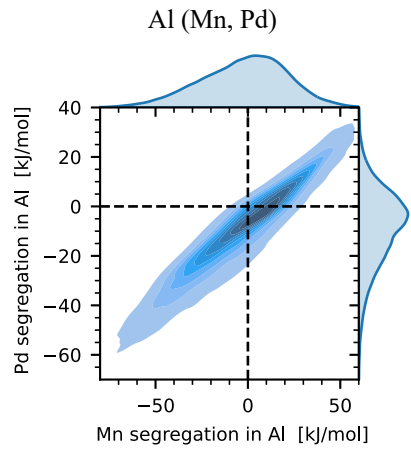

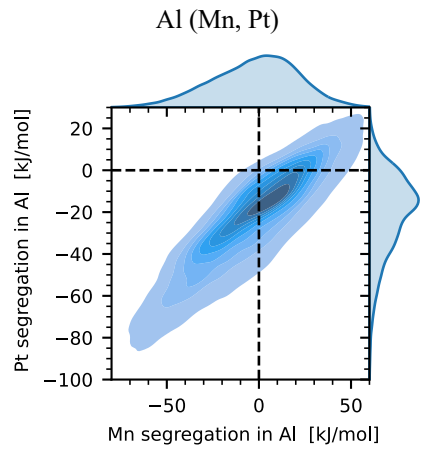

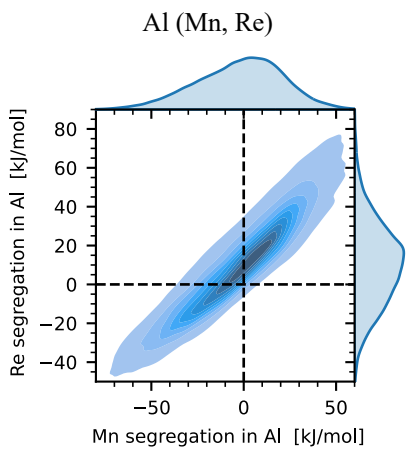

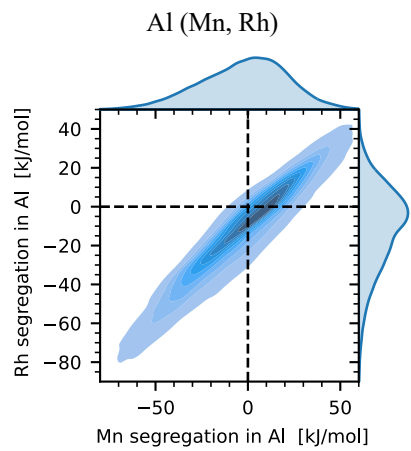

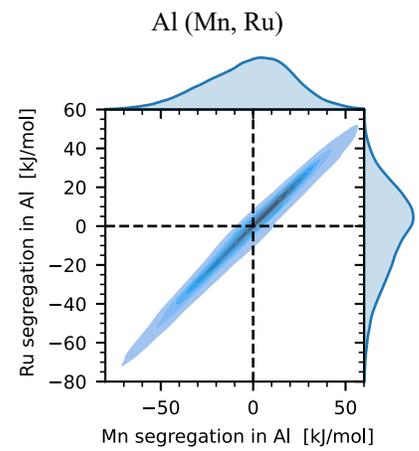

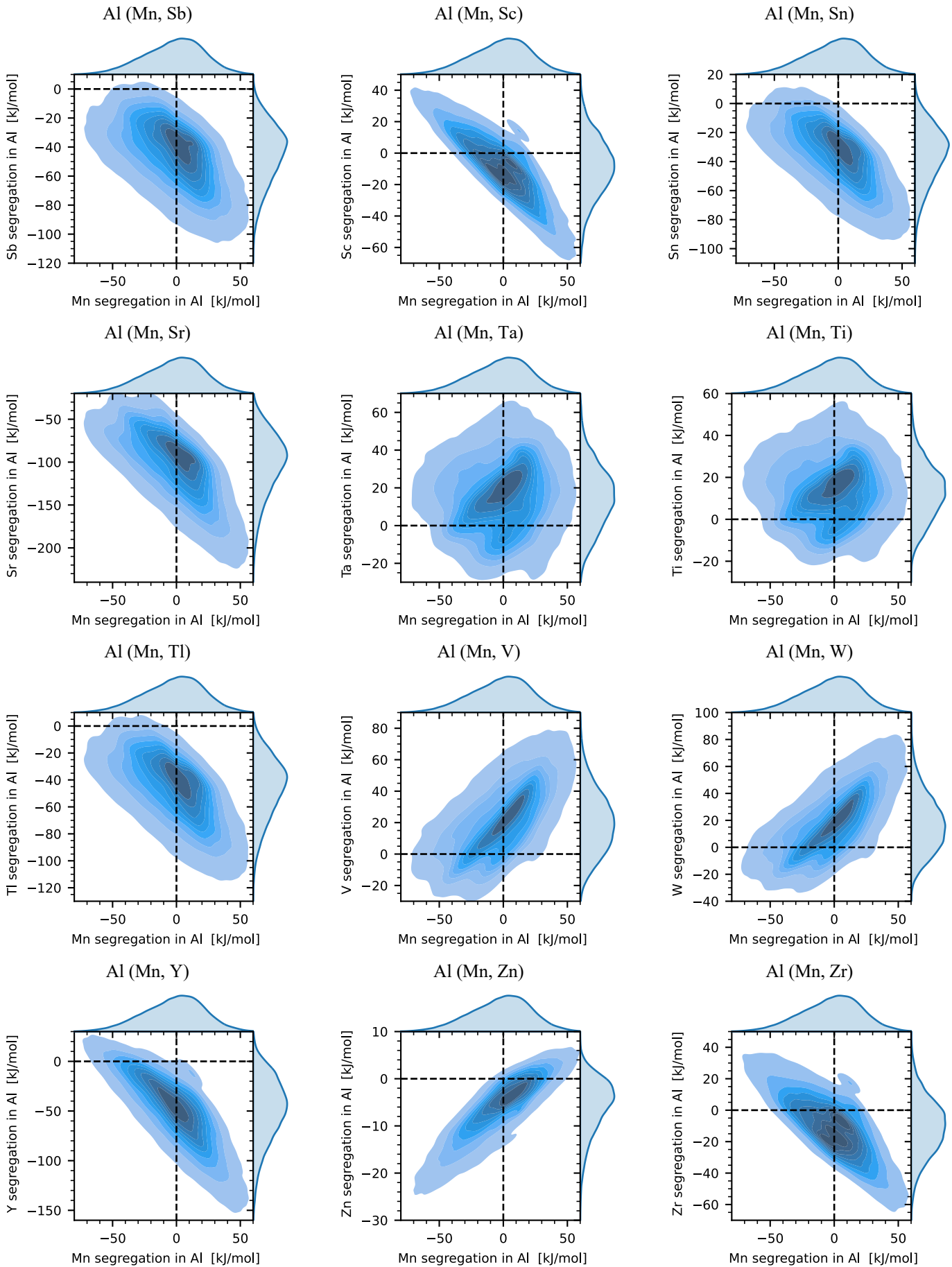

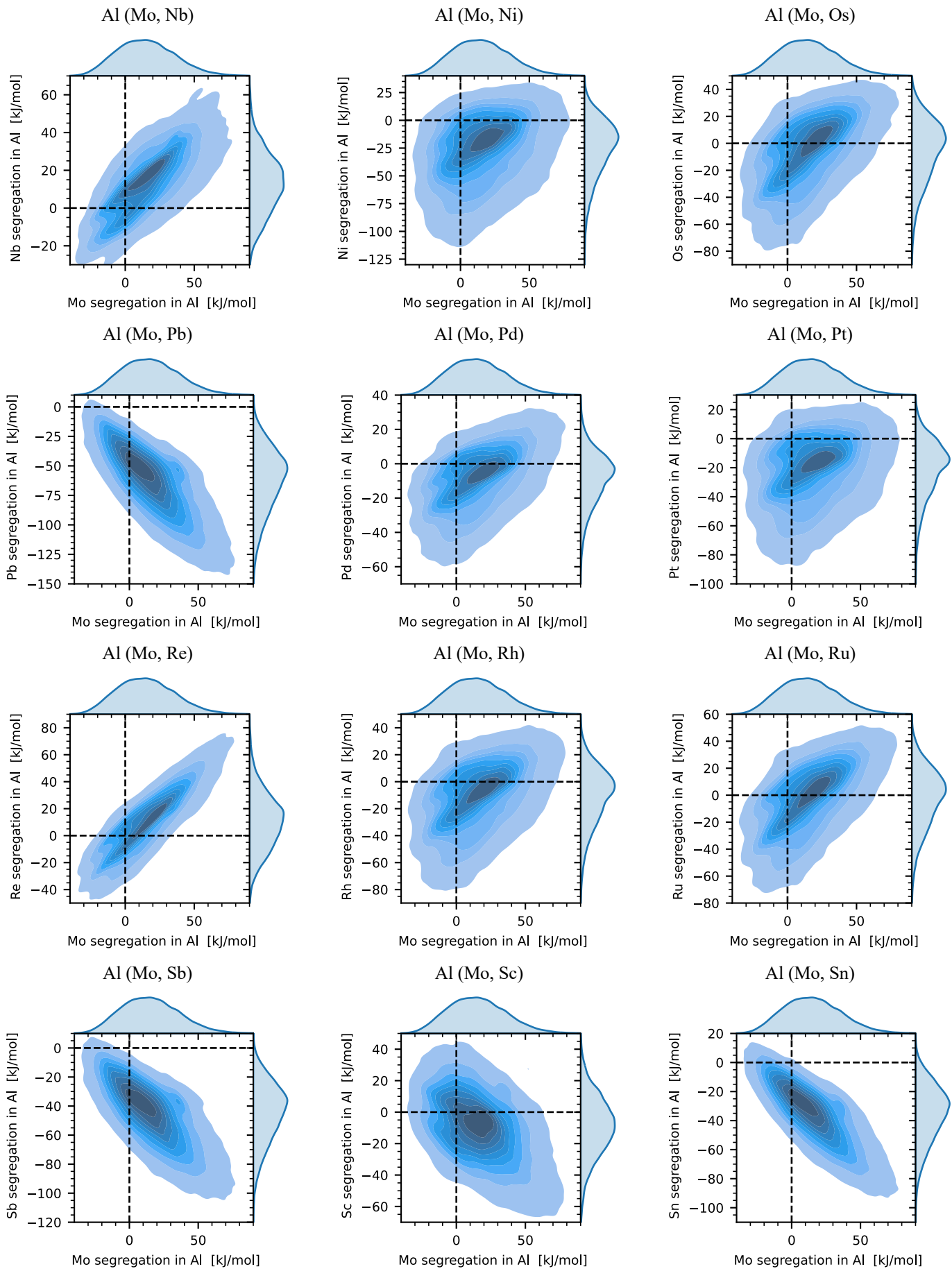

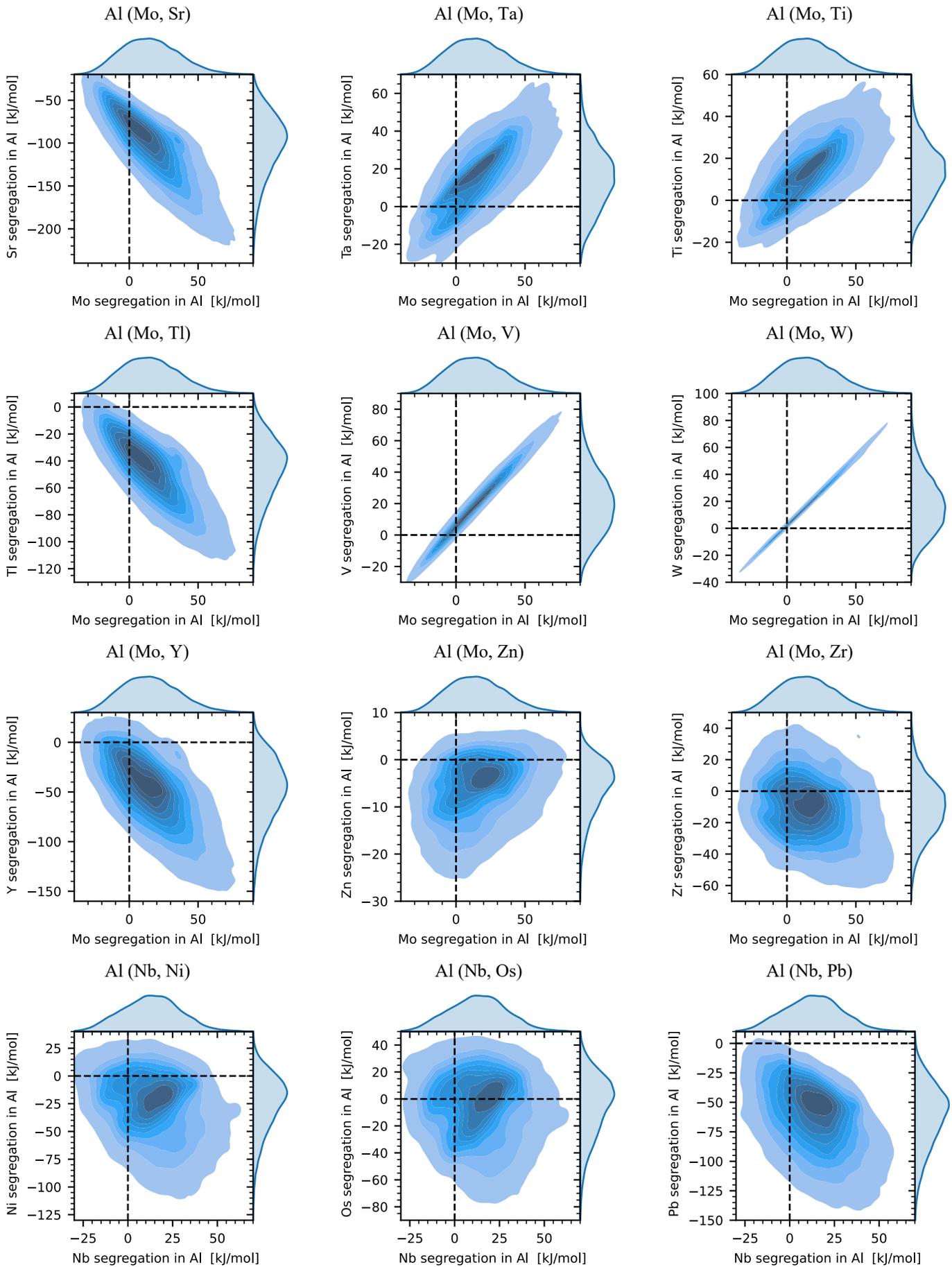

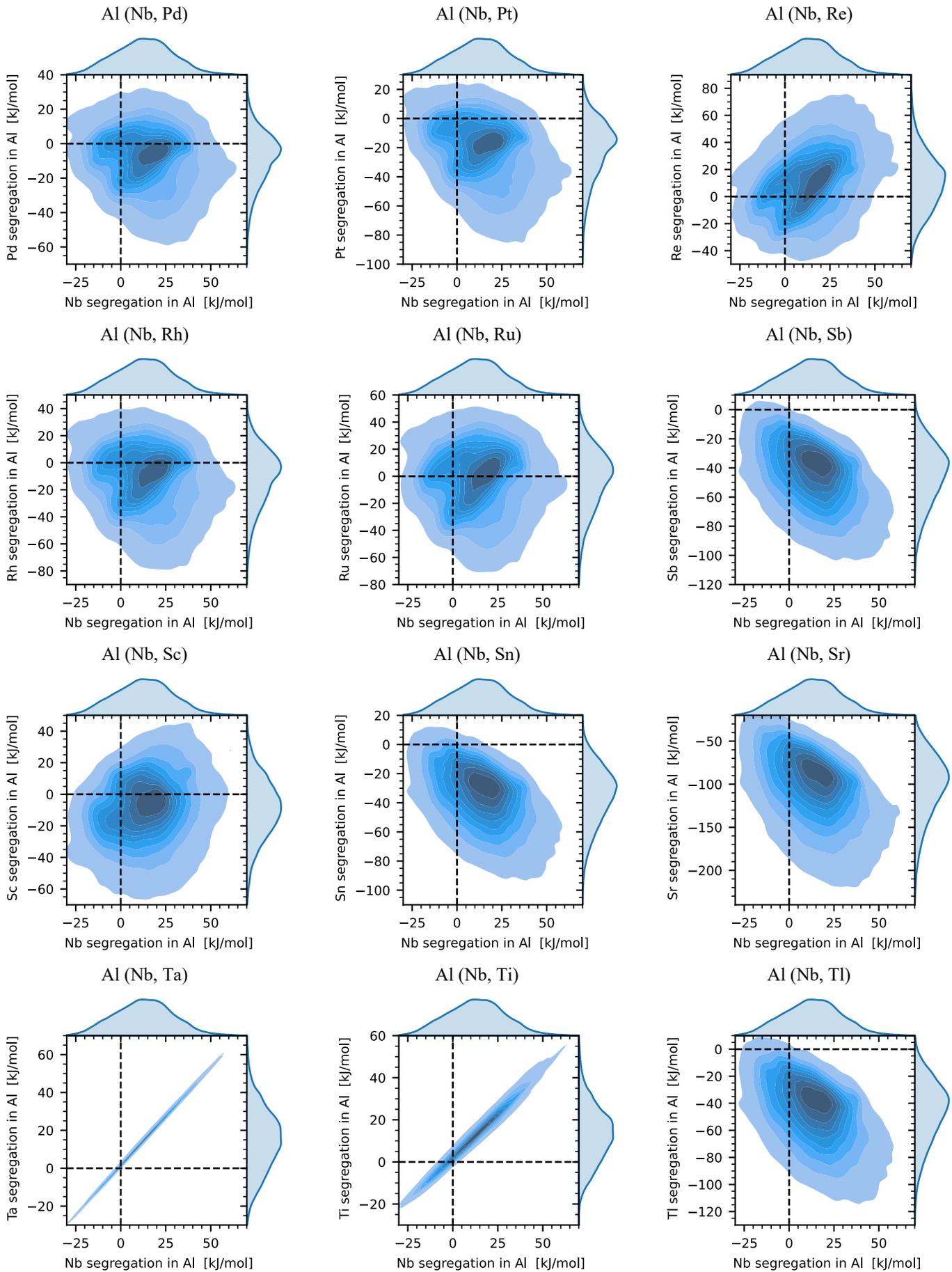

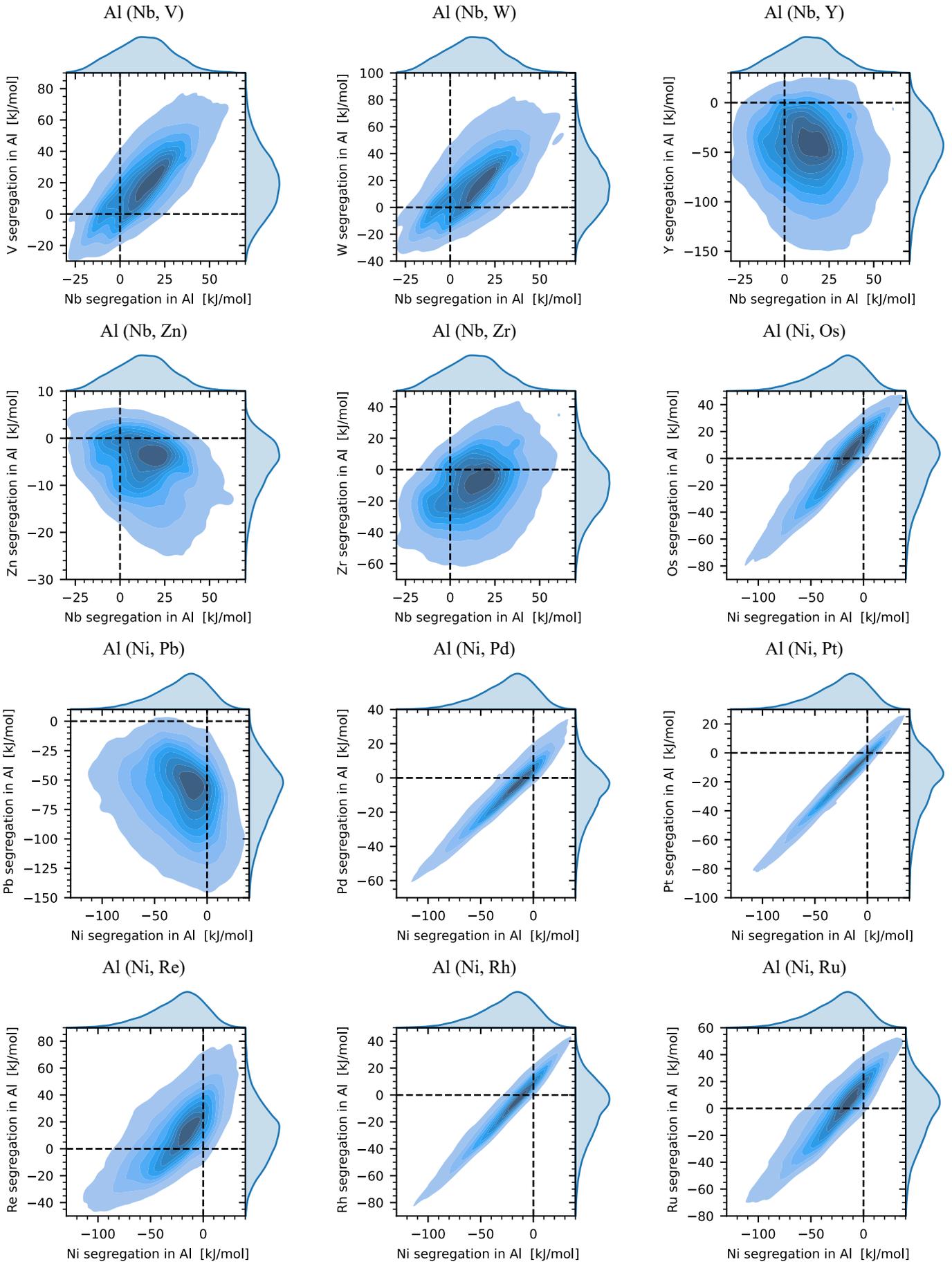

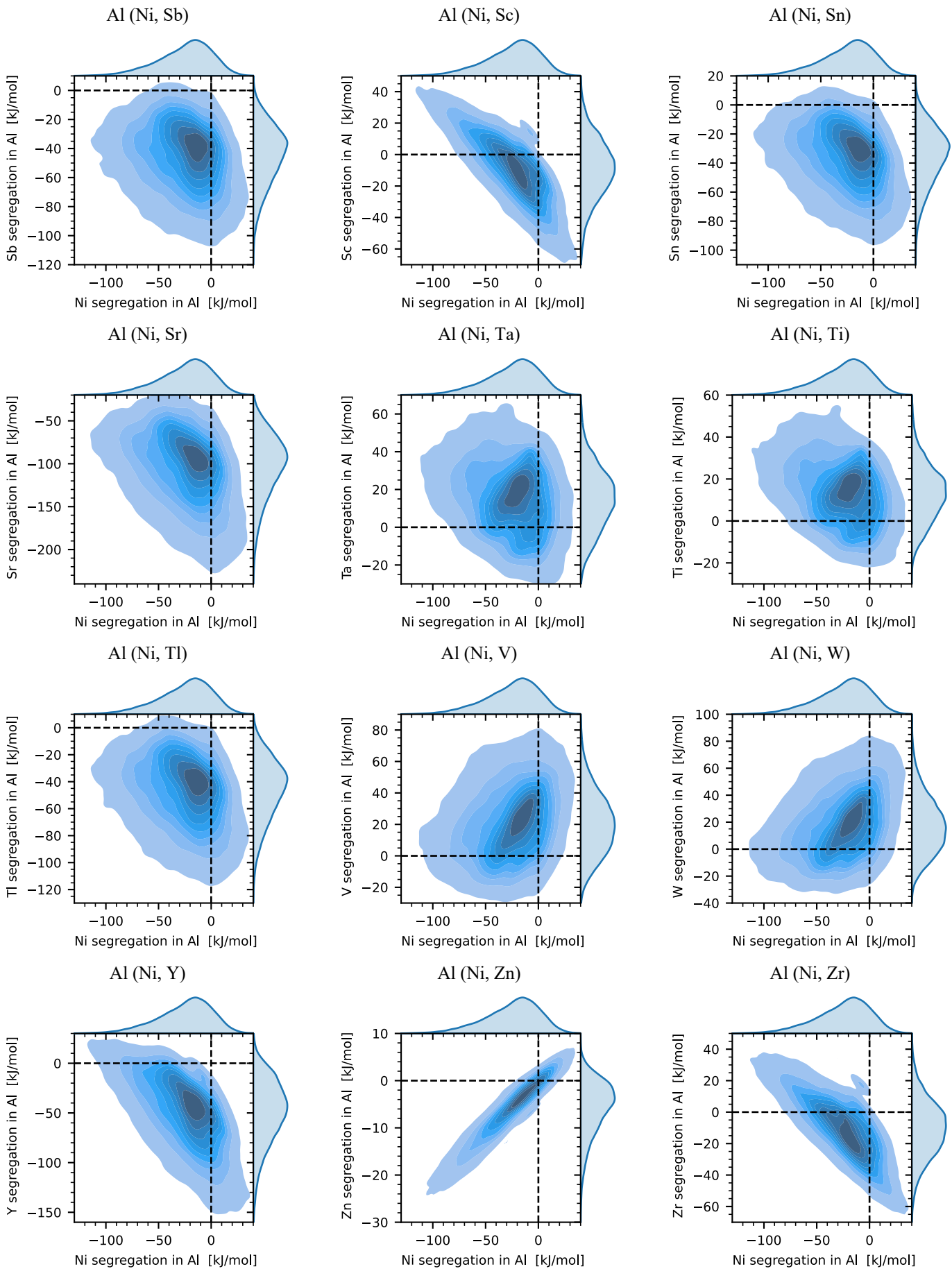

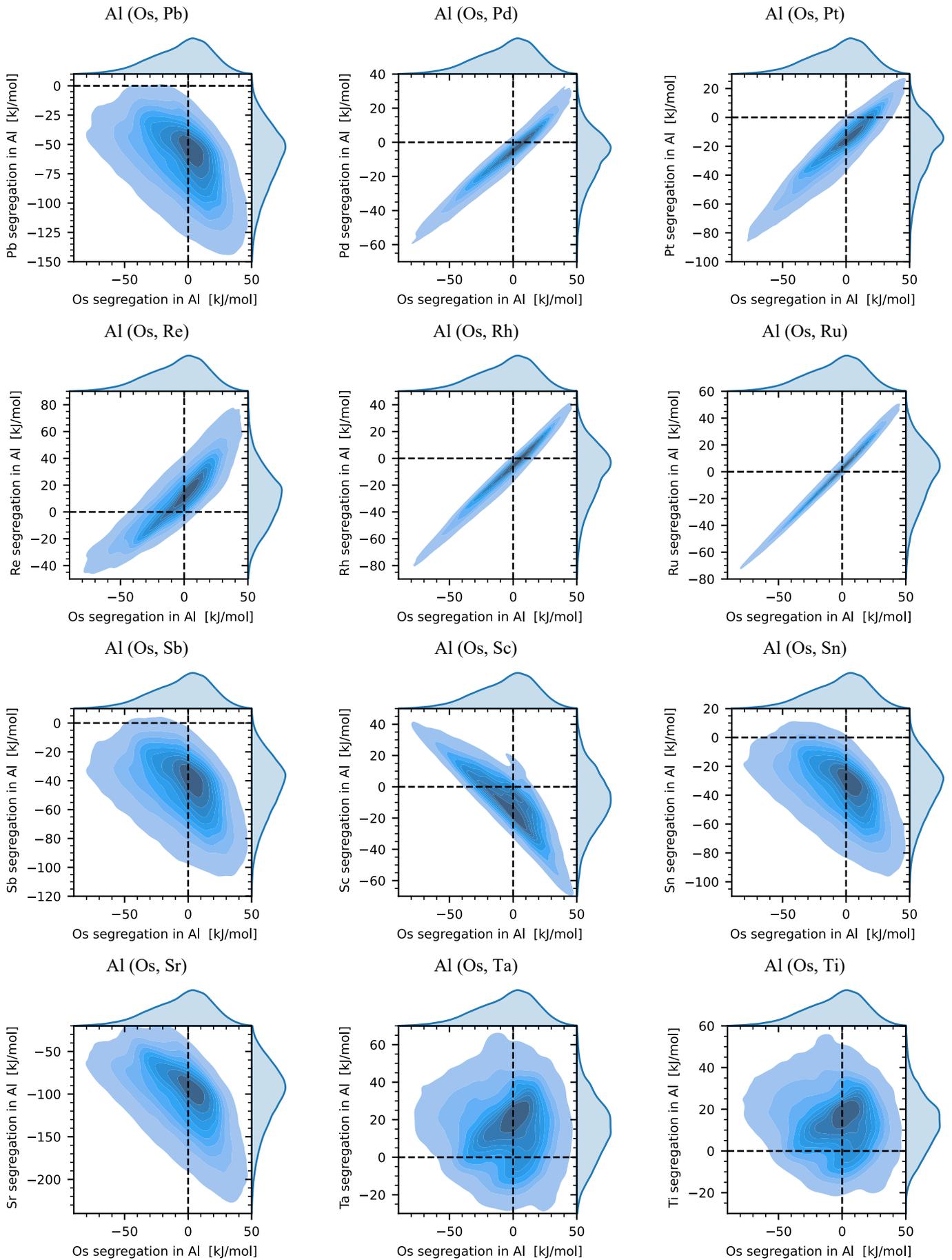

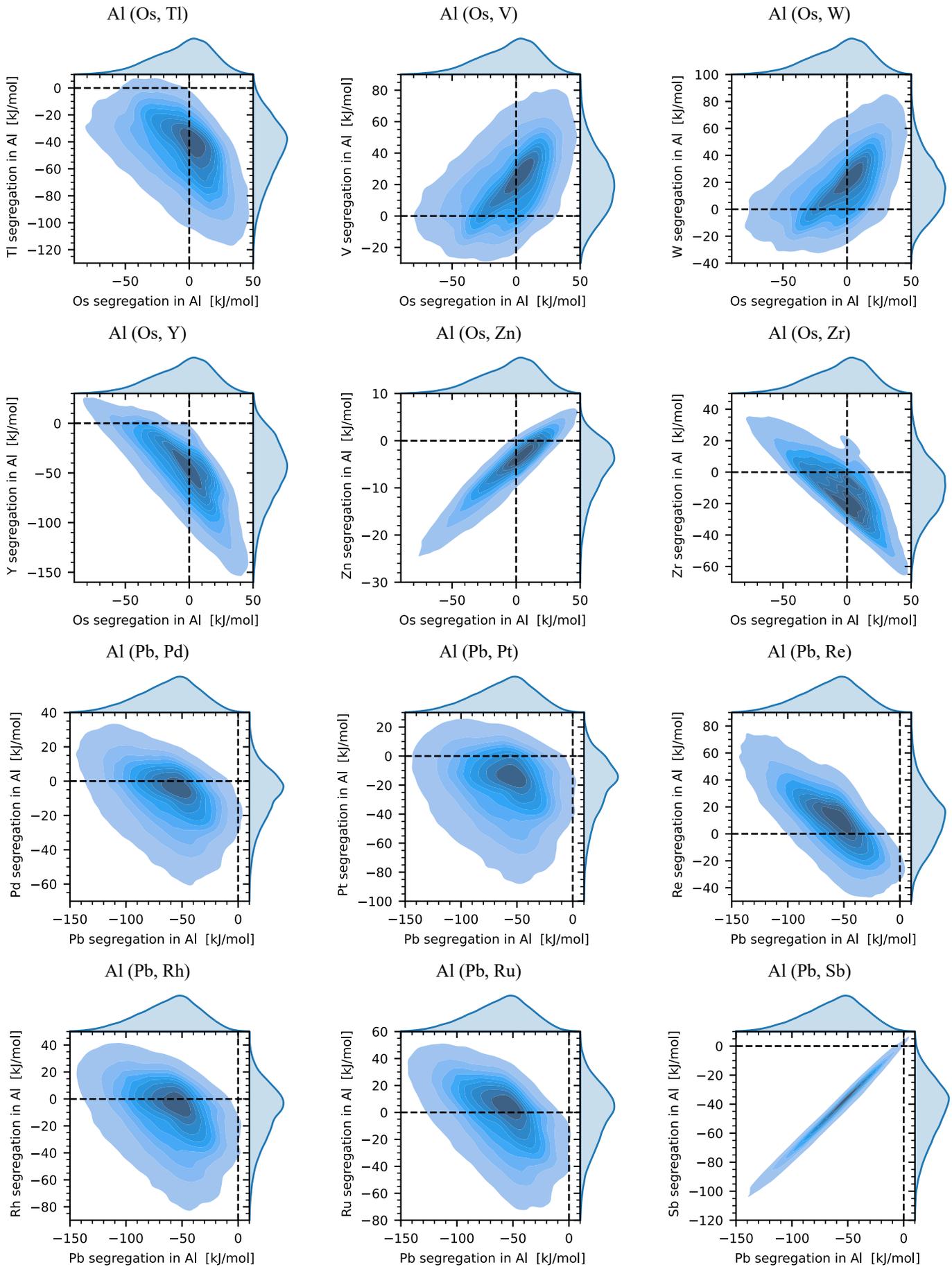

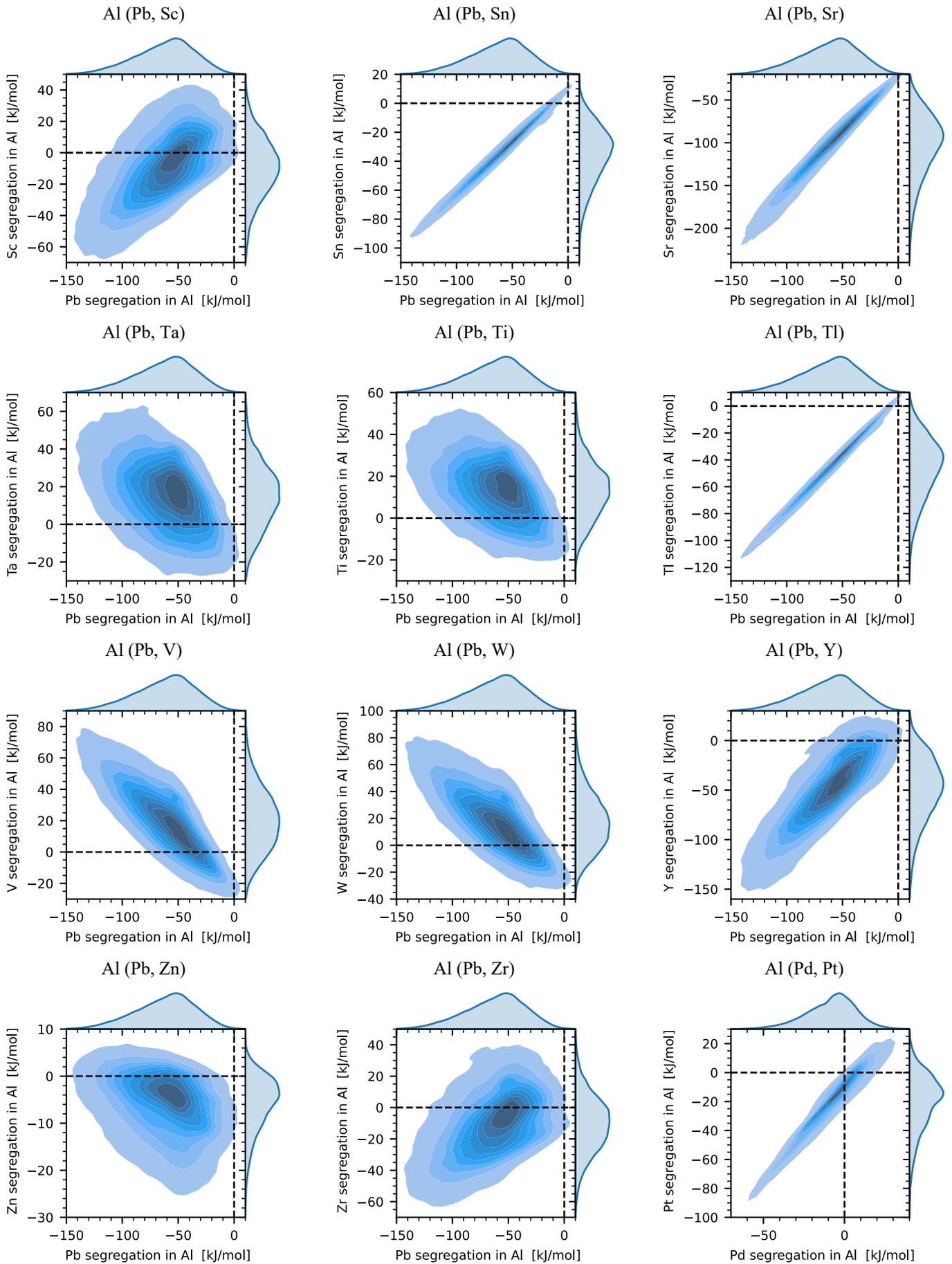

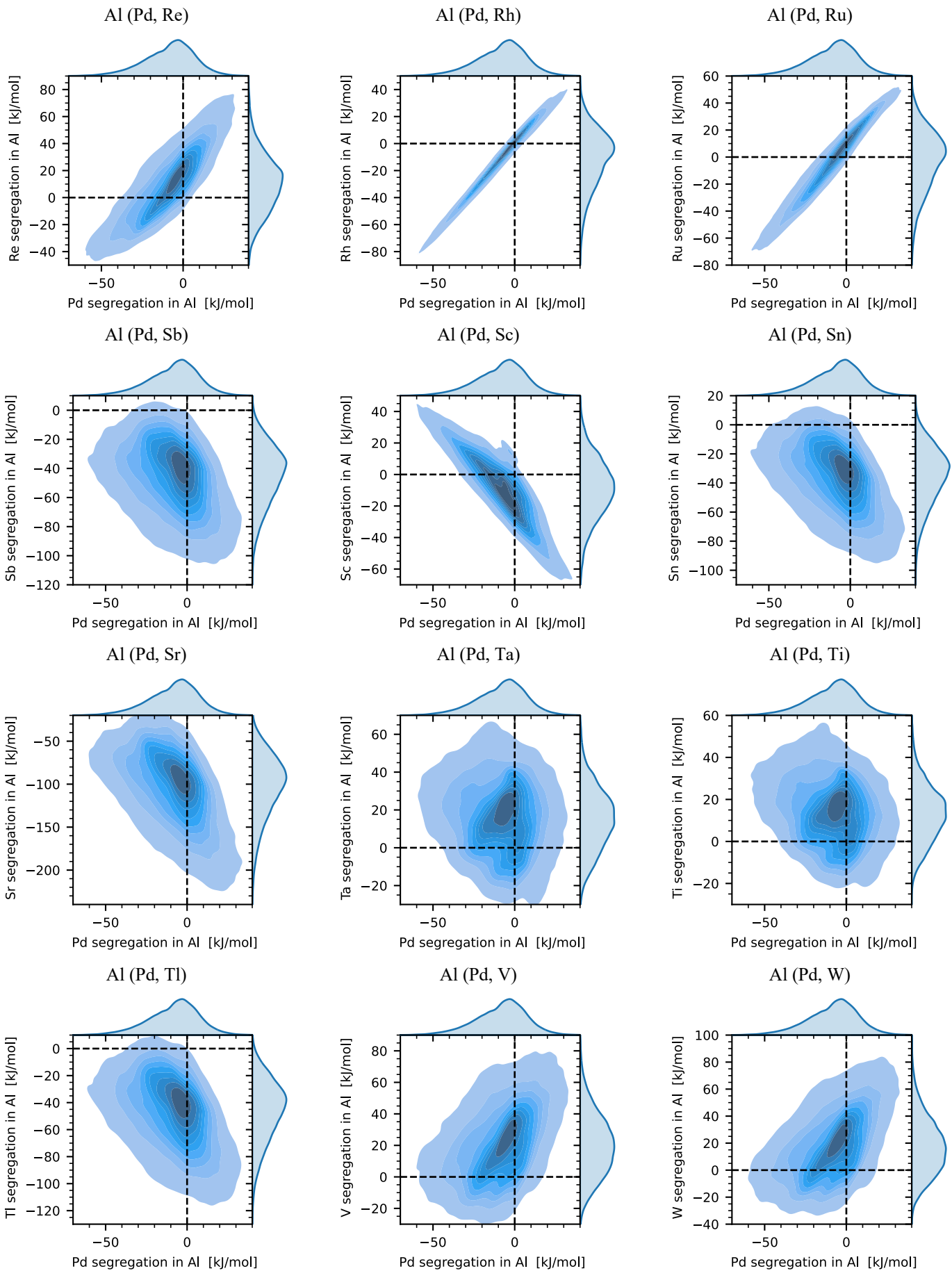

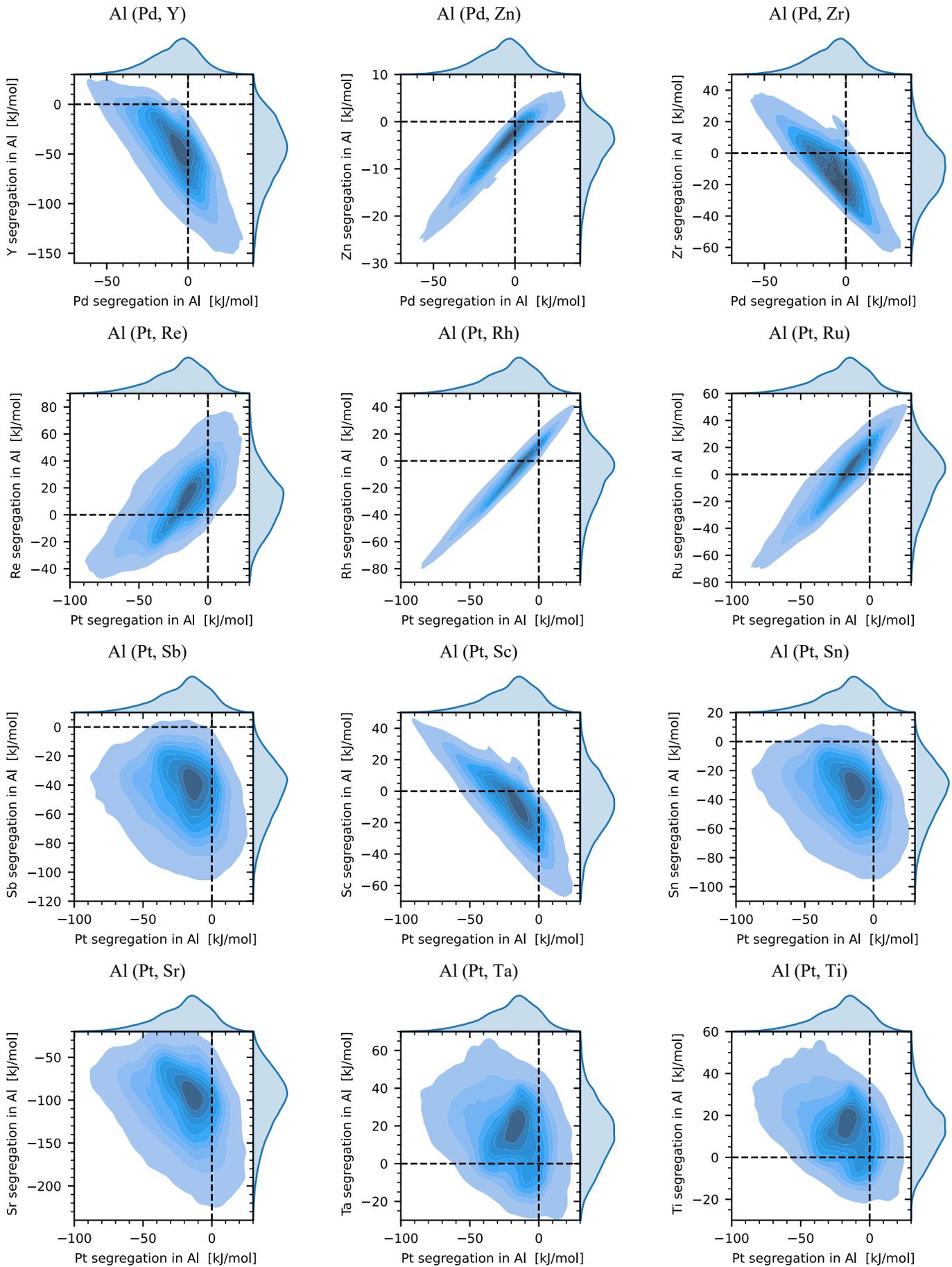

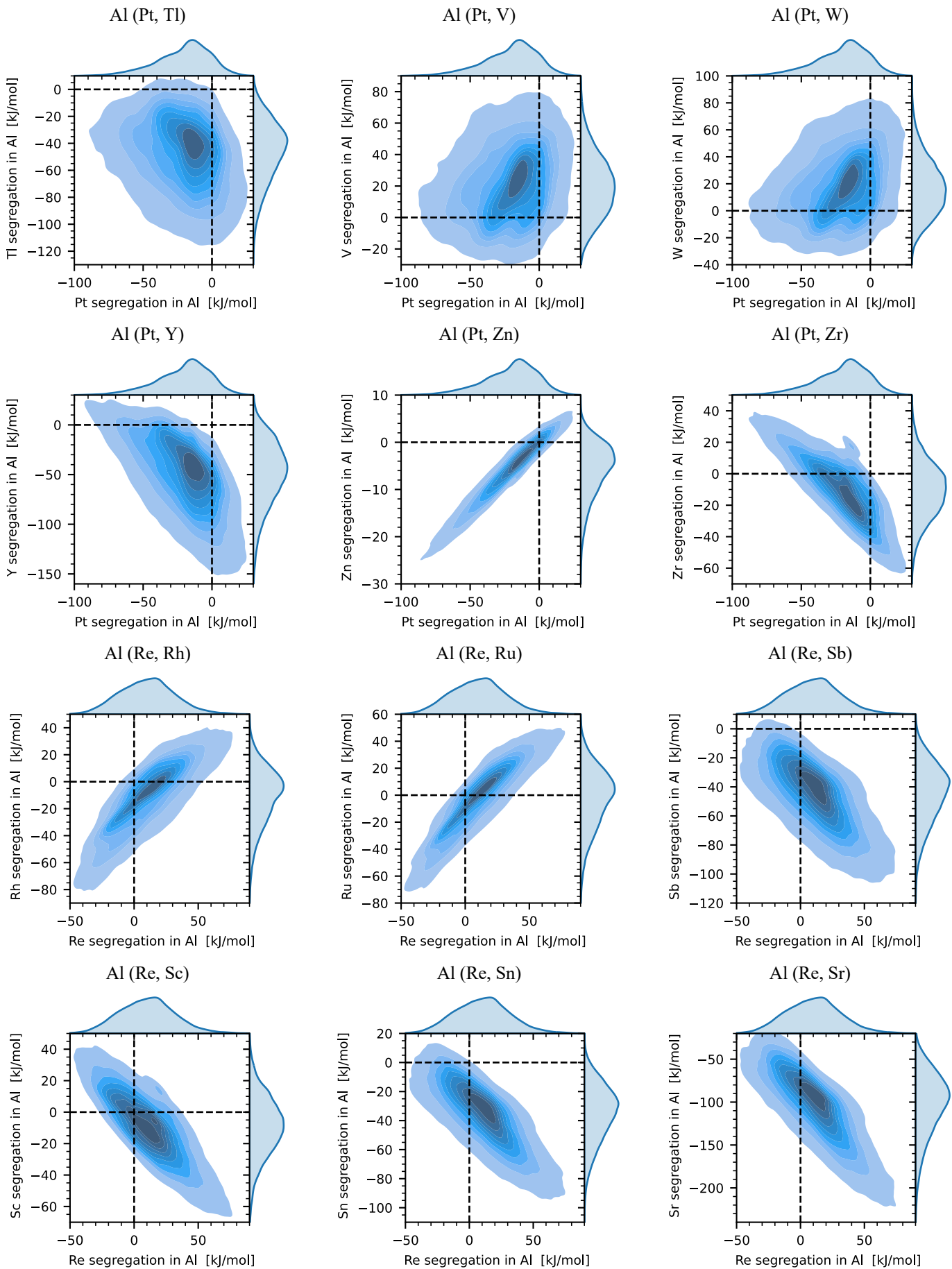

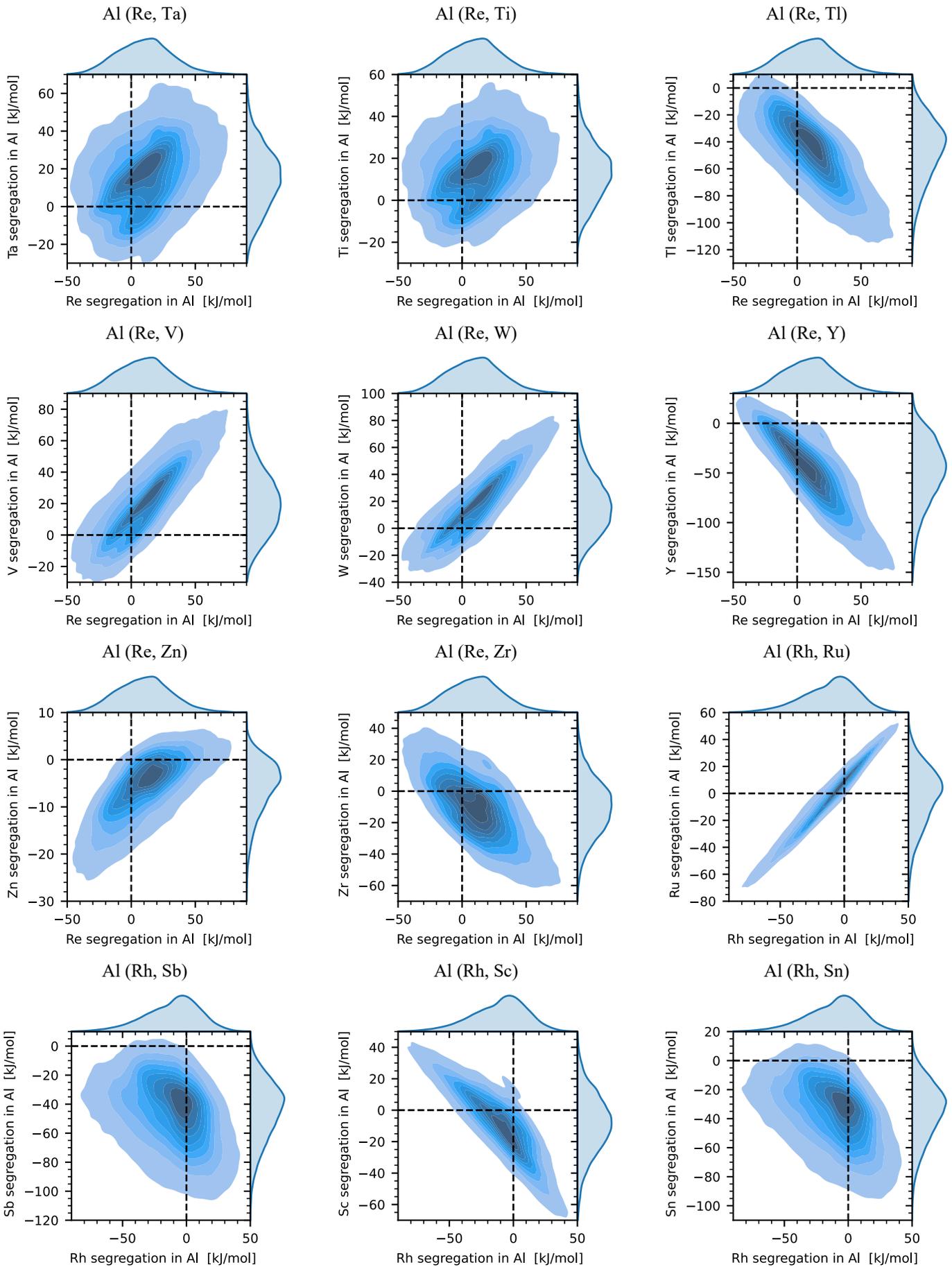

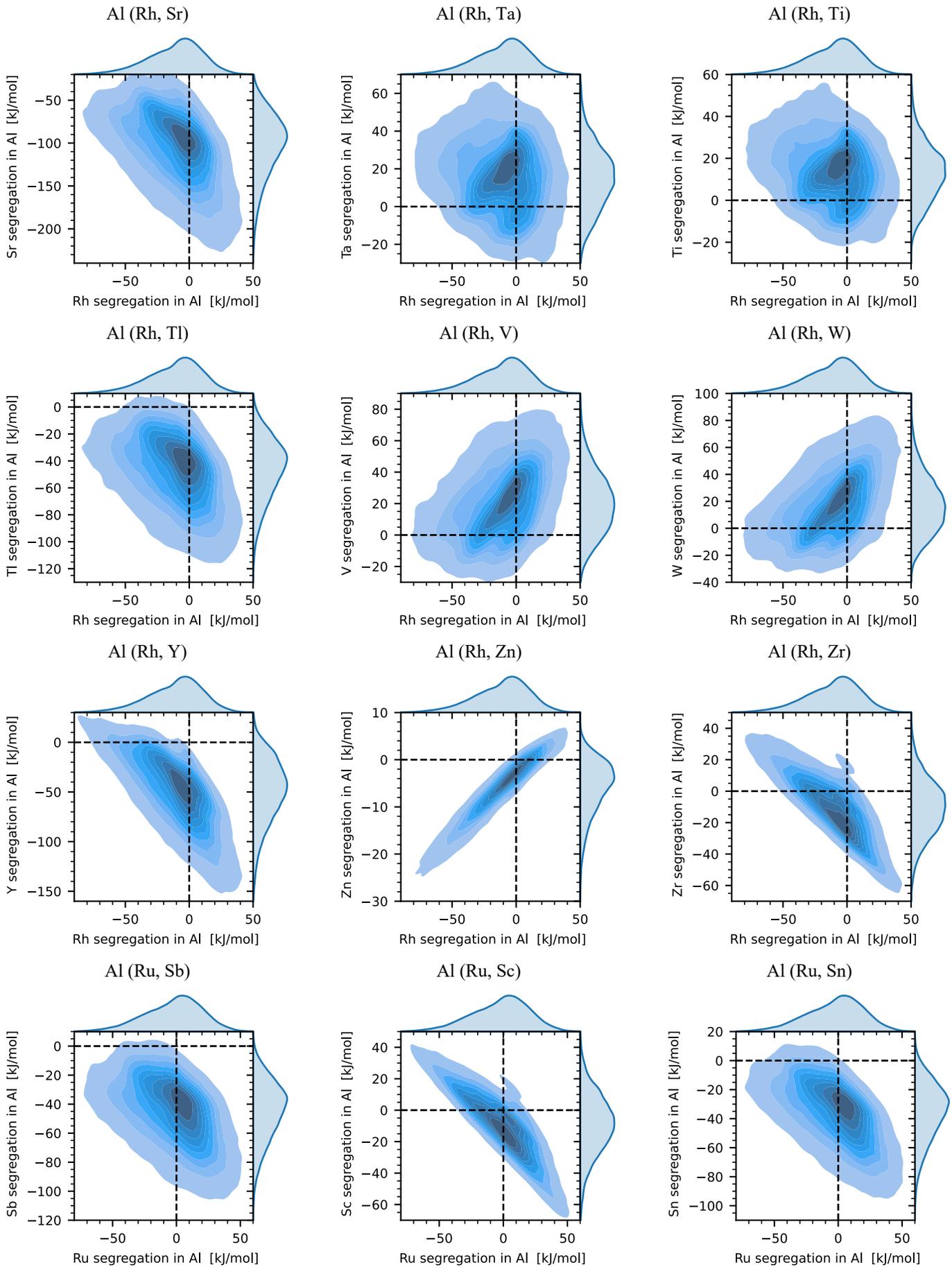

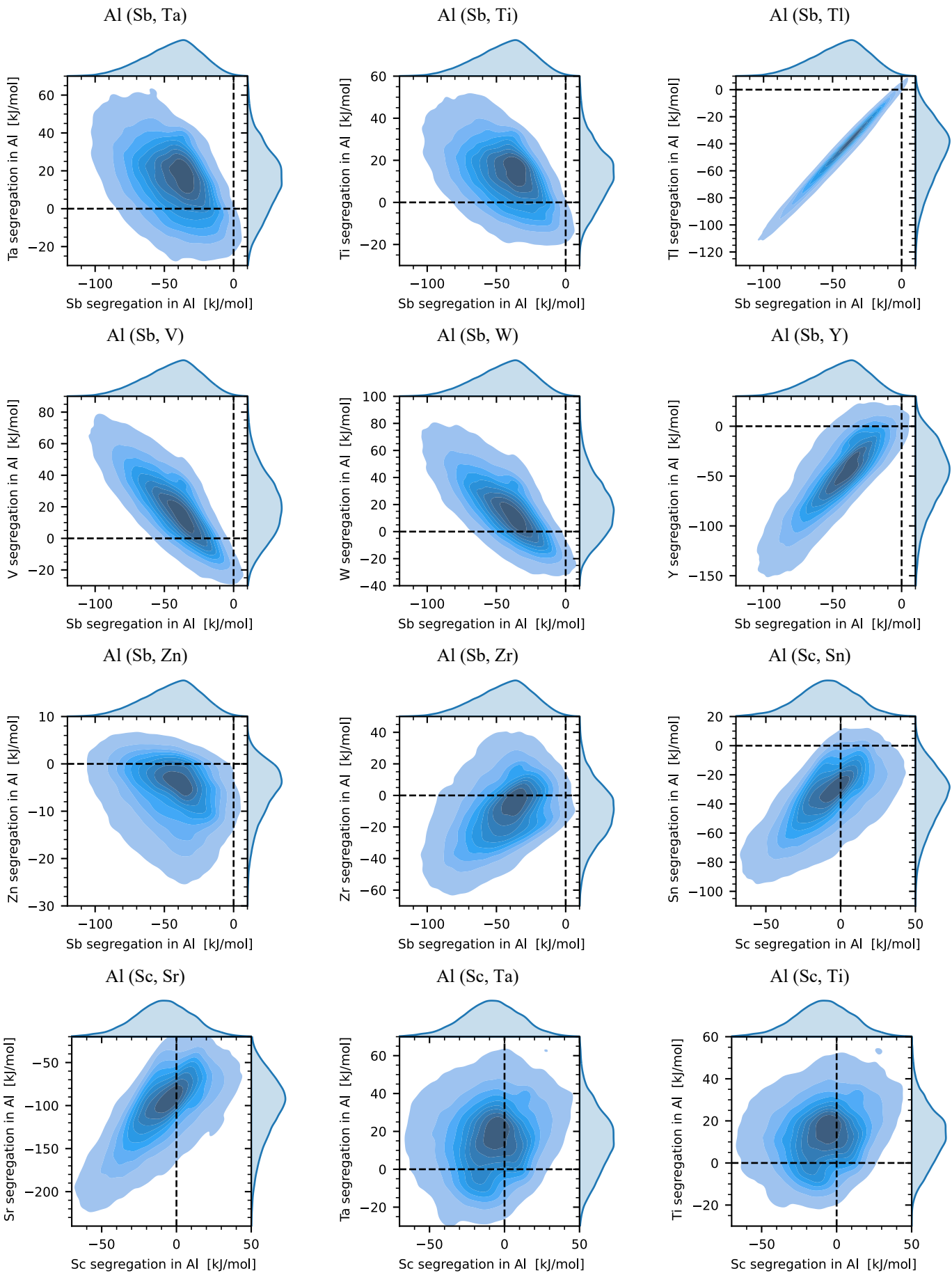

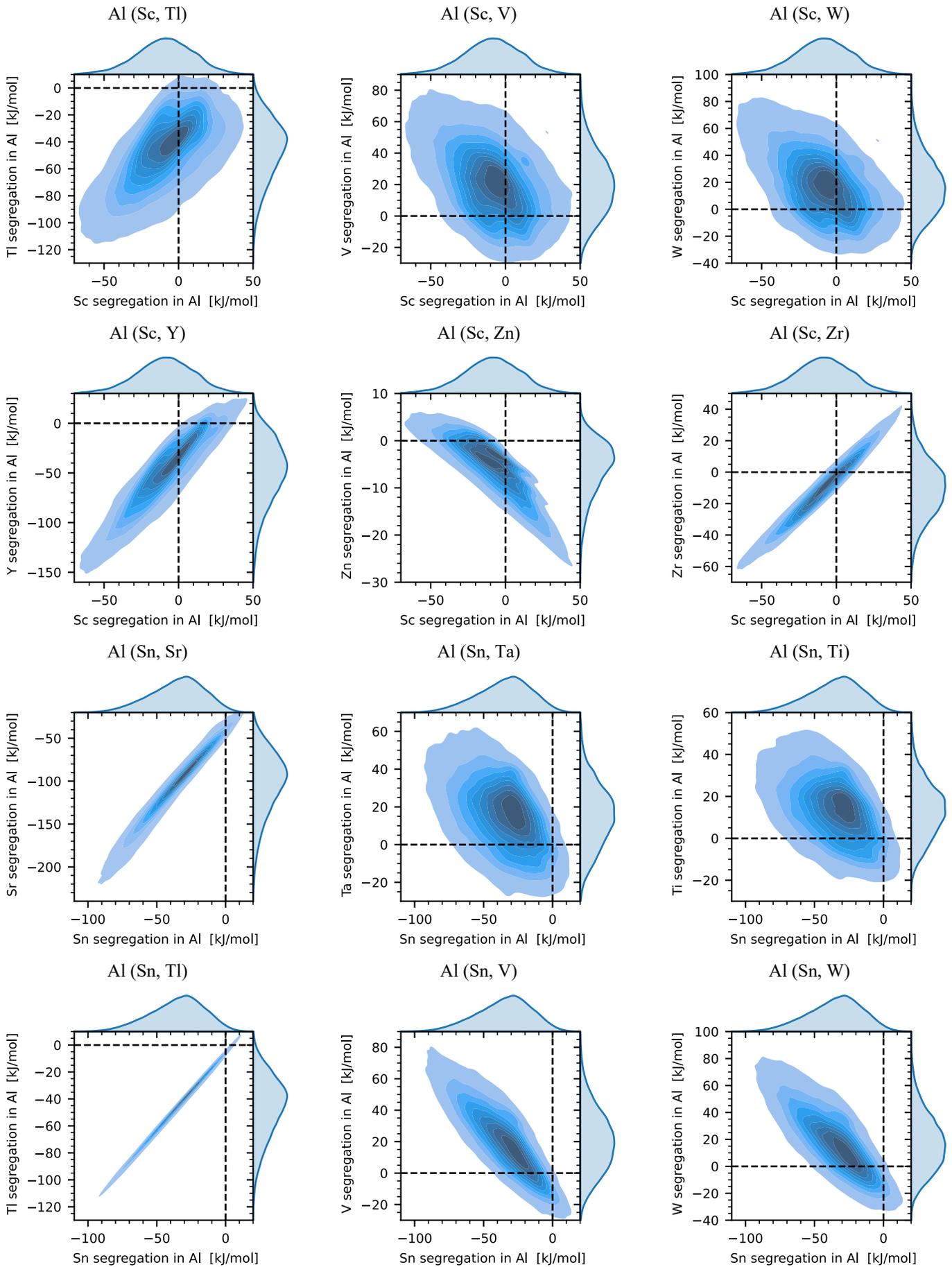

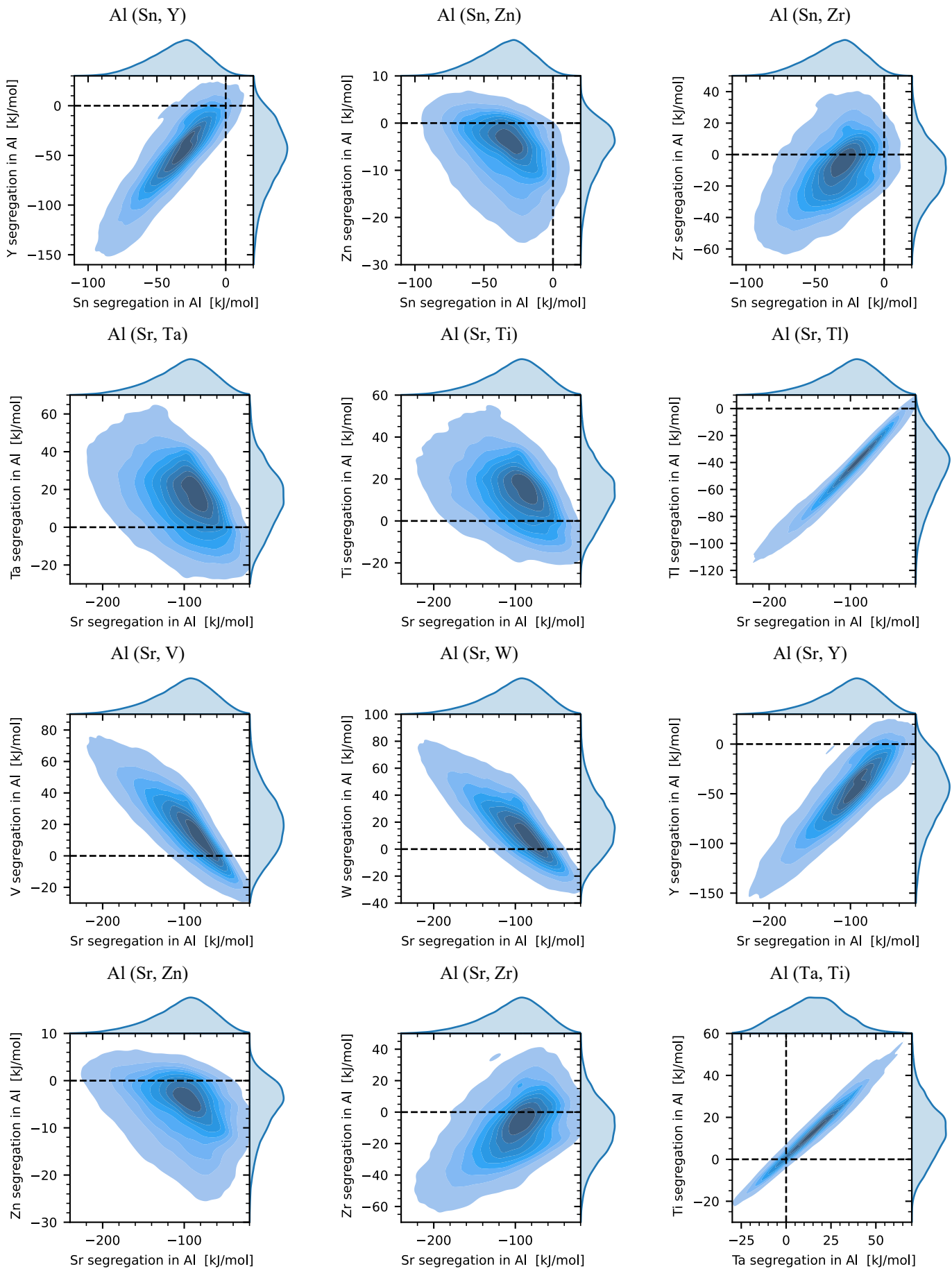

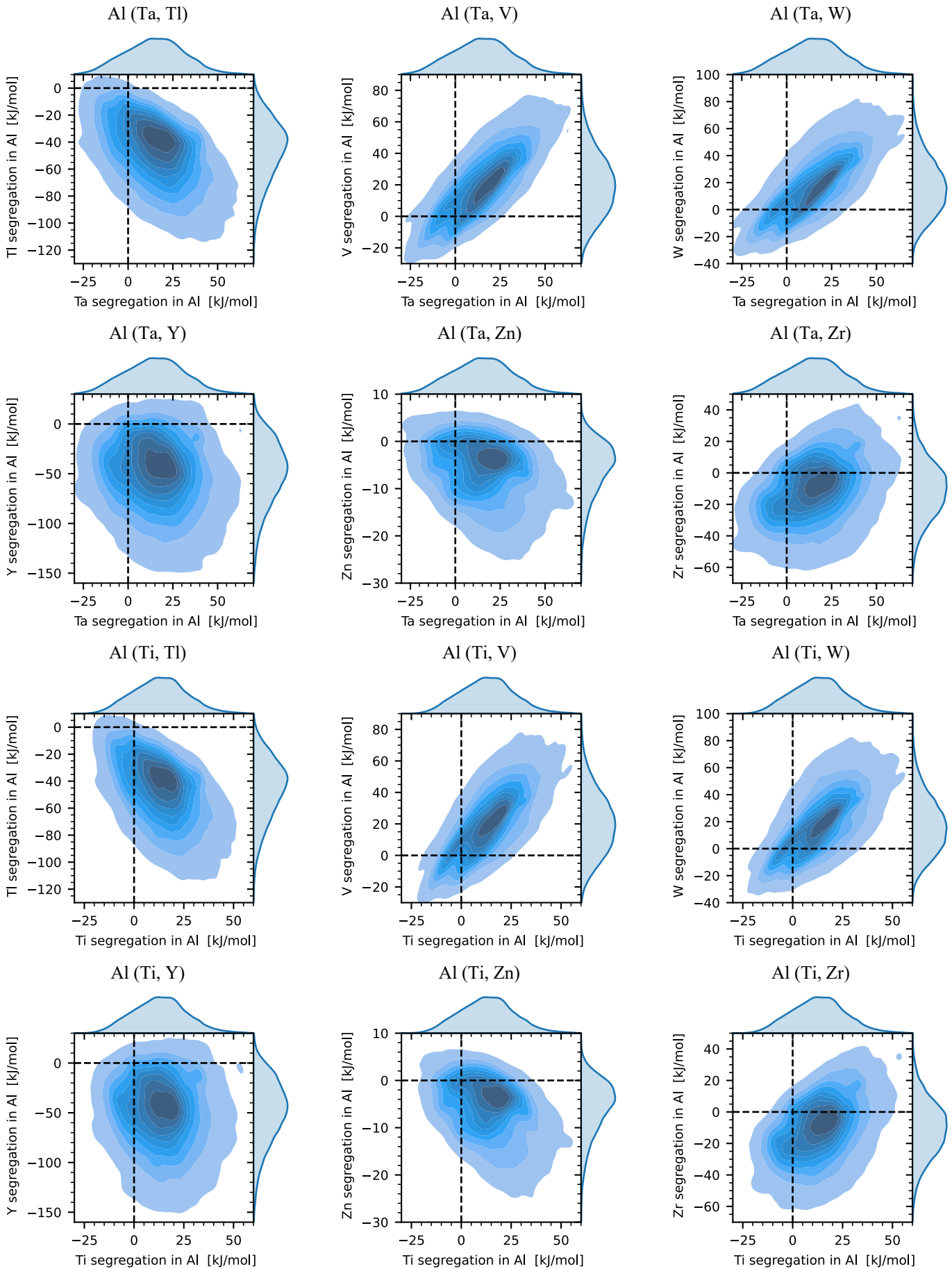

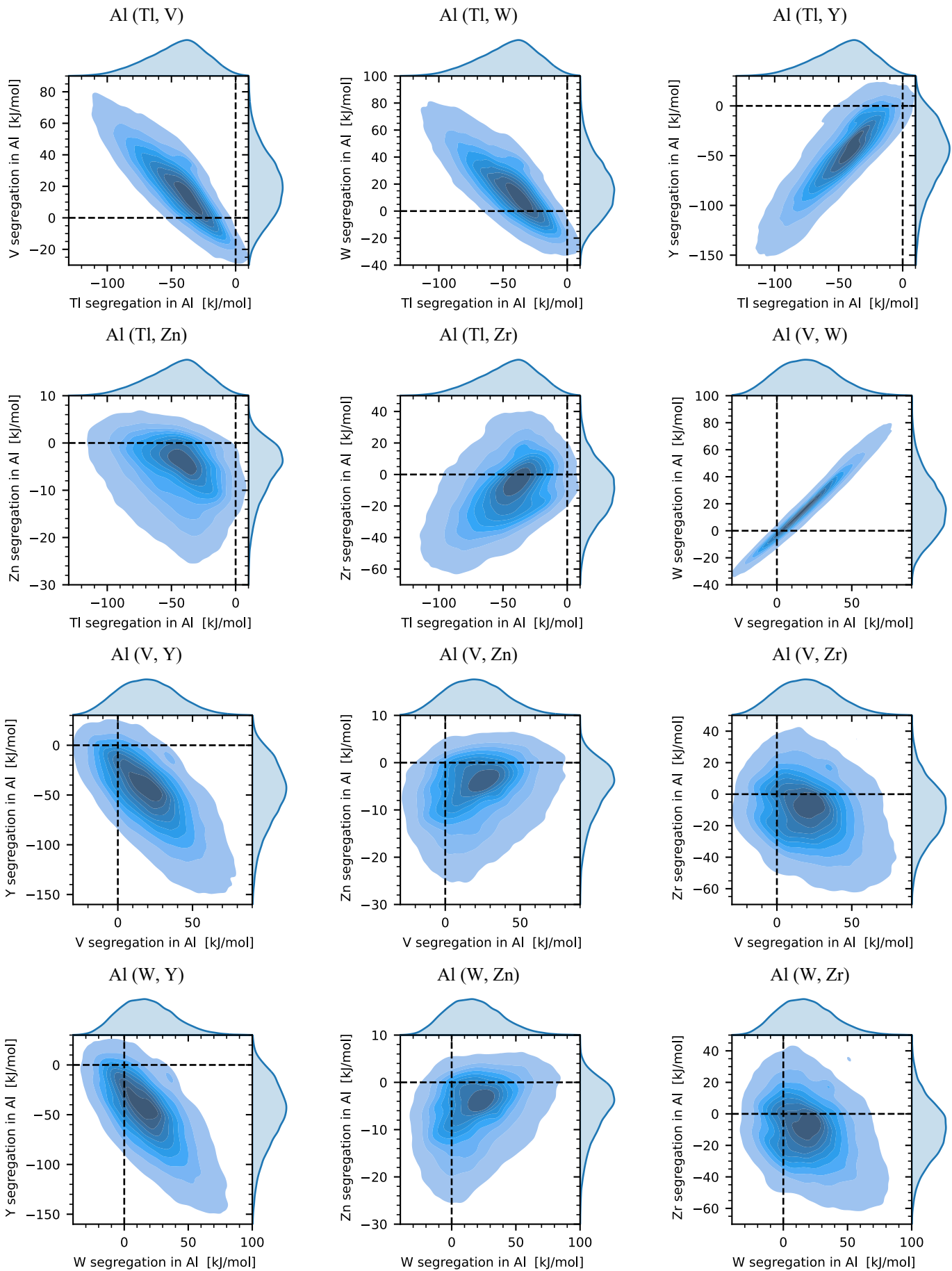

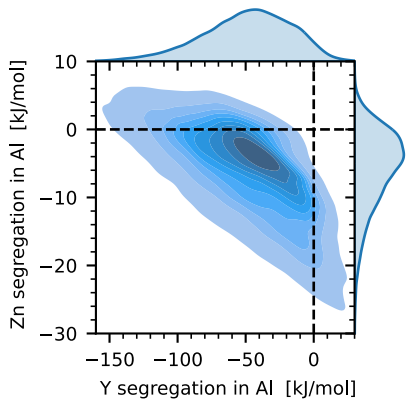
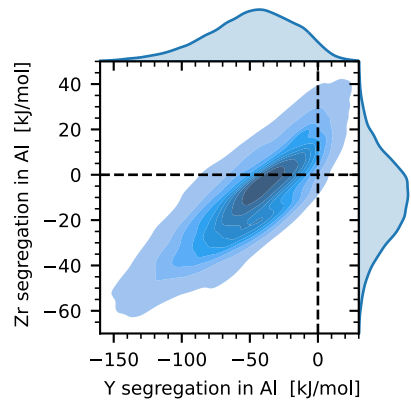
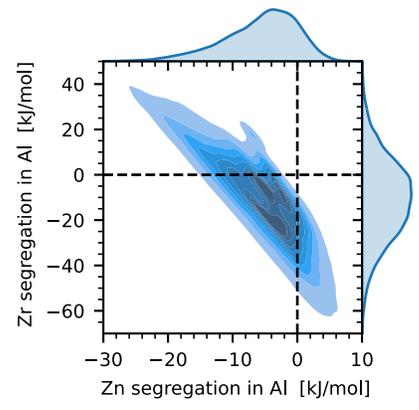